\documentclass[aps,10pt,twocolumn,tightenlines,nofootinbib,floatfix]{revtex4-1}
\usepackage{graphicx}
\usepackage[space]{grffile}
\usepackage{latexsym}
\usepackage{amsfonts,amsmath,amssymb}
\usepackage{url}
\usepackage[utf8]{inputenc}
\usepackage{hyperref}
\hypersetup{colorlinks=false,pdfborder={0 0 0}}
\usepackage{textcomp}
\usepackage{longtable}
\usepackage{multirow,booktabs}
\usepackage[utf8]{inputenc} 

\newcommand{\taon}{$\tau$-lepton }
\newcommand{\taons}{$\tau$-leptons }

\newcommand{\lsim}{\mathrel{\hbox{\rlap{\lower.75ex \hbox{$\sim$}} \kern-.3em \raise.4ex \hbox{$<$}}}}
\newcommand{\gsim}{\mathrel{\hbox{\rlap{\lower.75ex \hbox{$\sim$}} \kern-.3em \raise.4ex \hbox{$>$}}}}

\begin{document}

\title{Cosmic tau neutrino detection via Cherenkov signals from air showers from Earth-emerging taus}

\author{Mary Hall Reno}
\affiliation{Department of Physics and Astronomy, University of Iowa, Iowa City, IA 52242, USA}

\author{John F. Krizmanic}
\affiliation{CRESST/NASA Goddard Space Flight Center, Greenbelt, MD 20771, USA \\
University of Maryland, Baltimore County, Baltimore, MD 21250, USA}

\author{Tonia M. Venters}
\affiliation{NASA Goddard Space Flight Center, Greenbelt, MD 20771, USA}

\date{\today}
\begin{abstract}
We perform a new, detailed calculation of the flux and energy spectrum of Earth-emerging $\tau$-leptons generated from the interactions of tau neutrinos and antineutrinos in the Earth. A layered model of the Earth 
is used to describe the variable density profile of the Earth. Different assumptions regarding the neutrino charged- and neutral-current cross sections as well as the $\tau$-lepton energy loss models are used to quantify their contributions to the systematic uncertainty. A baseline simulation is then used to generate the optical Cherenkov signal from upward-moving extensive air showers generated by the $\tau$-lepton decay in the atmosphere, applicable to a range of space-based instruments.  We use this simulation to determine the neutrino sensitivity for $E_\nu \gsim$ 10 PeV for a space-based experiment with performance similar to that for the Probe of Extreme MultiMessenger Astrophysics (POEMMA) 
mission currently under study. 

\end{abstract}

\maketitle 

\section{Introduction}

The measurement of the spectrum of the very-high energy (VHE: $E_\nu \gsim 1$ PeV) neutrino  and antineutrino (hereafter denoted collectively as neutrinos) component of the cosmic radiation and its angular distribution on the sky provides a unique probe of high-energy astrophysical phenomena. A by-product of cosmic ray acceleration, astrophysical
neutrinos can reveal the environments of sources of ultra-high energy cosmic rays (UHECR) \cite{Gaisser:1994yf,IceCube2017}.
The results from IceCube \citep{Aartsen:2013jdh,Aartsen:2014gkd,IceCube2015} demonstrate the existence of an extra-solar system astrophysical neutrino flux with energies from above 10 TeV to potentially as high as 10 PeV.  A neutrino event detected by IceCube that appears correlated at $3\sigma$ with gamma-flaring data from an active galactic nuclei source has been reported \cite{IceCube:2018cha}.
Gamma ray bursts \cite{Murase:2007yt}, newborn pulsars \cite{Fang:2013vla}, active galactic nuclei
\cite{Murase:2015ndr}, galactic clusters with central sources
\cite{Murase:2008yt,Fang:2017zjf} and UHECR photodisintegration
within cosmic ray sources \cite{Unger:2015laa} are among candidates sources for the diffuse astrophysical flux of neutrinos.
Astrophysical neutrinos are key to the multi-messenger approach to understand sources of cosmic radiation \cite{Spurio:2018knn}. 

At the highest energies, neutrinos are anticipated from UHECR that attenuate through
interactions with photons in transit from sources \cite{Beresinsky:1969qj,Hill:1983mk,Engel:2001hd,Anchordoqui:2007fi,Kotera:2010yn,Decerprit:2011qe,Roulet:2012rv} with the Greisen-Zatsepin-Kuzmin (GZK) cutoff \cite{Greisen:1966jv,Zatsepin:1966jv} being a signal of this process.
The details of the cosmogenic neutrino spectrum would provide invaluable information to the cosmic ray acceleration process, source distribution, source evolution, and the UHECR nuclear composition \citep{Kotera:2010yn}. 
While the existence of the cosmogenic neutrinos is implied by the baryonic component in cosmic rays, the detection of these neutrinos has remained elusive \cite{nulimits}. 
It is one of the most important measurements in astroparticle physics. 

The weak neutrino interaction cross section makes neutrinos
a critical and unique component of multi-messenger astronomy and astrophysics. The neutrino horizon extends far beyond that for UHECR. Measurements of cosmogenic neutrinos have the potential to provide
information about sources much farther than those responsible for the  observed flux of UHECR. Measurements of astrophysical neutrinos probe the environments of cosmic accelerators.

While the weak interactions of neutrinos give a benefit in their transit to the Earth, their detection requires very large target volumes. Direct detection of neutrino interactions in, e.g., the IceCube instrumented volume, are also augmented by muons from $\nu_\mu$ charged current interactions that produce a muon outside the detector, thereby increasing the effective detection volume.  Tau neutrino production of \taons can also increase the effective volume at higher energies due to the Lorentz-boosted lifetime. Over cosmological distances, the $1:2$ ratio of $\nu_e+\bar{\nu}_e:\nu_\mu+\bar{\nu}_\mu$ produced in sources
of UHECR or in their transit to the Earth yields a nearly equal flux of electron, muon and tau neutrinos \cite{Learned:1994wg}.
This leads to characteristic upward-going tau neutrino induced signals in
underground detectors and ground-based detection of air showers   \cite{Halzen:1998be,Bertou:2001vm,Feng:2001ue,Lachaud:2002sx,Hou:2002bh,Tseng:2003pn,Aramo:2004pr,Dutta:2005yt,Asaoka:2012em,Fargion:2000iz,Bottai:2002nn,Fargion:2003kn,Fargion:2003ms,PalomaresRuiz:2005xw,Abreu:2012zz,Zas:2017xdj,Fang:2017mhl,Alvarez-Muniz:2018bhp,Ahnen:2018ocv,Otte:2018uxj,Aab:2019auo}.
The potential to use sub-orbital and space-based measurements of extensive air showers (EAS) induced from neutrino interactions either in the Earth \citep{Domokos:1997ve,Domokos2} or Earth's atmosphere has been recognized as a way to achieve even larger neutrino target masses, greater than $10^{13}$ metric tons for the atmosphere \citep{OWL}, for example.
In particular, the signals from upward-moving EAS that come from tau neutrino interactions within the Earth \cite{Fargion:2000iz,Bottai:2002nn,Fargion:2003kn,Fargion:2003ms,PalomaresRuiz:2005xw}
provide a path to measure the astrophysical and the cosmogenic neutrino fluxes above $\sim 1$ PeV, with a huge neutrino target mass \citep{Krizmanic2011, CHANT}.

Indirect detection techniques for measuring the characteristics of upward-going EASs from VHE neutrinos include: 1) the detection of the beamed optical Cherenkov radiation from EAS particles, and 2) the detection of the coherent radio radiation from the EAS. The latter comes from electric fields induced by separation of positive and negative EAS particles in the Earth's magnetic field (geomagnetic radiation) and from a time-varying net charge in showers initiated by VHE neutrinos in dense media such as ice (the Askaryan effect \cite{Askaryan:1962hbi}). Measurements of either type of signal, optical Cherenkov or radio, can be leveraged to determine the energy of the neutrino primary while having excellent angular resolution in the incident neutrino direction.
The long-duration balloon flights of ANITA have demonstrated the capability of using a sub-orbital instrument for detecting radio signals from downward-going and horizontally propagating UHECRs \cite{Hoover:2010qt,Schoorlemmer:2015afa,Gorham:2016zah}. Searches for radio signals from upward-going, neutrino-induced cascades in the Antarctic ice have led to limits on the diffuse flux of UHE neutrinos at energies above $10^{10}$~GeV \cite{ANITA2018,Gorham:2019guw}, culminating in stringent constraints \cite{ANITA} on ``top-down'' models of UHECR and UHE neutrino production via phenomena associated with relics from the Big Bang, from phase transitions or from physics near the Grand Unified scale in the early universe (e.g. Refs. \cite{Yoshida:1996ie,Berezinsky:1999az,Sigl:1998vz}).

Recently, the ANITA Collaboration reported the detection of two anomalous events with radio signal characteristics that appear to be compatible with upward-moving EASs.
The inferred EAS energies and the projected path lengths through the Earth imply that these events were likely not initiated by $\nu_\tau$ interactions in the Earth \cite{Gorham:2016zah, Gorham2018,Romero-Wolf:2018zxt}.
Nevertheless, these events have fueled speculation that they may signify physics beyond the Standard Model, such as Earth-interacting sterile neutrinos \cite{SterileNu,Huang:2018als},
the interactions or production of supersymmetric particles  \cite{Collins:2018jpg,2018arXiv180708892C,Fox:2018syq}, neutrino-induced supersymmetric
sphaleron transitions \cite{Anchordoqui:2018ssd}, or the decay of superheavy dark matter \cite{Anchordoqui:2018ucj}.

Several sub-orbital (EUSO-SPB2, \cite{SPB2}), space-based (CHANT, \cite{CHANT}; POEMMA, \cite{2017ICRC...35..542O}), and ground-based (Trinity, \cite{Otte:2018uxj}) instruments have been proposed to search for EASs from Earth-skimming tau neutrinos via optical Cherenkov radiation, which would allow for sensitivity to neutrino energies above $\sim 10$~PeV, even though the duty cycle for optical Cherenkov is $\sim 20$\% (compared to $\sim 100$\% for radio).
POEMMA, the Probe Of Extreme Multi-Messenger Astrophysics, is a NASA probe-class mission under study that will detect UHECRs and VHE neutrinos from space \cite{2017ICRC...35..542O}.
In this paper, we report on the development and the results from an initial, end-to-end calculation of the optical Cherenkov signal generated from upward-moving EAS from the decay of \taons sourced by tau neutrino interactions in the Earth. Many of the results presented here are not specific to the POEMMA concept. To perform a baseline simulation to calculate the neutrino sensitivity for a space-based experiment, we use the 525 km orbiting altitude, a 2.5 m$^2$ detector collecting area,
a photo-detection efficiency of 20\%,
and other detector characteristics
of POEMMA. 

In the next section, we give a brief overview of satellite and balloon-based neutrino detection starting with the geometry of the viewed surface of the Earth. Appendix A has many of the detailed relations between the various angles that describe the instrument viewing and the emerging showers. The treatment of tau neutrino interactions to produce \taons and the subsequent tau energy loss and/or decay in the Earth to produce a lower energy neutrino, are discussed in Sec. \ref{sec:two}. We show the flux independent and flux dependent $\nu_\tau\to \tau$ transmission results.

In Sec. \ref{sec:three}, we review \taon  decay probabilities in the atmosphere and
the geometry of decays as a function of altitude. We also describe how
air showers in the atmosphere are modeled to get the photon density to arrive at a detector like POEMMA at 525 km altitude.

Sec. \ref{sec:four} includes results that incorporate detection by a POEMMA-like optical Cherenkov detector as an application of our transmission results. We evaluate the aperture and flux independent sensitivity from our simulations. We also estimate the number of events that could be detected for two flux models for the isotropic cosmogenic flux \cite{Kotera:2010yn}.

We summarize our conclusions and discuss uncertainties and potential improvements to the simulation in Sec. \ref{sec:five}. In Appendix A, we collect further details of the geometry of the Earth, as seen at altitude. We display tables
for the cumulative probability functions for $\nu_\tau\to \tau$ as a function of outgoing \taon energy fraction of the incident neutrino energy, for selected incident neutrino energies and angles, in Appendix B. Finally, in Appendix C we outline in more detail inputs to the Monte Carlo simulation of the detection probabilities as a function of energy and angle.

\section{Neutrino and tau propagation in Earth}\label{sec:two}

\begin{figure}[ht]
\centering
\includegraphics[width=0.8\columnwidth, trim = 12cm 9cm 7cm 1cm, clip]{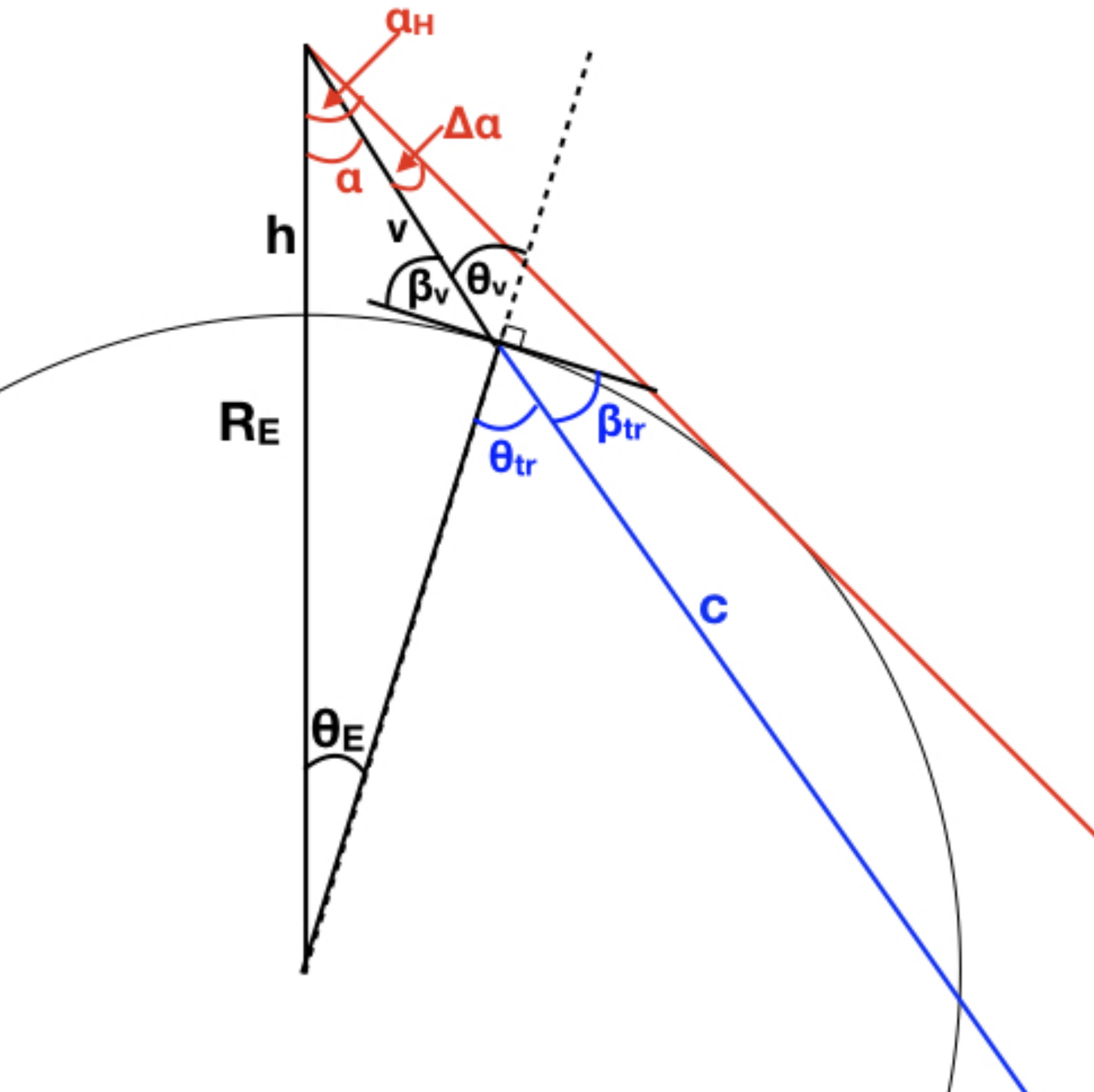}
\caption{\small Geometry for detecting an EAS from an upward moving tau neutrino. The angles $\beta_v$ and $\theta_v$ label the elevation angle
and local zenith angle for the point along the line of sight.
Angle $\beta_{\rm tr}$ and $\theta_{\rm tr}$ describe the
emerging tau trajectory.} 
\label{fig:geometry-poemma}
\end{figure}

The Earth-skimming technique for detecting tau neutrinos relies on using the Earth as a neutrino converter. The feasibility of this technique depends on the details of tau neutrino propagation along trajectories through the Earth that are determined by the detection geometry, shown in Fig.~\ref{fig:geometry-poemma}. 
We begin with a summary of our notation in Sec. \ref{subsec:geometry}. Along the
chord $c$ through the Earth, a neutrino interacts to produce a tau,
then the tau loses energy and may decay with a neutrino in the final state. This regeneration process can occur multiple times, depending on the energy and angle. In Secs. \ref{subsec:nuprop}, \ref{subsec:tauprop}
and \ref{subsec:taudecay}, we describe how we incorporate these elements in our simulation. Sec. \ref{subsec:fluxindep} shows results for tau exit  probabilities, and tau energy distributions, for a fixed tau neutrino incident energy and angle. Transmission functions, 
the ratios of outgoing taus to incoming tau neutrinos, as functions energy and angle are shown in Sec.
\ref{subsec:fluxdep}. The results presented in Sec. \ref{sec:two} are related to tau neutrino and tau propagation in the Earth. Sec.
\ref{sec:three} is focused on tau neutrino induced air showers from
tau decay and their detection.

\subsection{Geometry of Neutrino Trajectories and Column Depth Through the Earth}\label{subsec:geometry}

For neutrino
energies above a PeV, the neutrino and antineutrino interaction cross sections are essentially equal, and their interaction length is
shorter than the column depth along the diameter of the Earth
\cite{Gandhi:1995tf,Gandhi:1998ri}. Consequently,
neutrino detectors at altitude $h$ above the Earth need to point near the limb. The limb is at a viewing angle $\alpha_H$ away from the nadir in our notation. A given point on the Earth's surface 
in view of the detector
is described by the angle $\theta_E$, as shown in Fig. \ref{fig:geometry-poemma}.
It corresponds to a detector viewing angle $\alpha$ away from the satellite nadir. The line of sight from the point to the detector is at an elevation angle $\beta_v$ and angle $\theta_v$ from the local zenith. The distance $v$ is the distance along the line of sight.
Relations between angles and distances are listed in  Appendix~\ref{sec:app_a}.

In order for the detector to receive Cherenkov radiation from a neutrino-induced EAS, the $\tau$-lepton must emerge 
with a trajectory with an elevation angle $\beta_{\rm tr}$ ($\theta_{\rm tr}$ from the local zenith) that is within a factor of the Cherenkov angle $\theta_{\rm Ch}$ from the line of sight with the detector, namely, $\beta_{\rm tr} = \beta_v \pm \theta_{\rm Ch}$.
Hence, at any given location within the field of view of the detector, only those neutrinos with specific trajectories through the Earth will be detectable. These trajectories determine the column depth of material neutrinos must traverse as they propagate through the Earth.
Similar arguments can be made for instruments that make use of the radio detection technique.

The Cherenkov angle in air at sea level is $\theta_{\rm}=1.5^\circ$, so
as 
Fig. \ref{fig:geometry-poemma} shows, $\beta_{\rm tr}\simeq 
\beta_v$ and $\theta_{\rm tr}\simeq \theta_v$. 
The geometry will be discussed in more detail in Sec. \ref{subsec:aperture}. 

Assuming a spherical Earth, the chord length $c$ of the path through the Earth for a neutrino trajectory with elevation angle $\beta_{\rm tr}$ is given by
\begin{equation}
c = 2 R_E \sin\beta_{\rm tr}\ ,
\end{equation}
where $R_E = 6371$ km is the radius of the Earth. The column depth
through the Earth for a given trajectory with $\beta_{\rm tr}$ is given by
\begin{equation}
X\left(\beta_{\rm tr}\right) = \int_{0}^{c} \rho\left(s',\beta_{\rm tr}\right)ds'\,
\end{equation}
where $s$ is the distance along the path traversed through the Earth and $\rho\left(s,\beta_{\rm tr}\right)$ is the density of the Earth along the path length at $s$ for the trajectory. We distinguish between column depth and chord length because the neutrino interaction length and the energy losses for the produced $\tau$-lepton depend on the column depth whereas the probability of the $\tau$-lepton decaying prior to emerging through the surface depends on the remaining distance to the surface.

\begin{figure}[htb]
\centering		
\includegraphics[width=0.9\columnwidth]{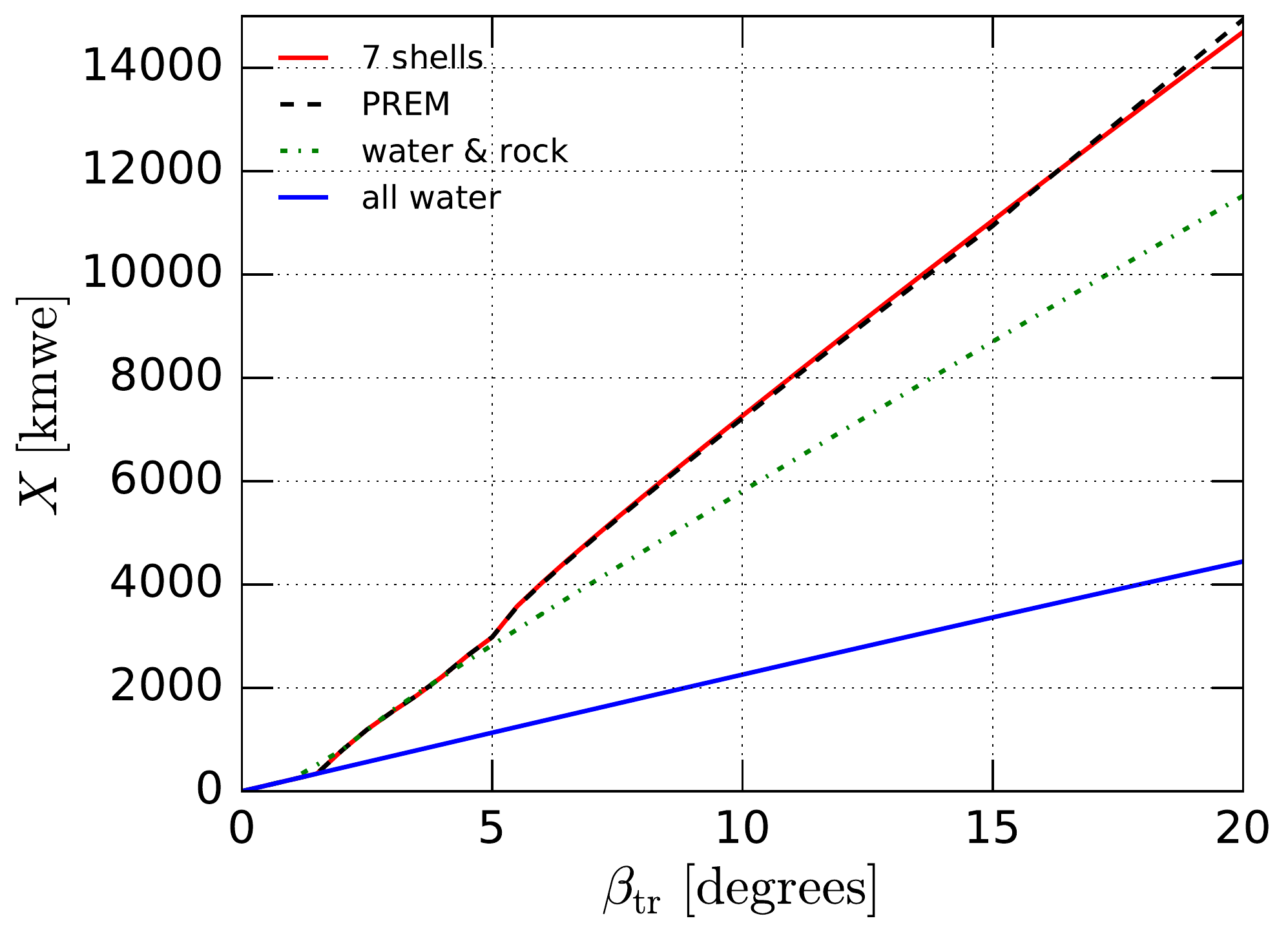}	
\includegraphics[width=0.9\columnwidth]{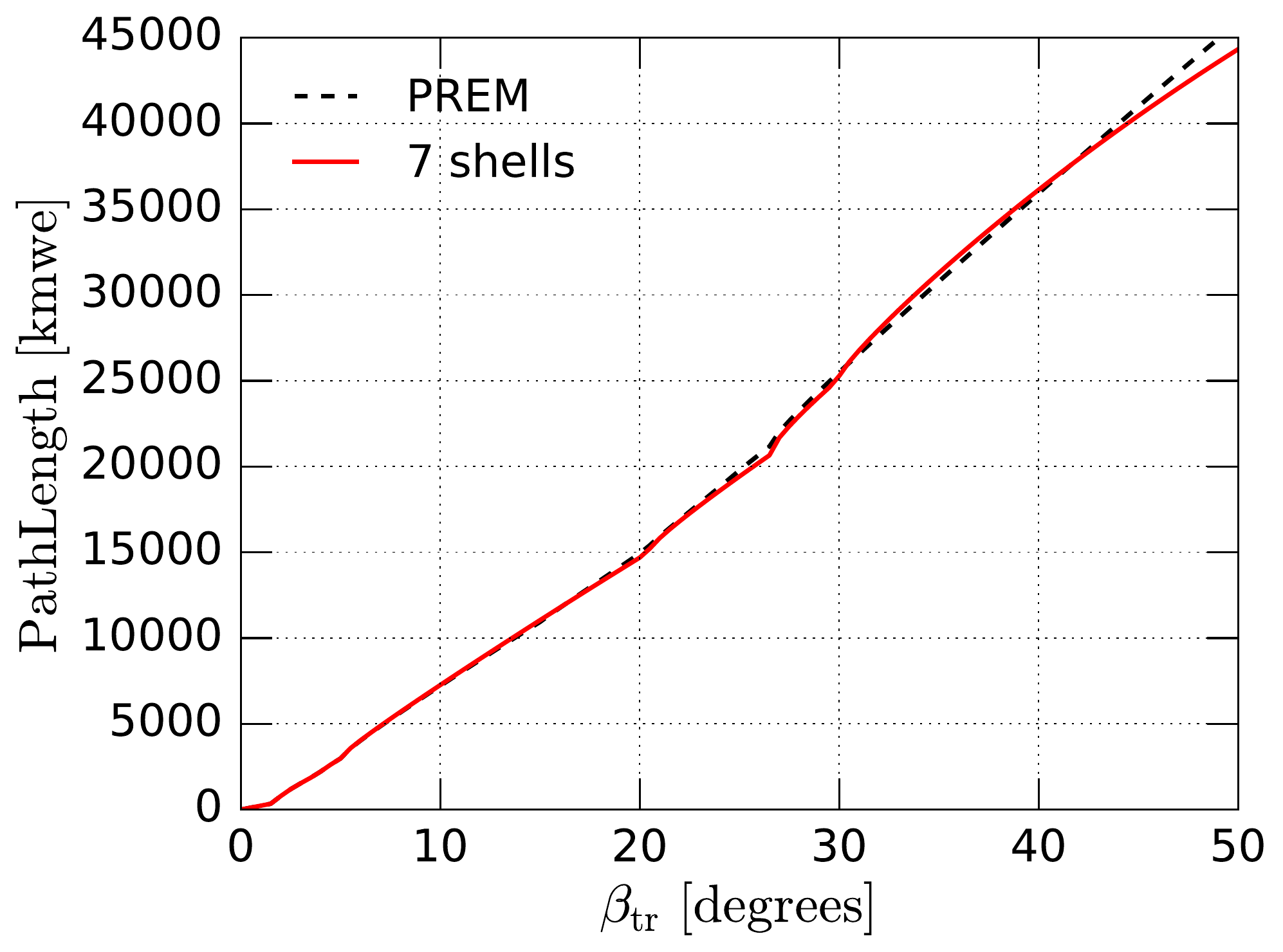}	
	\caption{\small \textit{Upper}: The column depth as a function of $\beta_{\rm tr}$ following the PREM parametrization (black dashed) and using the seven-shell density model (solid red) for trajectory angles $0^{\circ} \leq \beta_{\rm tr} \leq 20^{\circ}$. For reference, the column depth for a sphere of radius $R_E$ composed entirely of water (solid blue line) or water and rock (dot-dashed green) are also plotted. \textit{Lower}: The column depth for the PREM and seven-shell
	density models, for trajectory angles $\beta_{\rm tr} \leq 50^{\circ}$.}
	\label{fig:columndepth}
\end{figure}

For the density of the Earth, we use a multi-shell model based on the average radial profile given by the Preliminary Reference Earth Model (PREM) parameterization in Ref.~\cite{DZIEWONSKI1981297}. For neutrino trajectories with angles $\beta_{\rm tr}\leq 50^\circ$, we find that a model consisting of seven shells of constant density is a good approximation to
the PREM column depth, as shown in Fig.~\ref{fig:columndepth} with
solid and dashed lines, respectively.  The density parameters are listed in Table~\ref{table:density}
as a function of vertical depth $d=R_E-r$, where $r$ is the radial distance from the center of the Earth. To evaluate
the column depth, $r$ is related to $s$ and $\beta_{\rm tr}$ by the law of cosines.

\begin{table}[t]
\begin{center}
\begin{tabular}{|c|c|}
\hline
$\rho \ [{\rm g/cm}^3]$ & Depth $d = R_E-r$ \\
\hline
\hline
1.02 &  $d\leq 3.0\ {\rm km}$\\
\hline
2.6 &  $3.0\ {\rm km }<d\leq15\ {\rm km}$\\
\hline
2.9 &  $3.5\ {\rm km }<d\leq 24.4\ {\rm km}$\\
\hline
3.4 & $24.4 \ {\rm km }<d \leq 400\ {\rm km}$ \\
\hline
3.8 & $400\ {\rm km }<d \leq 670\ {\rm km}$ \\
\hline
4.4 & $670\ {\rm km }<d \leq  850\ {\rm km} $ \\
\hline
4.8 & $850\ {\rm km} < d$\\
\hline
\end{tabular}
\caption{Density $\rho$ as a function of vertical depth $d=R_E-r$ below the Earth's surface for $\beta_{\rm tr}\leq 50^\circ$ based on the PREM parameterization \cite{DZIEWONSKI1981297} of the Earth's density.}
\label{table:density}
\end{center}
\end{table}

\subsection{Neutrino interactions}\label{subsec:nuprop}

The first step in the evaluation of skimming tau neutrinos is their interactions via charged current or neutral current weak interactions. The high energy neutrino cross section depends on the small-$x$ behavior of the structure functions of the nucleon targets. 
There is an extensive literature on the high energy extrapolations of the neutrino-nucleon cross section
\cite{Gandhi:1995tf,Gandhi:1998ri,Reno:2004cx,Jeong:2010za,Connolly:2011vc,CooperSarkar:2011pa,Arguelles:2015wba}. 
We use as a standard the leading order neutrino cross section
evaluated with next-to-leading order (NLO) parton distribution functions (PDFs) for free protons provided by the nCTEQ group (nCTEQ15-1)
\cite{Kovarik:2015cma}, adjusted for an isoscalar target.  
It has been shown that the NLO matrix element
squared changes the cross section by less than 5\% compared to using the leading order matrix element squared with NLO PDFs, a much smaller correction than the uncertainty associated with the choice of small-$x$ extrapolations of the PDFs
\cite{Jeong:2010za}.

Alternate neutrino cross sections can be used to assess uncertainties. Two neutrino cross sections 
are evaluated with small-$x$ extrapolations based on the Abramowicz et al. (ALLM) extrapolation \cite{Abramowicz:1991xz,Abramowicz:1997ms} of the electromagnetic
structure function $F_2(x,Q^2)$ and the Block et al. (BDHM) extrapolation \cite{Block:2013nia,Block:2013mia,Block:2014kza} 
of $F_2$. As discussed below, these two 
expressions for the electromagnetic structure function will be used 
for the photonuclear energy
loss of the \taon. Details of the correspondence between tau energy loss and
the tau neutrino cross section appear in ref. \cite{Jeong:2017mzv}. 

With the cross section and differential energy
distributions, we
use the column depth determined from the Earth density shells to find the interaction point of the neutrino. High densities
are favorable for neutrino conversions to taus, however, tau energy loss in dense materials promotes tau decay.

\begin{figure}[ht]
\centering
	\includegraphics[width=0.9\columnwidth]{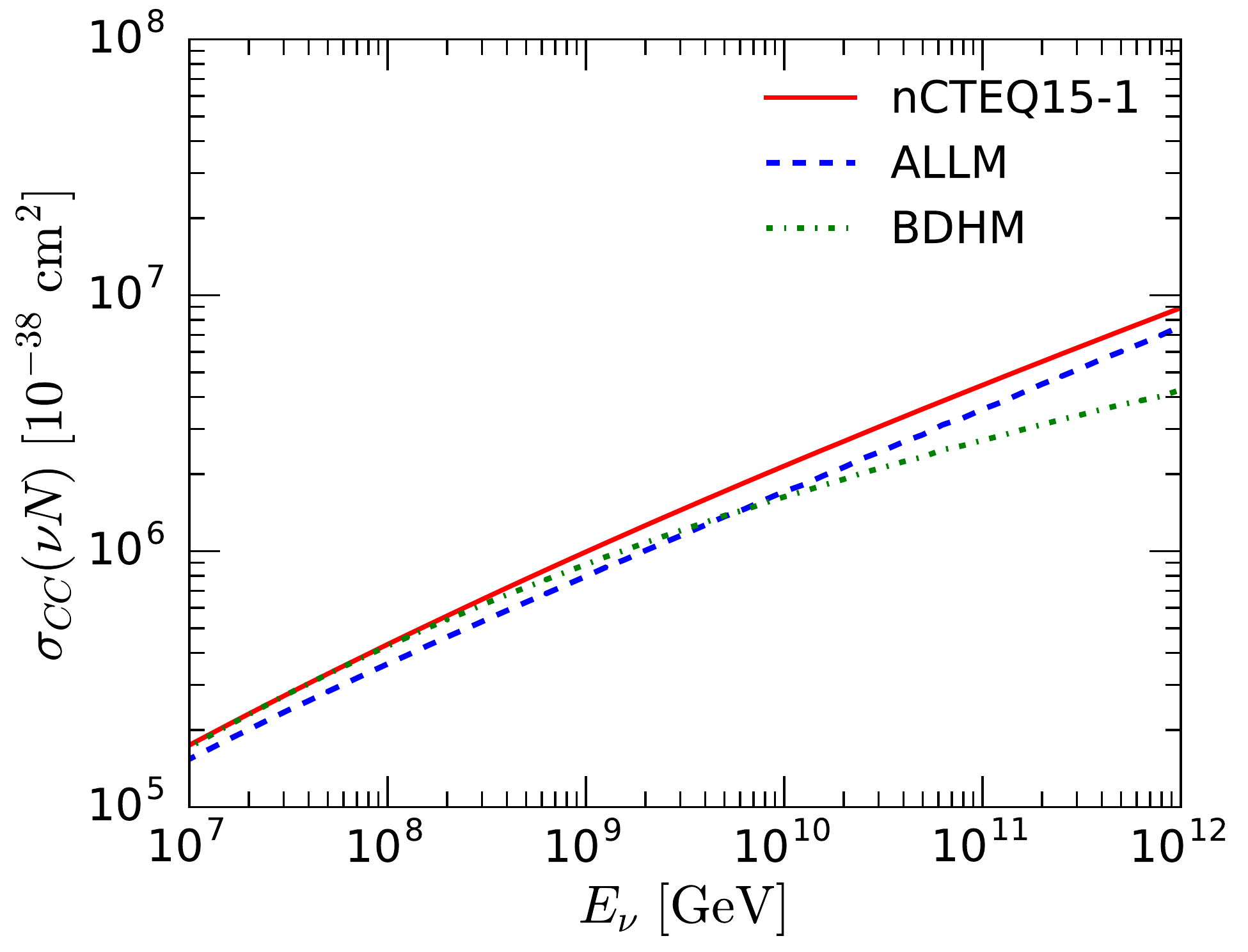}	
	\caption{\small As a function of incident neutrino energy, the charged current cross sections evaluated using ALLM and BDHM small-$x$ extrapolations and 
	using next-to-leading order QCD with a power law extrapolation at small-$x$ \cite{Jeong:2010za}.}
	\label{fig:nu-new-cc}
\end{figure}

\begin{figure}[ht]
\centering
	\includegraphics[width=0.9\columnwidth]{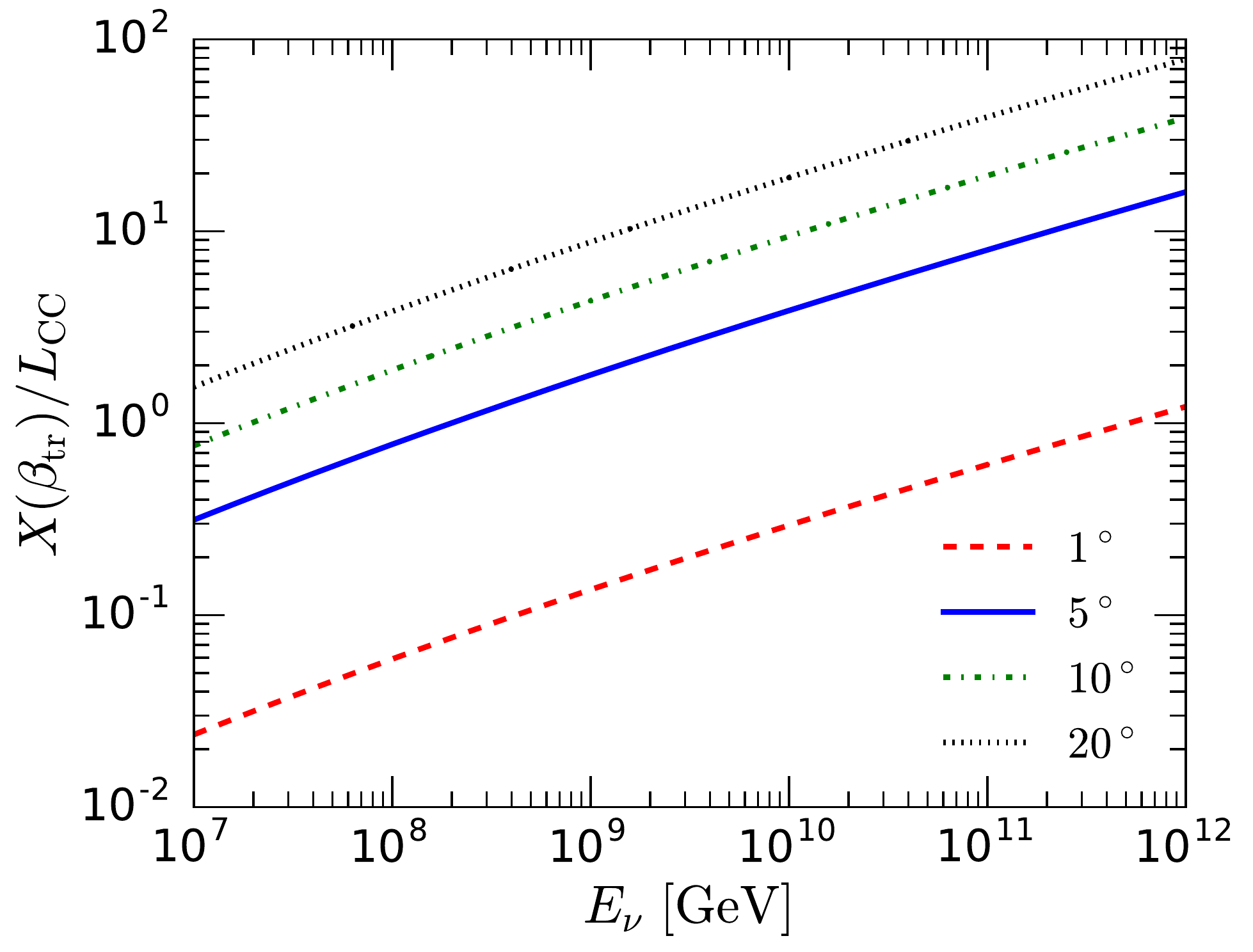}	
	\caption{\small As a function of incident neutrino energy, at fixed neutrino angle relative to the horizon 
	$\beta_{\rm tr}=1^\circ,\ 5^\circ ,\ 10^\circ$ and $20^\circ$, the
	ratio of the column depth $X(\beta_{\rm tr})$ to the neutrino charged current interaction length evaluated using the nCTEQ15-1 PDFs.}
	\label{fig:pathlength-ratio}
\end{figure}

For reference, in Fig. \ref{fig:nu-new-cc} we show the neutrino charged current cross sections using our standard
nCTEQ15-1 \cite{Kovarik:2015cma} cross section.
and the two alternate cross sections. 
Fig. \ref{fig:pathlength-ratio} shows the ratio of the column depth for fixed angles, $X(\beta_{\rm tr})$, to the nCTEQ15-1 neutrino charged current interaction length as a function of neutrino energy. This shows the importance of attenuation at large angles. 

Our Monte Carlo program propagates the neutrino along the chord
with repeated interactions of neutral currents as applicable, until a tau is produced (or not) over the neutrino trajectory. The outgoing lepton energy is determined by the differential weak interaction
cross section $d\sigma/dy$ where $y=(E_{\nu_\tau}-E_\ell)/E_{\nu_\tau}$ where $\ell = \nu_\tau$ or $\ell=\tau$ for neutral current
and charged current interactions, respectively.

\subsection{Tau propagation}\label{subsec:tauprop}

Taus produced by charged current interactions of tau neutrinos with nucleons then propagate along the trajectory through the Earth. 
The lifetime,
including the $\gamma$-factor, influences the distance traveled, as does the electromagnetic energy loss. For reference, the time dilated
decay length of a tau of energy $E_\tau$ is $\gamma c\tau \simeq 5\ {\rm km} [E_\tau/(10^8\ {\rm GeV})]$ given $c\tau=87.11$ $\mu$m and $m_\tau=1.776$ GeV/$c^2$.

Electromagnetic energy loss by charged leptons $\ell$ in transit through materials
comes from ionization, bremsstrahlung, electron-positron pair production and photonuclear interactions
\cite{Lohmann:1985qg,Lipari:1991ut,Dutta:2000hh,Dutta:2005yt,Armesto:2007tg,Bigas:2008ff,Koehne:2013gpa,Alvarez-Muniz:2017mpk}. 
The average energy loss per unit column depth $X$ for a charged lepton $\ell$, here with $\ell=\tau$,  is written
\begin{equation}
\Biggl\langle \frac{dE_\tau}{dX}\Biggr\rangle = -(a_\tau+b_\tau E_\tau)\ .
\label{eq:dedx}
\end{equation}
At high energies, the ionization loss characterized by $a_\tau$ is small. The energy loss parameter $b_\tau$
is written as 
\begin{equation}
b_\tau = \sum_i b_\tau^i=b_\tau^{\rm brem}+b_\tau^{\rm pair} + b_\tau^{\rm nuc}\ ,
\label{eq:betai}
\end{equation}
where the $b_\tau^i$ are each calculated in terms of the differential cross section for tau electromagnetic
scattering
\begin{equation}
\label{eq:betadef}
b_\tau^i(E) = \frac{N}{A}\int dy\, y\, \frac{d\sigma_{\tau A}^i(y,E)}{dy}
\end{equation}
weighted by the inelasticity parameter $y = (E_\tau-E_\tau ')/E_\tau$ for incident lepton energy $E$ and outgoing energy $E'$.
The $b_\tau^i$ that characterize $\langle dE/dX\rangle$ depend weakly on energy.
It can be shown that $b_\ell^{\rm brem}$
scales with lepton mass as $\sim 1/m_\ell^2$, while for pair production and photonuclear interaction,
$b_\ell^{\rm pair,nuc}\sim 1/m_\ell$ \cite{Tannenbaum:1990ae,Reno:2005si}. For muons, the three $\beta_\mu^i$ are similar in scale
\cite{Lohmann:1985qg}, however,
for tau energy loss, our interest here, only pair production and photonuclear interactions are important.

We can use the $b_\tau$ in eq. (\ref{eq:dedx}), written in terms of the column depth, to understand qualitative features of tau propagation in materials. The density along chord $c$ accounts for most of the differences in materials of interest here (7 density shells), however, there are some material dependent corrections. For pair production, dominated by coherent scattering on the nucleus, the differential cross
section scales as the nuclear charge $Z^2$, normalized by $A$ in the definition of $b_\tau^{\rm pair}$. For photonuclear
interactions, the scaling is with $Z/A$, so to a good approximation $b^{\rm nuc}_\tau$ is independent of material.
We use two sets of parameters for the calculation below: one set of parameters for water, and a second set evaluated with standard rock $Z$ and $A$ to use for rock and the other density shells. Of course, for tau propagation, the density of each shell is included in the column depth for that shell.

\begin{figure}[t]
\centering
	\includegraphics[width=0.9\columnwidth]{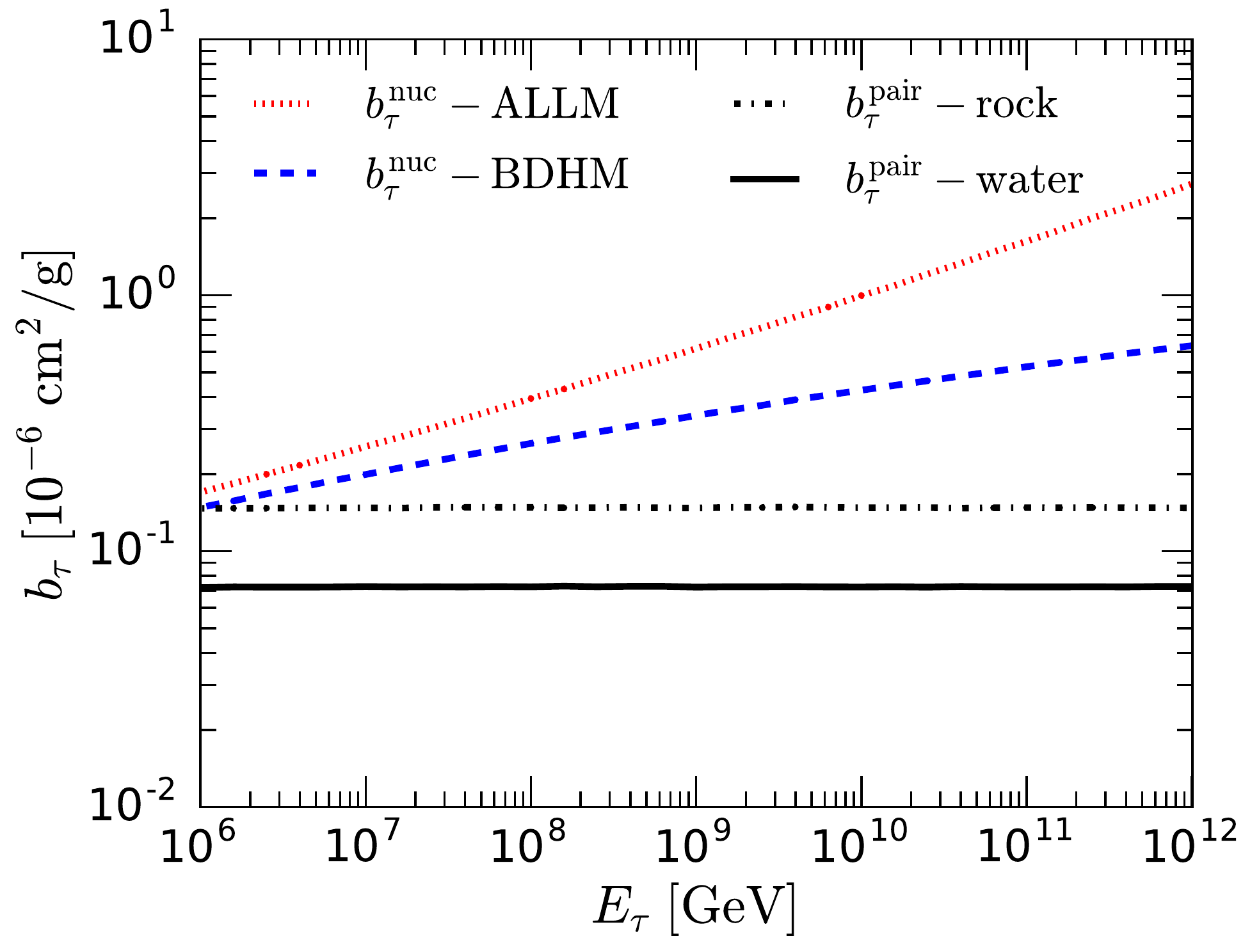}	
	\caption{\small The energy loss parameters $b_\tau^{\rm nuc}$ and $b^{\rm pair}$ for tau energy loss in $\langle dE/dX\rangle$ as a function of initial tau energy $E$. The dotted, dashed ($b_\tau^{\rm nuc}$) and dot-dashed ($b_\tau^{\rm pair}$) lines
	are for standard rock, and the solid line shows $b_\tau^{\rm pair}$ for water. 	Two separate curves are shown for photonuclear energy loss, labeled with ALLM and BDHM.
	The photonuclear energy loss parameters for water are approximately equal to those for rock.
}
	\label{fig:beta}
\end{figure}

We remark that electromagnetic energy loss of taus by photonuclear interactions, at high energies, is tied to the small $x$,  and the $Q\sim 1$ GeV regime of the electromagnetic structure function $F_2(x,Q)$. As with the high energy neutrino cross section, 
small-$x$ extrapolations of $F_2$, as yet unconstrained by experiment, are required. Different extrapolations  
give different values for $b_\tau^{\rm nuc}$. In Fig. \ref{fig:beta}, we show how two extrapolations of $F_2$ translate to
$\beta_\tau^{\rm nuc}$ to characterize theoretical uncertainties in the electromagnetic energy loss. We show the Abramowicz et al. (ALLM) form
\cite{Abramowicz:1991xz,Abramowicz:1997ms}, and the Block, Durand, Ha and McKay (BDHM) parameterization
\cite{Block:2013mia,Block:2013nia,Block:2014kza}. The ALLM extrapolation of $F_2$ gives a more strongly energy dependent $b_\tau^{\rm nuc}$ than
the BDHM parameterization, however, in both cases, the growth with energy is only (approximately) logarithmic in tau energy. The values of $b_\tau^{\rm pair}$ for rock and water are also shown in Fig. \ref{fig:beta}. A more complete discussion of connection between tau energy loss and the small-$x$ extrapolations of $F_2$ appears in, e.g., ref. \cite{Jeong:2017mzv}.

It has been known for some time that stochastic energy loss effects are important, especially for muons \cite{Lipari:1991ut}. 
At high energies, stochastic losses from 
bremsstrahlung and the photonuclear interaction dominate muon energy loss in IceCube \cite{Aartsen:2015nss}. For \taons, $\beta_\tau^{\rm brem}\ll \beta_\tau^{\rm nuc},\beta_\tau^{\rm pair}$.
As Fig. \ref{fig:beta} shows, $b_\tau^{\rm nuc}>b_\tau^{\rm pair}$, so we propagate \taons stochastically.
We account
for stochastic effects using the procedure outlined in Refs. \cite{Lipari:1991ut,Antonioli:1997qw,Dutta:2000hh,Jeong:2017mzv}, in a one-dimensional approximation. Since we are at high energies,
a one-dimensional approach is appropriate. The procedure involves rewriting
\begin{eqnarray}
\label{eq:betaycut}
b_\tau^i(E) &=&  \frac{N}{A}\int_0^{y_{\rm cut} }
dy\, y\, \frac{d\sigma_{\tau A}^i(y,E)}{dy}\\ \nonumber
& + & \frac{N}{A}\int_{y_{\rm cut}}^1
 dy\, y\, \frac{d\sigma_{\tau A}^i(y,E)}{dy} \ .
\end{eqnarray}
We account for energy loss with a continuous term for $y\leq y_{\rm cut}$ and simulate tau energy loss 
stochastically for $y_{\rm cut}<y\leq 1$. We take $y_{\rm cut}=10^{-3}$,
a value that reliably includes stochastic energy loss for muons
\cite{Antonioli:1997qw,Chirkin:2004hz}, and by extension, $\tau$-leptons \cite{Dutta:2000hh}. 
It has been shown that an effective $b_\tau$ that describes the average
energy of emerging taus from stochastic propagation is typically larger than what one obtains from eq. (\ref{eq:betadef}) \cite{Dutta:2005yt}.

\subsection{Tau decays and regeneration}\label{subsec:taudecay}

Over large column depths, 
tau neutrino regeneration from $\nu_\tau\to \tau$ charged current conversion and $\tau\to \nu_\tau$ decays is an important effect
\cite{Halzen:1998be,Beacom:2001xn,Fargion_2004,Alvarez-Muniz:2017mpk}.
The tau neutrino from a tau decay has approximately one-third of the tau energy. This neutrino then has the potential to interact
with the remaining column depth, producing a tau. Each step in the regeneration process reduces the tau energy.

\begin{figure}[ht]
\centering
	\includegraphics[width=0.9\columnwidth]{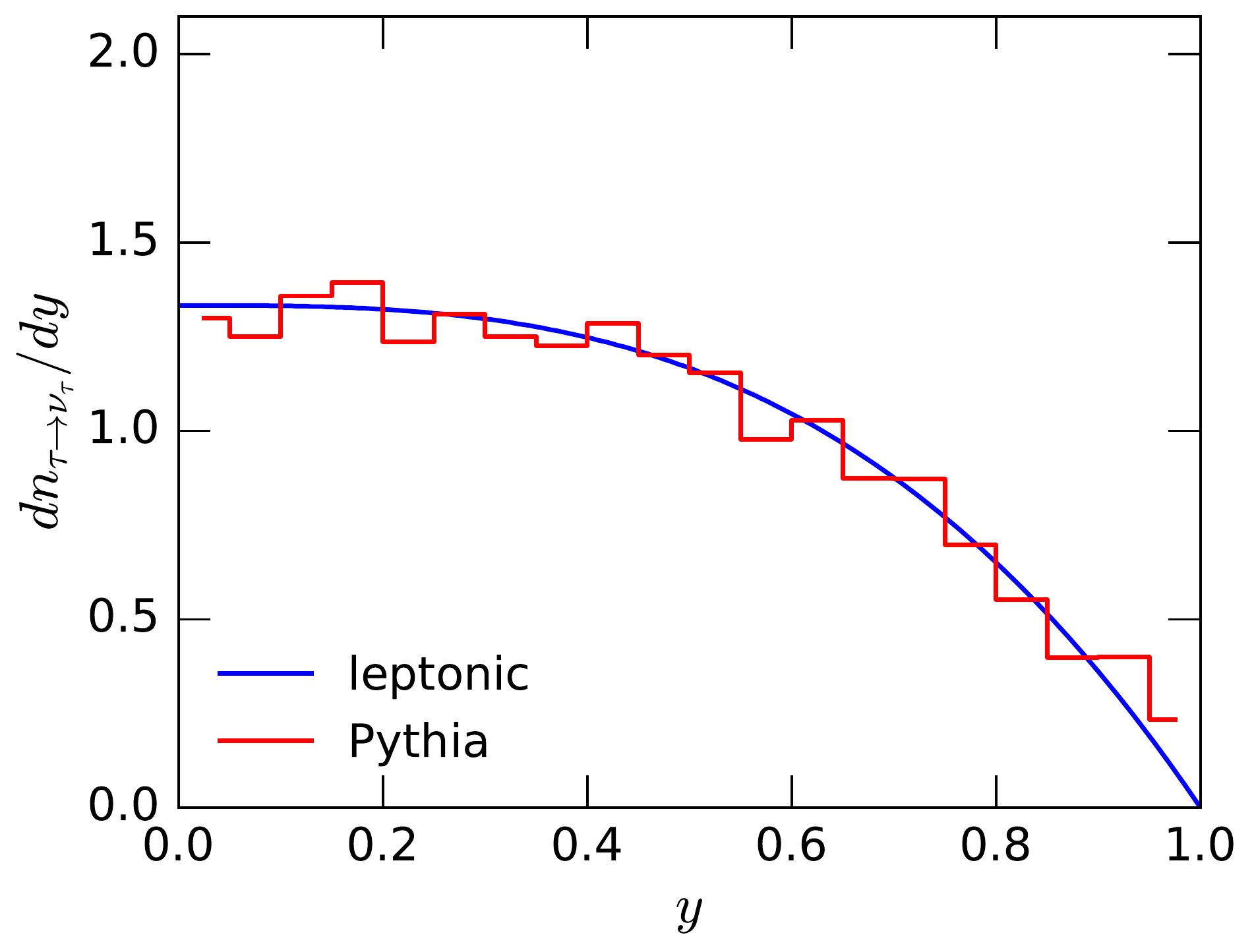}	
	\caption{\small Tau decay distribution (eq. (\ref{eq:dndy})) normalized to one for $\tau\to\nu_\tau X$ from PYTHIA (histogram)
	and $\tau\to \nu_\tau\ell\nu_\ell$ (solid curve). The neutrino
	energy fraction is $y=E_{\nu_\tau}/E_\tau$.}
	\label{fig:dndy}
\end{figure}

One can make an approximate model of the  energy distribution of neutrinos from tau decays 
for the two- and three-body decays 
$\tau\to \nu_\tau \ell \nu_\ell,\ \tau\to \pi\nu_\tau,\ \tau\to \rho \nu_\tau$ and $\tau \to a_1\nu_\tau$, accounting for $\sim 85\%$ of the tau decay width \cite{Barr:1988rb,Pasquali:1998xf}. The energy distribution of the tau neutrino can be evaluated directly where the form of
the distribution is  \cite{Gaisser:2016uoy}:
\begin{equation}
\frac{dn_{\tau\to \nu_\tau}}{dy} = g_0(y) -  g_1(y)
\label{eq:dndy}
\end{equation}
for $y\equiv E_{\nu_\tau}/E_\tau$, given that the taus produced by neutrino interactions are left-handed
and antineutrinos produce right-handed $\tau^+$. The detailed formulas for $g_0$ and $g_1$ appear in, e.g.,  Ref. \cite{Gaisser:2016uoy}
and the appendix of Ref. \cite{Bhattacharya:2016jce}. For three-body leptonic decays,
\begin{eqnarray}
\label{eq:3body}
g_0(y) &=& \frac{5}{3} - 3y^2 +\frac{4}{3} y^3\\
\nonumber
g_1(y) &=& \frac{1}{3}-3 y^2+\frac{8}{3} y^3\ .
\end{eqnarray}
In fact, the expression for the three-body purely leptonic decay is a good representation of 
the energy distribution of tau neutrinos from tau decays evaluated in PYTHIA 8 \cite{sjo15}.
In Fig. \ref{fig:dndy}, we show with solid line histogram the sum of the decay channels
from PYTHIA. The solid curve shows just the decay $\tau\to \nu_\tau \ell \nu_\ell$, 
normalized to one.
Because of the correspondence shown in Fig. \ref{fig:dndy} between PYTHIA and the analytic leptonic decay distribution, we approximate the full decay distribution of the tau by the semileptonic distribution.
For reference, $\langle y\rangle=\langle E_{\nu_\tau}/E_\tau\rangle\simeq 0.4$
in Fig. \ref{fig:dndy}.
Depolarization can in principle occur through multiple scattering of the $\tau$'s, suppressing the term $g_1(y)$ in Eq. (\ref{eq:dndy}). We neglect this effect here. The spectrum of regenerated taus is not very sensitive to the details of the energy distribution.

To model regeneration in the propagation of neutrinos and taus in the Earth for angles relative to the horizon of
$\beta_{\rm tr}\leq 25^\circ$, we account for up to
five tau decays to neutrinos, where at each step, the tau neutrino is propagated to determine whether or not it reinteracts to produce a tau. This means we simulate up to six neutrino charged current interactions. Any number of neutrino neutral current interactions are included.

The number of interactions becomes progressively more important as a function of increasing energy and angle $\beta_{\rm tr}$. Alvarez-Muniz et al. 
\cite{Alvarez-Muniz:2017mpk} have
evaluated the number of charged current events for fixed initial tau neutrino energy. They find that the average number of
charged current interactions for $E_\nu=10^{10}-10^{12}$ GeV is less than $\sim 3$ for $\beta_{\rm tr}\leq 40^\circ$. 
We checked that 5 regeneration steps are sufficient for our purposes here: the tau exit probabilities are not significantly changed by including the last regeneration step for angles less than $20^\circ$. For angles larger than those accessible by a POEMMA detector,
the final regeneration step does not have a large impact on the exit probability, even at high energy. The exit probability for $E_{\nu_\tau}=10^{10}$ GeV changes by 2\% for $\beta_{\rm tr}=25^\circ$ and by 7\% for $\beta_{\rm tr}=35^\circ$.

\subsection{Flux independent \texorpdfstring{$\nu_\tau\to\tau$}{nutotau} results}
\label{subsec:fluxindep}

\begin{figure}[t]
\centering
	\includegraphics[width=0.9\columnwidth]{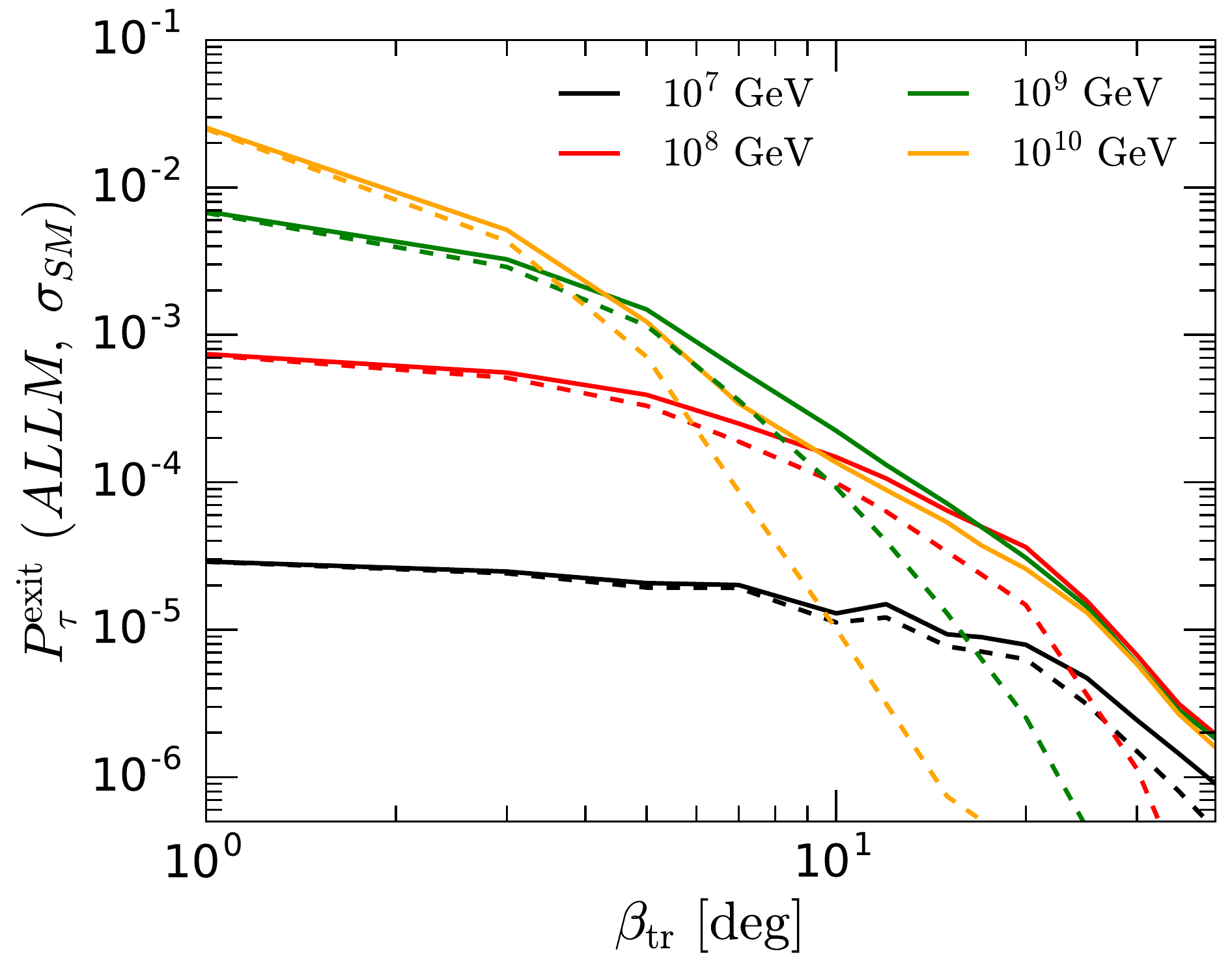}	
	\caption{\small Tau exit probability as a function of $\beta_{\rm tr}$ for incident $E_{\nu_\tau}=10^7$, $10^8$, $10^9$ and $10^{10}$ GeV evaluated using the 
	ALLM photonuclear energy loss and the nCTEQ-1 neutrino cross section ($\sigma_{SM}$). The dashed lines are without regeneration, and the solid
	lines include regeneration. }
	\label{fig:proballm}
\end{figure}

To illustrate the effects of regeneration, we show the tau exit probability for fixed incident tau neutrino energies: $10^7$, $10^8$, $10^9$ and $10^{10}$ GeV, as a 
function of angle $\beta_{\rm tr}$. The dashed lines Fig. \ref{fig:proballm} show the exit probability neglecting regeneration, while the solid lines include regeneration.
As has recently been emphasized by Alvarez-Muniz et al. \cite{Alvarez-Muniz:2017mpk}, regeneration is important for all but the smallest values of
$\beta_{\rm tr}$.

The details of the tau exit probabilities are difficult to model simply, but the case without regeneration follows from a few factors. The 
probability for the tau to exit the Earth
depends on the attenuation of the neutrino flux and the ratio of the tau range to the neutrino interaction length. This approximation works well for low energies (at or below $E_{\nu_\tau} \sim 10^8$ GeV) where the range is essentially the time dilated decay length. For higher energies, tau energy loss reduces the range relative to the decay length, and regeneration enhances the large angle exit probabilities, albeit with lower energy taus. Further details of the enhancement of exit probabilities with regeneration are discussed in Appendix B.
For a range of 12.5 km, $\rho=\rho_w$ for $\beta_{\rm tr}\leq 14^\circ$, and for 22 km, trajectory is solely in water for $\beta_{\rm tr}\leq 8^\circ$.

The BDHM energy loss model yields somewhat higher exit probabilities
at high energies. For $E_{\nu_\tau}=10^8$ GeV, the enhancement is only a few percent. At this energy, the tau range is only slightly smaller than the decay length. The enhancement of the BDHM evaluation relative to the ALLM electromagnetic energy loss model
is of order 25\% for $E_{\nu_\tau}=10^9$ GeV and of order 50\%
for $10^{10}$ GeV. These enhancements are only due to a smaller $b_{\tau}^{\rm nuc}$ relative to that of the ALLM energy loss model. The neutrino cross section is the same in these comparisons, set to the standard model evaluation with the nCTEQ15-1 PDFs.

\begin{figure}[!h]
\centering
	\includegraphics[width=0.9\columnwidth]{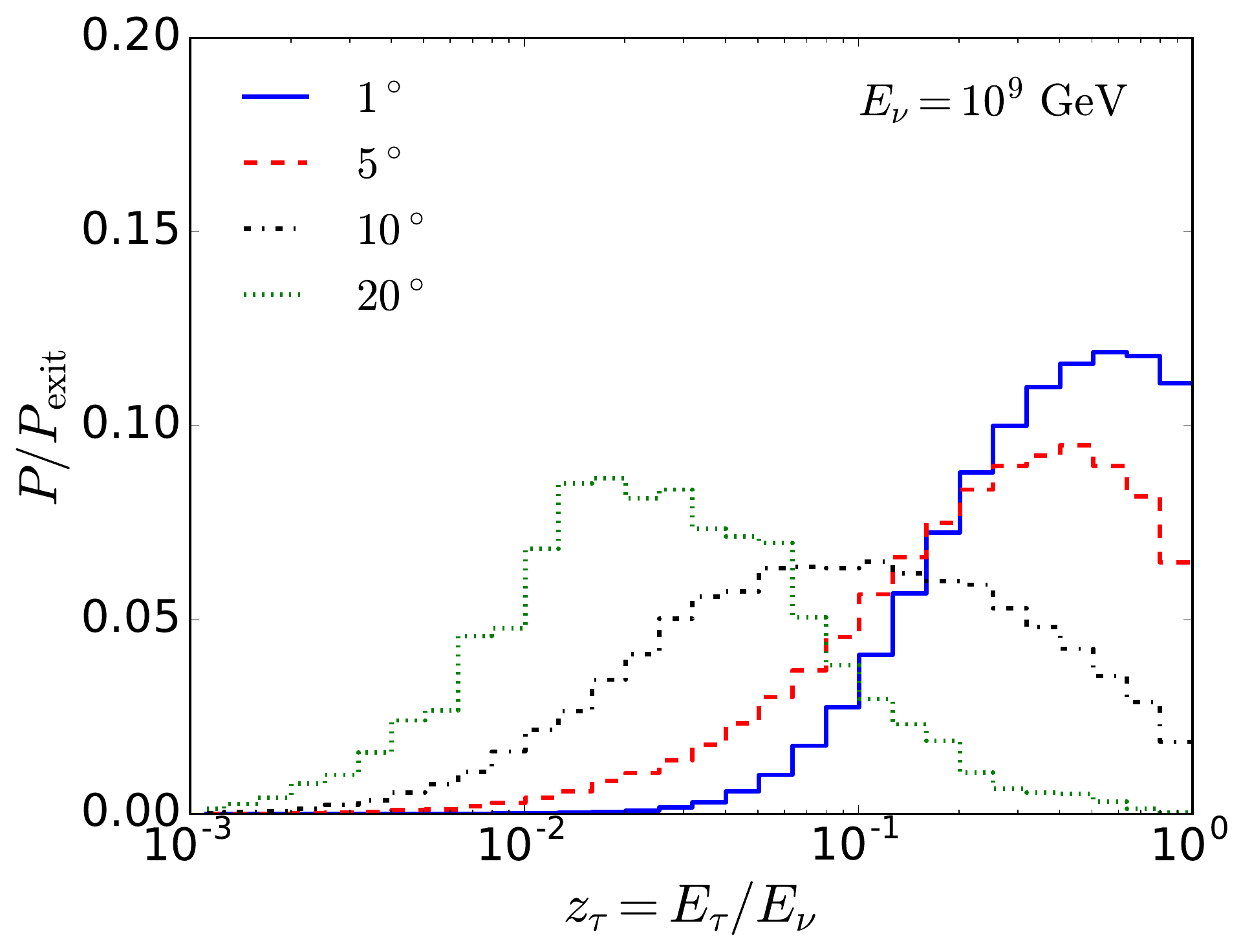}	
	\includegraphics[width=0.9\columnwidth]{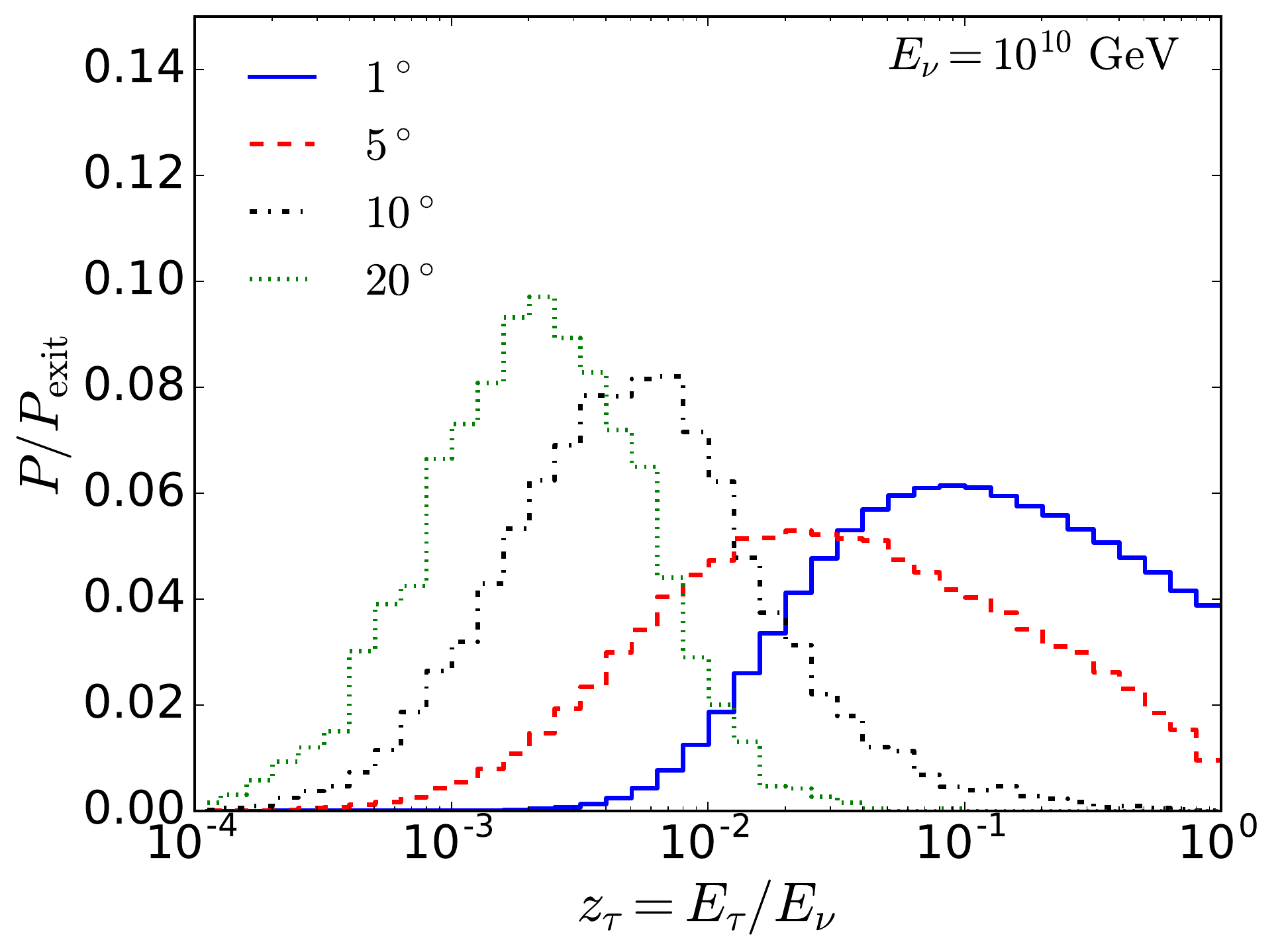}		
	\caption{\small \textit{Upper:} Relative probability of a tau emerging on a trajectory at an angle of $\beta_{\rm tr}=1^\circ,\ 5^\circ,\ 10^\circ$ and $20^\circ$ through the Earth for  $E_\nu=10^{9}$ GeV.
	\textit{Lower:} As above, with $E_\nu=10^{10}$ GeV. The figures are plotted as a function of $z_\tau=E_\tau/E_\nu$. }
	\label{fig:gridsN}
\end{figure}

The energy distributions of directly produced taus and regenerated taus are combined in the curves in Fig. \ref{fig:gridsN}
for $E=10^{9}$  and $10^{10}$ GeV. Plotted are the relative probabilities, the probability for a given bin normalized by the total exit probability shown in Fig. \ref{fig:proballm},  for taus to emerge as a function of energy fraction $z_\tau=E_\tau/E_{\nu_\tau}$. For low angles, the contributions are mainly from direct
neutrino production of taus that emerge, while at larger angles, the lower tau energies reflect the fact that regenerated taus dominate the number of taus that exit the Earth. The tau energy fraction upon exit depends on both the energy of the initial tau neutrino and the angle. For example, for $\beta_{\rm tr}=20^{\circ}$, the typical tau energy fraction is a few per cent of the tau neutrino energy for $E_\nu=10^9$ GeV, but only a few tenths of a percent for
$E_\nu=10^{10}$ GeV, even though the overall tau exit probabilities for these two energies differ by less than a factor of 2. 
Tables of tau exit probabilities and the relative probability for an
exiting tau with energy $E_\tau$ given the incident
neutrino energy $E_{\nu_\tau}$ and angle $\beta_{\rm tr}$ are used in the Monte Carlo evaluation of the aperture and sensitivity independent of neutrino flux. Representative tables are included in Appendix B.

\subsection{Flux dependent \texorpdfstring{$\nu_\tau\to\tau$}{nutotau} results}
\label{subsec:fluxdep}

The exit probabilities $P_\tau^{\rm exit}$ shown in Fig. \ref{fig:proballm} are for fixed neutrino energies, integrated over all exiting tau energies. We are also interested in the spectrum of exiting taus given an incident spectrum of neutrinos. 
As Fig. \ref{fig:gridsN} illustrates, the taus exiting with energy $E_\tau$ come from neutrinos with higher energies. We can also evaluate the ratio of the flux of emerging taus to an incident isotropic tau neutrino flux.
To begin, we show
the ratio $F_\tau(E)/F_\nu(E)$, the ratio of out-going taus to incident neutrinos, each at the
same energy $E$.  We denote the ratio by the terminology ``transmission function'' to emphasize that it includes not just the conversion of tau neutrinos to $\tau-$leptons, but also requires that the \taons exit the Earth before decaying. We use the cosmogenic neutrino 
fluxes of Kotera et al. in Ref. 
\cite{Kotera:2010yn} to exhibit the energy and angular dependence of the tau transmission functions.

\begin{figure}[ht]
\centering
\vskip 0.1in
\includegraphics[width=0.9\columnwidth]{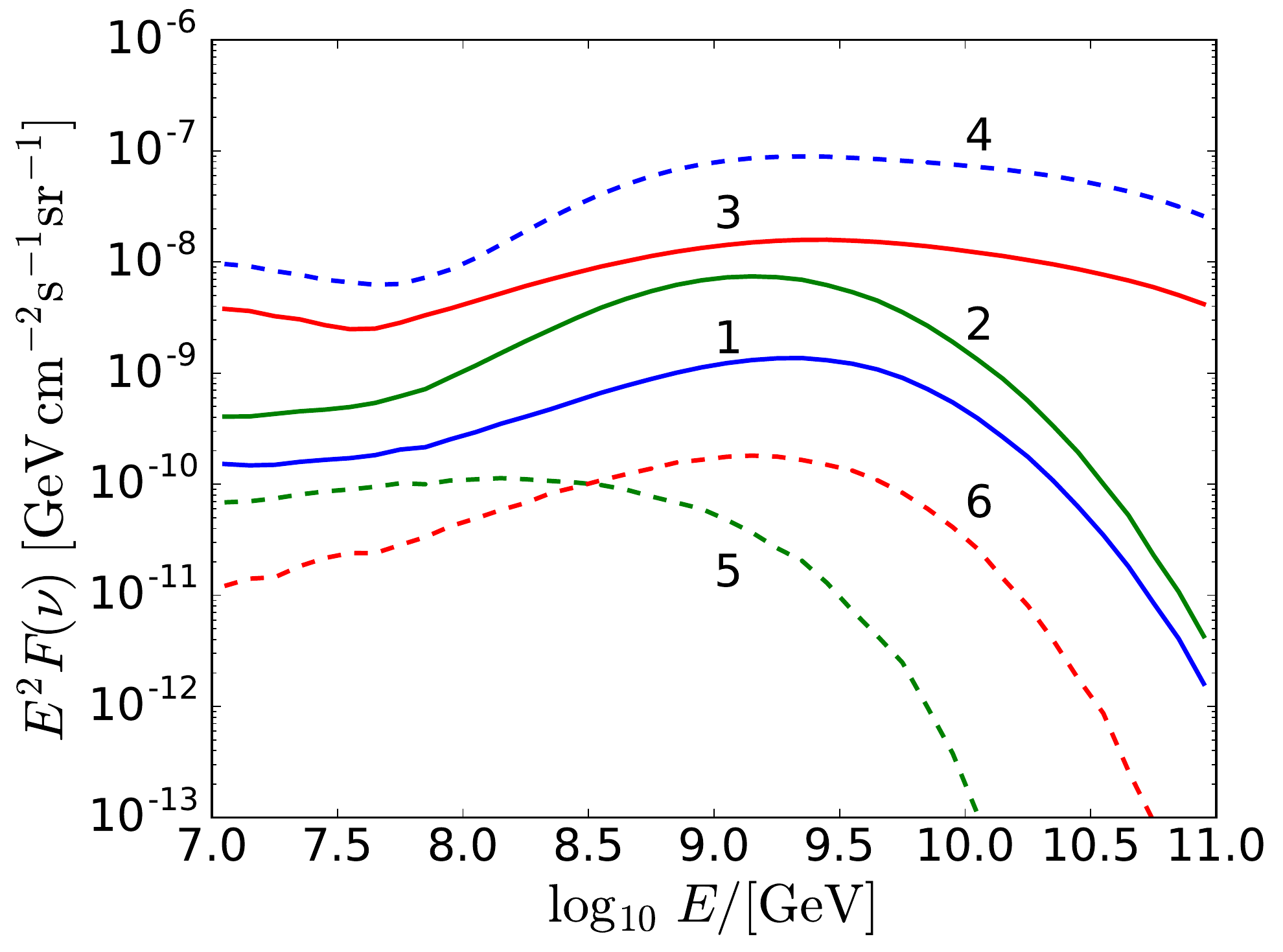}
\caption{\small Neutrino plus antineutrino flux summed over flavors, scaled by the neutrino energy squared, from Ref. \cite{Kotera:2010yn}. The curves represent 1: uniform evolution, mixed composition, $E_{p{\rm max}}=10^{11}$ GeV; 2: 
SFR1 evolution, mixed composition, $E_{p{\rm max}}=10^{11}$ GeV; 3: SFR1 and GRB2 evolution, protons, dip model,  $E_{p{\rm max}}=10^{12.5}$ GeV; 4: FR II  evolution, protons, $E_{p{\rm max}}=10^{12.5}$ GeV; 5: uniform evolution, an iron rich composition, $E_{p{\rm max}}=10^{10}$ GeV; 6: uniform evolution, iron, $E_{p{\rm max}}=10^{11}$ GeV. See text for details.}
\label{fig:kotera-neutrino-flux}
\end{figure}

The range of cosmogenic neutrino fluxes in
Ref. \cite{Kotera:2010yn} follow from reasonable inputs to the evolution of the source emissivity, maximum acceleration energy, chemical 
composition and galactic to extragalactic transition model. Injection spectra and overall normalizations for the cosmic rays were adjusted to best fit the Auger data, then used to predict the associated neutrino spectra \cite{Kotera:2010yn}. The neutrino fluxed from 
six combinations of these inputs are shown in Fig. \ref{fig:kotera-neutrino-flux}. Plotted is the sum over all flavors of neutrinos plus antineutrinos from all sources,
with the differential flux $F(\nu)$ scaled by the neutrino energy squared.
The tau neutrino plus antineutrino flux is one third of the flux shown in Fig. \ref{fig:kotera-neutrino-flux}.

In Fig. \ref{fig:kotera-neutrino-flux}, curves 1, 5 and 6 use a uniform source emissivity (no evolution of the sources) up to redshift $z=8$. Curves 2 and 3
use a star formation rate (SFR1) according to Hopkins and Beacom \cite{Hopkins:2006bw}. Curve 3 has an adjusted gamma ray burst evolution (GRB2) following  Le and Dermer \cite{Le:2006pt}. Curve 4 in Fig. \ref{fig:kotera-neutrino-flux} uses the evolution of Faranoff-Riley type II galaxies (FRII) of Wall et al. in Ref. \cite{Wall:2004tg}. A mixed cosmic ray composition is used in curves 1 and 2, pure protons in 3 and 4, an iron rich composition in curve 5 and pure iron in curve 6. The maximum proton acceleration energy is
$E_{p,{\rm max}} = 10^{11}$ GeV in curves 1, 2 and 6, $10^{10}$ GeV for curve 5, and $10^{12.5}$ GeV in  curves 3 and 4. For mixed composition
$E_{Z,{\rm max}}=Z E_{p,{\rm max}}$.
Curves 3 and 4 use a dip model for the transition from galactic to extragalactic sources.

\begin{figure}[ht]
\centering
	\includegraphics[width=0.9\columnwidth]{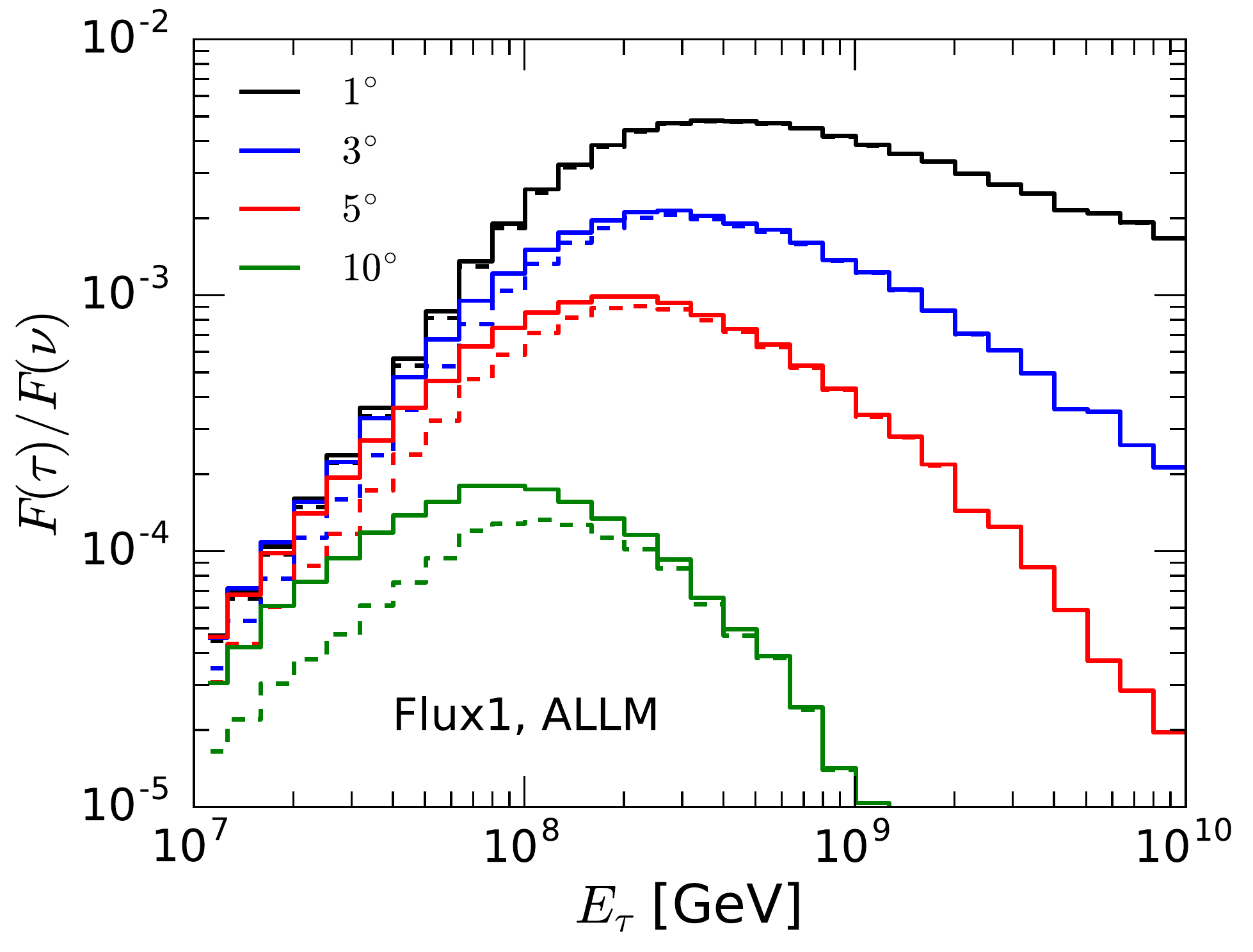}	
	\includegraphics[width=0.9\columnwidth]{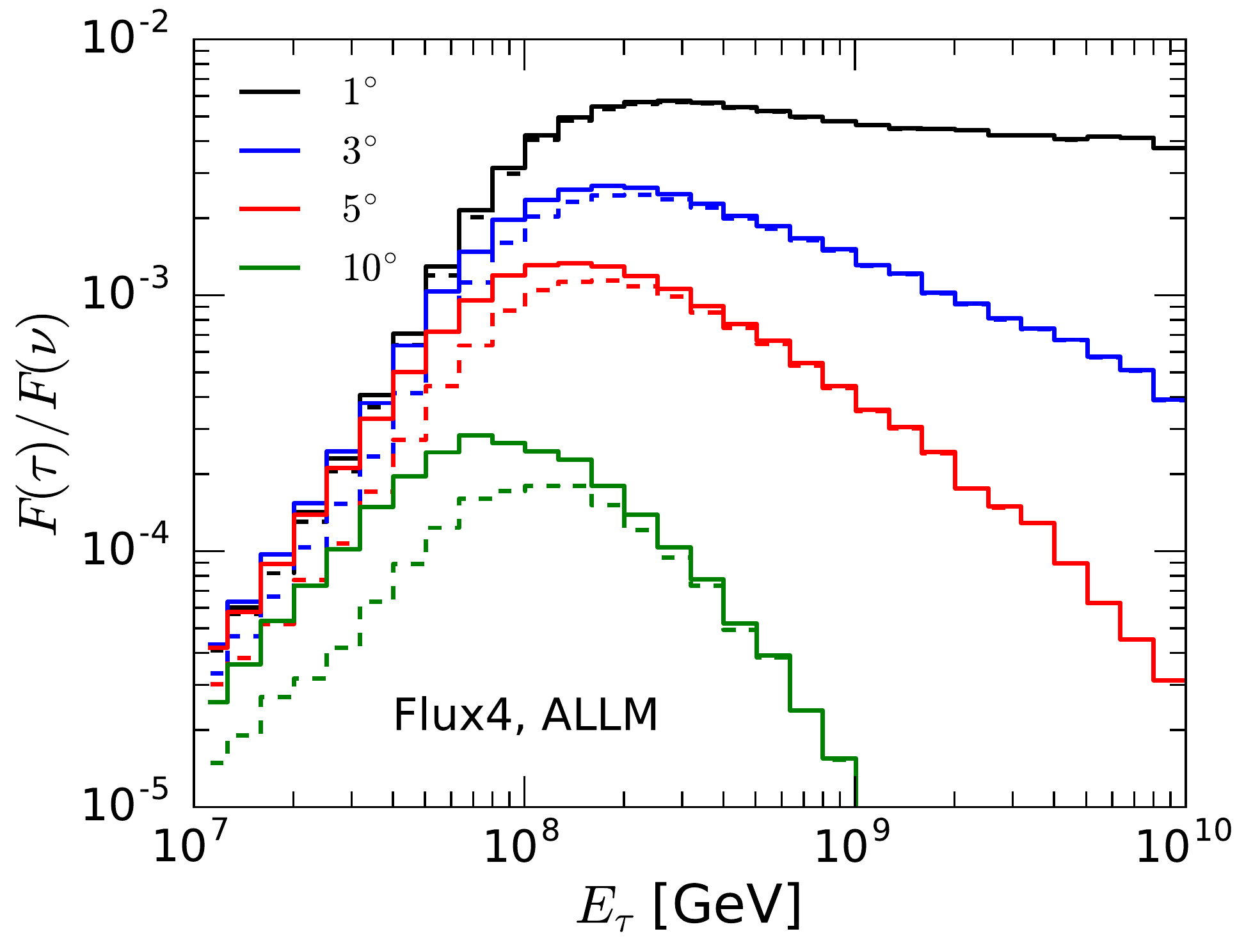}		
	\caption{\small \textit{Upper:} The ratio of the outgoing tau flux to the incident neutrino flux, at the same energies, for fixed values of the angle of the trajectory
	relative to the horizon $\beta_{\rm tr}$
	for cosmogenic flux 1 \cite{Kotera:2010yn}. The ALLM tau energy loss model is used, along with the standard model neutrino cross section.
	 The solid histograms include regeneration, while the dashed histograms do not. \textit{Lower:} As in the upper plot, for
	 flux 4.}
	\label{fig:ratios}
\end{figure}

For the results presented here, we use two of these representative cosmogenic neutrino fluxes, those labeled by 1 and 4, to evaluate transmission functions that depend on angle. 
Flux 4 is excluded by Auger in the energy range of $E_{\nu}\sim 4\times 10^8-4\times 10^9$ GeV
\cite{Zas:2017xdj,Aab:2019auo}. We use it here to illustrate the effect of a harder high energy cosmogenic
spectrum than that of flux 1. 
Fig. \ref{fig:ratios} shows the transmission ratios for fluxes 1 and 4, with the ALLM energy loss model and the standard model neutrino cross section. 
The dashed histograms show the transmission functions without regeneration, while the solid histograms include the effects of regeneration. Regeneration effects increase as $\beta_{\rm tr}$
increases, as expected. 

The transmission functions for the two fluxes show similar features. The increase in transmission function for $E=10^7-10^8$ GeV
is largely due to the $\gamma$-factor in the decay length and the increase in the cross section with energy. Some regeneration effects are evident already at $\beta_{\rm tr}=5^\circ$. At higher energies, the flattening in the ratio occurs because the tau range saturates. Eventually, attenuation of the neutrino flux
cuts off the transmission ratio. For $E_\tau$ larger than a few $\times 10^8$ GeV, the transmission ratio is primarily from a single charged-current interaction. 
Multiple interactions
in regeneration feed down the neutrino flux to a lower emerging tau flux.
The harder neutrino spectrum of flux 4 at high energies yields a larger ratio of the tau flux to the incident neutrino flux 4 compared to the ratio for flux 1 for energies above a few times $10^7$ GeV to a few times $10^8$ GeV, and for small angles $\sim 1^\circ-5^\circ$, above $E\sim 10^9$ GeV.  This can be seen in a comparison of the upper and lower panels of 
Fig. \ref{fig:ratios}.

\begin{figure}[ht]
\centering
	\includegraphics[width=0.9\columnwidth]{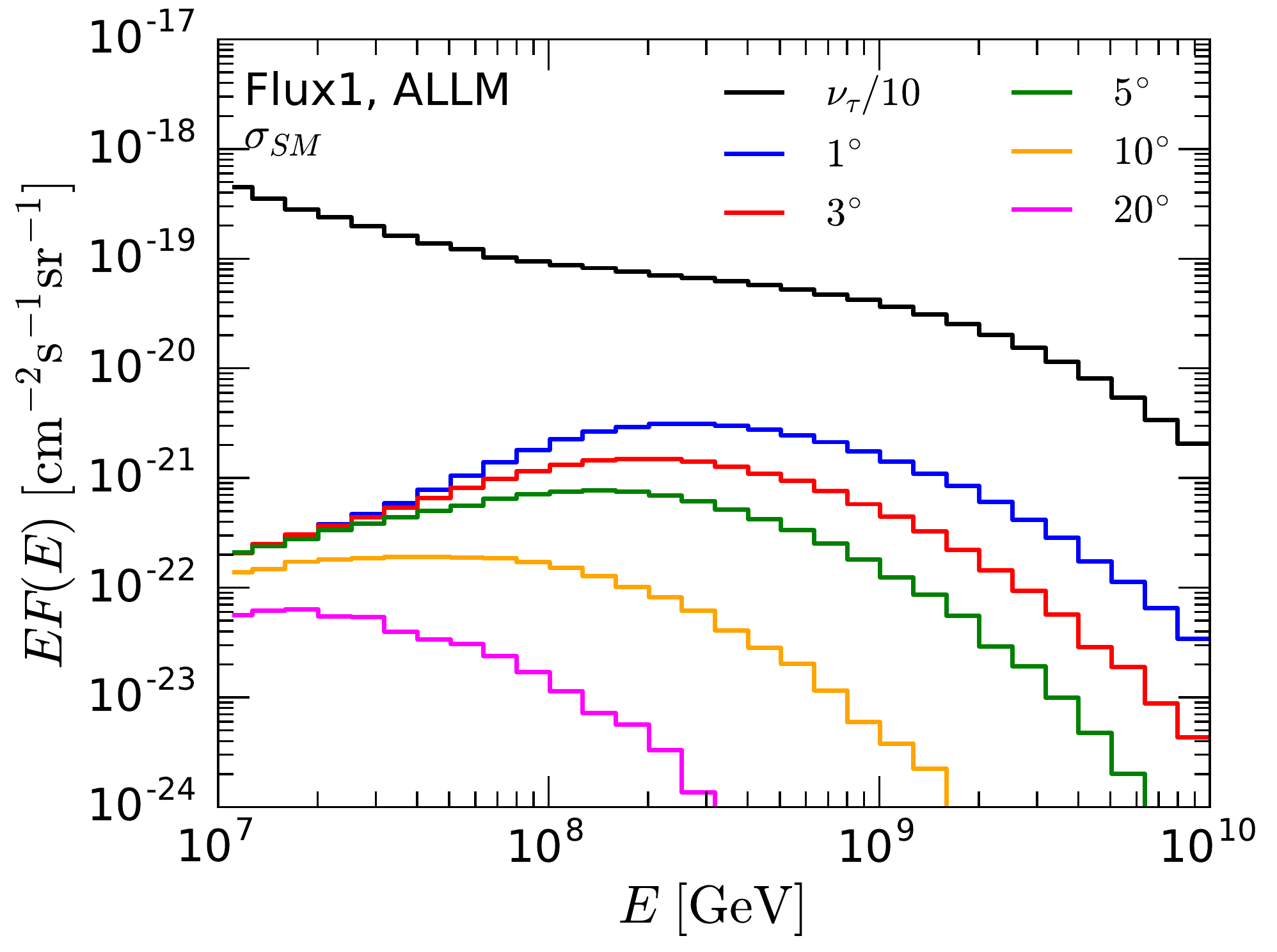}	
	\caption{\small The five lower histograms show the exiting tau flux scaled by energy as a function of tau energy for cosmogentic neutrino flux 1 \cite{Kotera:2010yn} and for fixed values of the angle of the trajectory
	relative to the horizon $\beta_{\rm tr}$. The ALLM tau energy loss model is used, along with the standard model neutrino cross section.
	 The upper-most histogram shows the incident tau neutrino flux scaled by a factor of $1/10$. 
	}
	\label{fig:taunorm}
\end{figure}

In Fig. \ref{fig:taunorm}, we show $EF_\tau(E)$ rather than the transmission function for flux 1 
to illustrate the difference in the energy behavior of exiting \taons compared to incident tau neutrinos. The figure comes from using the ALLM energy loss model, 
again for fixed angles $\beta_{\rm tr}$ relative to the horizon.
The much larger incident isotropic tau neutrino flux is scaled by a factor of 1/10.

\begin{figure}[ht]
\centering
	\includegraphics[width=0.9\columnwidth]{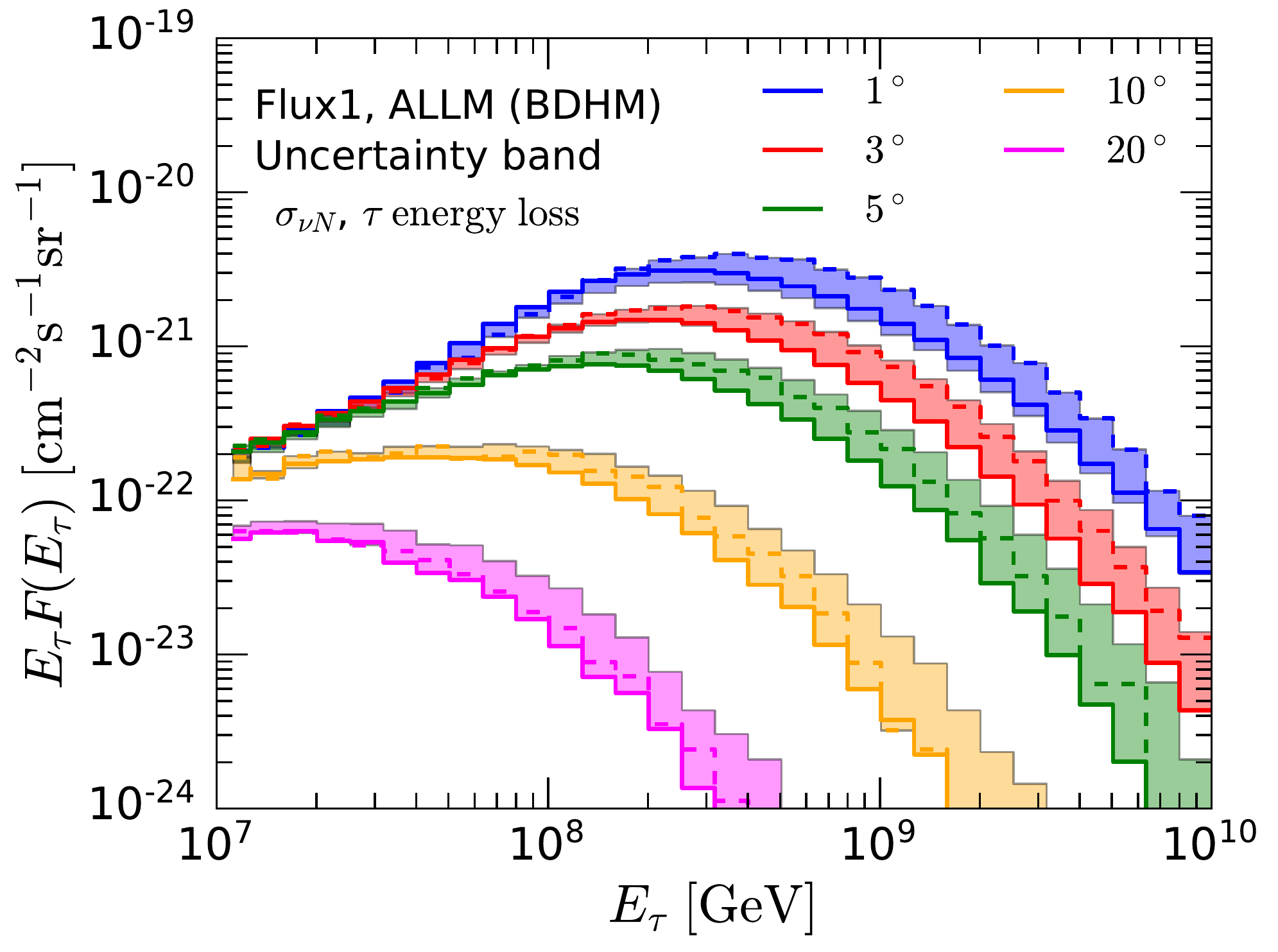}			
	\caption{The exiting tau flux scaled by energy
	as a function of tau energy for flux 1 \cite{Kotera:2010yn}, for fixed values of the angle of the trajectory
	relative to the horizon $\beta_{\rm tr}$. The ALLM tau energy loss model is shown with the solid histograms, while the BDHM energy loss model is
	shown with the dashed histograms, in both cases with the neutrino cross section taken to be $\sigma_{SM}$. 
	The band shows the minimum and maximum values of the energy-scaled flux when the 
	BDHM energy loss and neutrino cross section, and the ALLM energy loss and neutrino cross sections, are also considered.}
	\label{fig:taunormcompare}
\end{figure}

The energy loss model makes some difference in the predictions. In 
Fig. \ref{fig:taunormcompare}, the ALLM model results are shown with the solid histograms while the dashed histograms are results using the BDHM model for tau electromagnetic energy loss, both with standard model (SM) neutrino-nucleon cross section. The parameter $b_\tau^{\rm nuc}(E)$ evaluated using BDHM is smaller than for ALLM, so tau energy loss at high energies is smaller for BDHM than ALLM evaluations. This effect accounts for the difference
at high energies.
We note, however, that we use stochastic energy loss
rather than $\langle dE_\tau/dX\rangle=-b_\tau E$ for the tau energy loss to better model the exiting tau energy after transport through the column depth $X$.

Below $E_\tau=10^8$ GeV, there is little difference in the exiting tau fluxes for a fixed incident neutrino flux 
because the main feature
is that taus are produced in the final few kilometers before exiting the Earth.  The predicted tau fluxes with the two energy loss models differ by $\sim \pm 15\%$ for energies below $E_\tau=10^8$ GeV. Between $E_\tau=10^8-10^9$ GeV,
the ratio of the flux prediction from BDHM energy loss grows to a factor of $\sim 1.7$ relative to the ALLM energy loss prediction, increasing further, by more than a factor of 2, as the tau energy increases.

By changing the cross section for neutrino interactions, the variation in the predictions at high energies is wider than if our default SM  neutrino-nucleon
cross section is used, as shown with the error band in
Fig. \ref{fig:taunormcompare}. The error band shows the minimum and maximum exiting tau flux where we also use 
BDHM extrapolations for both the energy loss and the neutrino cross section,  and ALLM extrapolations for both the energy loss and the neutrino cross section.

\section{Tau air showers}\label{sec:three}

\subsection{Tau decays in the atmosphere}

\begin{figure}[ht]
\centering	
\includegraphics[width=0.9\columnwidth]{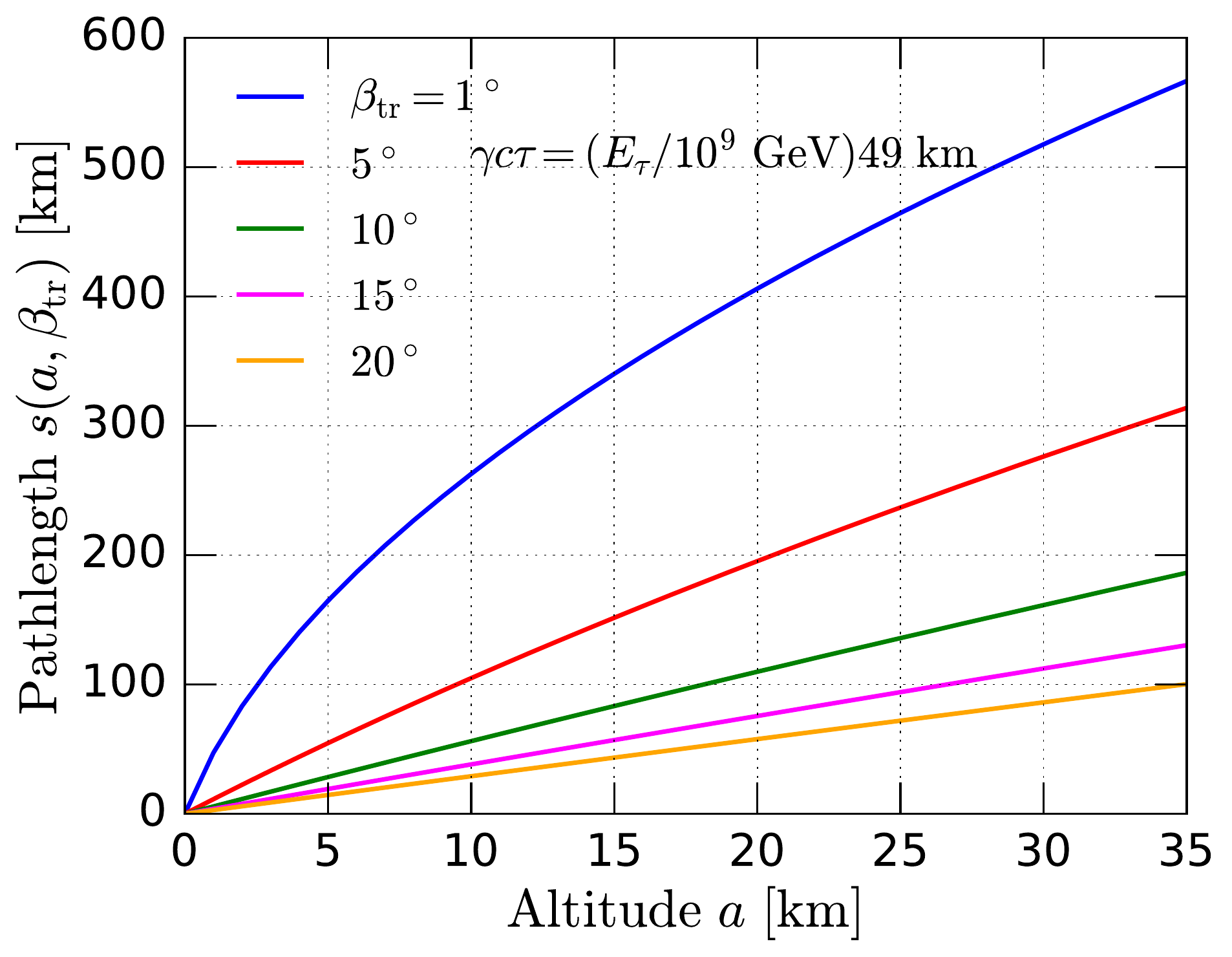}	
	\caption{\small Pathlength $s$ at altitude $a$ given a trajectory that emerges from the surface of the Earth at angle
	$\beta_{\rm tr}$ relative to the horizon.}
	\label{fig:pathlength}
\end{figure}

\begin{figure}[htb]
\centering	
\includegraphics[width=0.9\columnwidth]{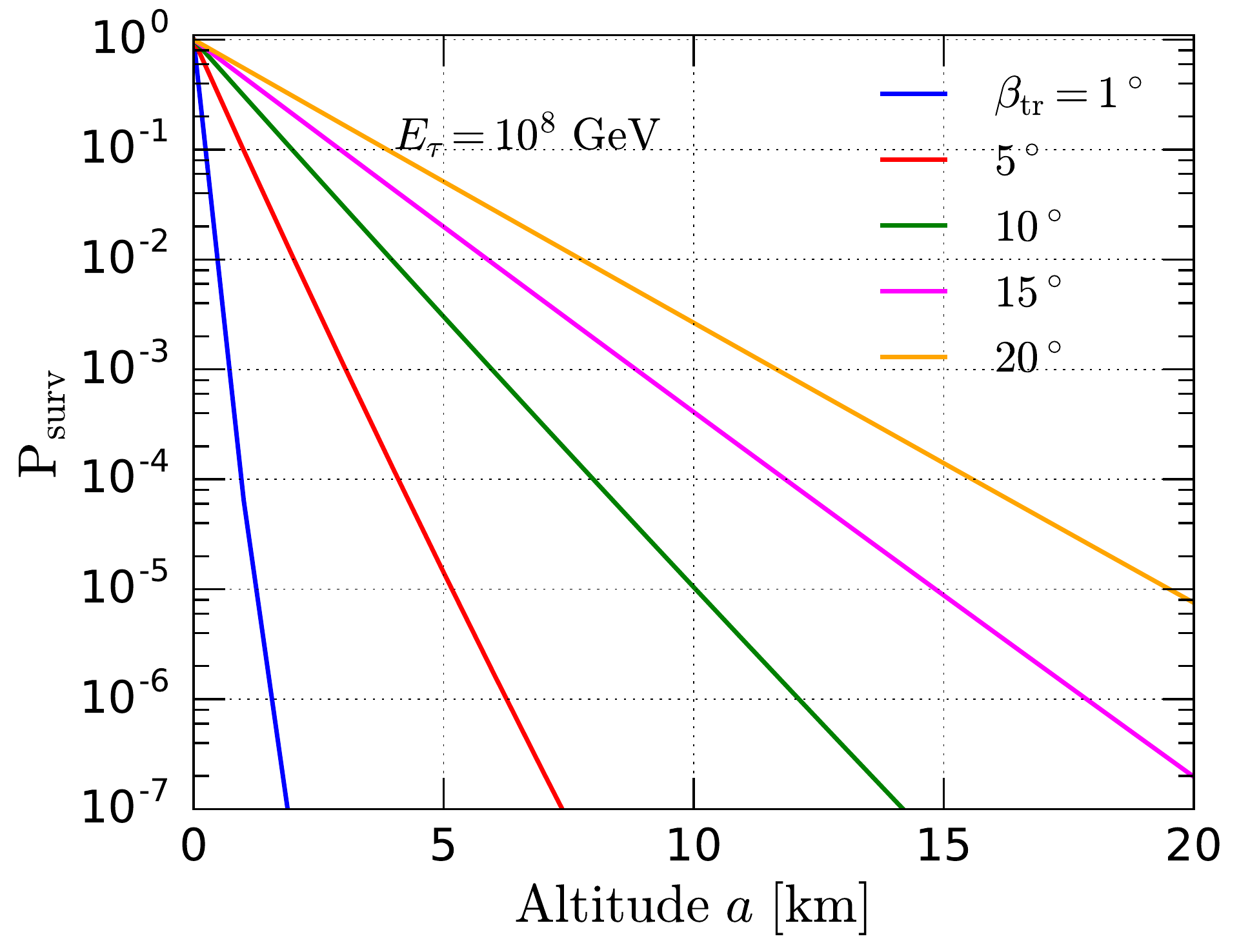}	
\includegraphics[width=0.9\columnwidth]{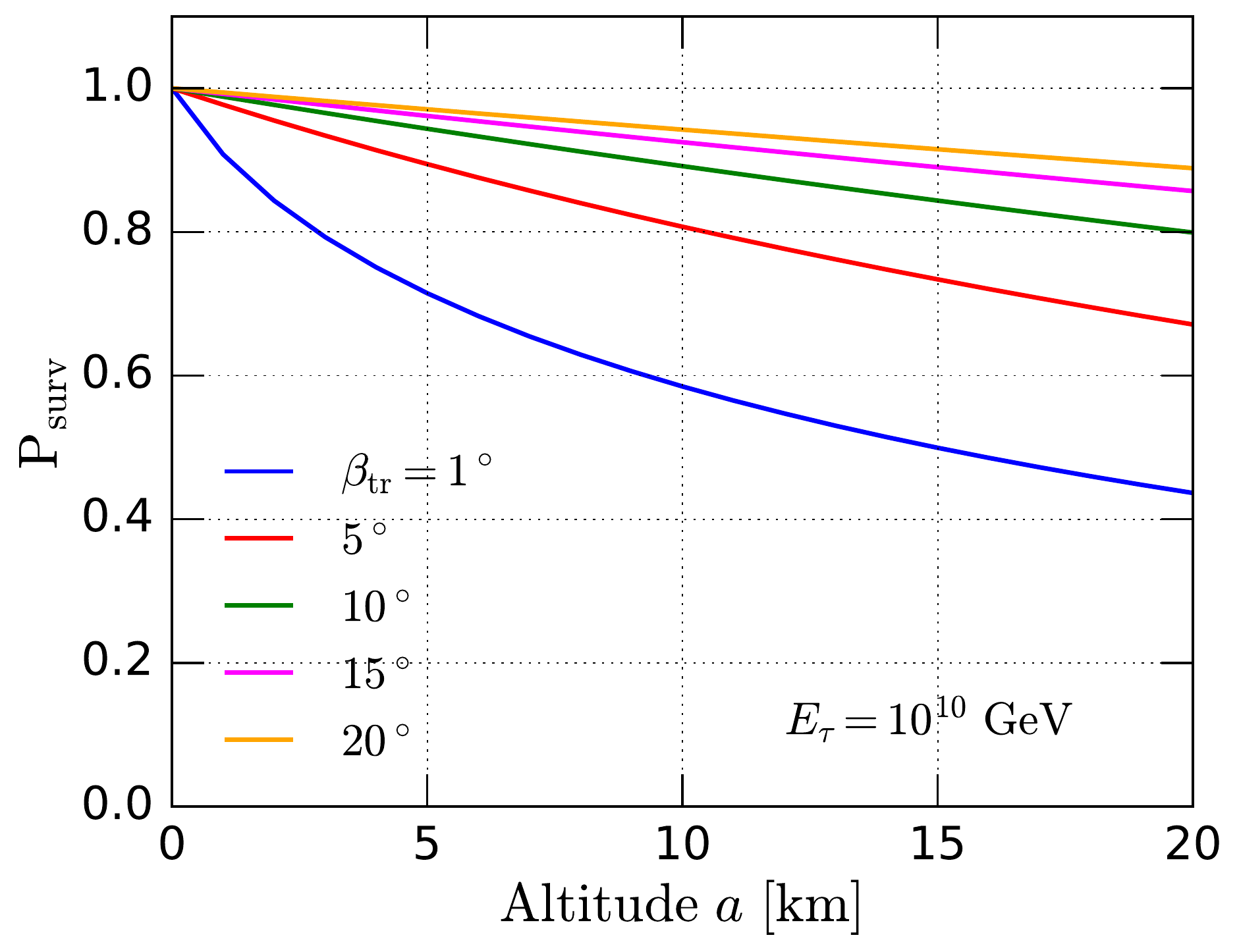}	
	\caption{\small \textit{Upper:} The tau survival probability as a function of altitude $a$ for $E_\tau=10^8$ GeV. \textit{Lower:} 
	As above, for $E_\tau=10^{10}$ GeV.}
	\label{fig:geo-decay}
\end{figure}

The signals that could be detected
by sub-orbital or space-based instruments come from air showers produced by tau decays in the atmosphere. The characteristics of the air shower discussed below in Sec. \ref{subsec:air_shower_model} depend on the tau emergence angle and the altitude at which the tau decays and the shower begins. For a given emergence angle, $\beta_{\rm tr}$, the  probability $P_{\rm decay}$ that a tau with energy $E_{\tau}$ will decay at an altitude, $a$, is given by
\begin{eqnarray}
{P}_{\rm surv} &=& \exp\Biggl(-\frac{s\left(a,\beta_{\rm tr}\right)}{\gamma c\tau}\Biggr)\\
{P}_{\rm decay} &=& 1-P_{\rm surv}=\int_0^{ s} p_{\rm decay}(s') ds' \ ,
\end{eqnarray}
where 
$s\left(a,\beta_{\rm tr}\right)$ is its path length through the atmosphere as a function of $a$ and $\beta_{\rm tr}$
and $p_{\rm decay}(s) = \exp(-s/\gamma c\tau)/(\gamma c\tau)$. 
Tau energy loss in the atmosphere will be small, so we neglect it here.
Fig. \ref{fig:pathlength} shows the path length as a function of altitude, derived from geometry
and
$\beta_{\rm tr}$
to be 
\begin{eqnarray}
\nonumber
    s(a,\beta_{tr}) &=& \sqrt{R_E^2\sin^2\beta_{\rm tr} + ((R_E+a)^2-R_E^2)}\\
    &-& R_E\sin\beta_{\rm tr}\ ,
\end{eqnarray}
for selected emergence angles $\beta_{\rm tr}$ between
$1^\circ-20^\circ$.
Fig. \ref{fig:geo-decay} shows the tau survival probability as a function of altitude for two tau energies: $10^8$ GeV (upper figure) and $10^{10}$ GeV (lower figure). Note that at higher energies, the survival probability changes slowly with altitude, so the $y$-axis in the lower figure is linear, whereas it is logarithmic in the upper figure.

In addition to the emergence angle and the altitude, the characteristics of the air shower also depend on the amount of energy that goes into the shower, $E_{\rm shr}$. The \taon is massive enough to produce quarks through its decay, giving rise to a variety of possible final states with substantial variations in the amounts of energy being given to daughter particles that will produce 
extensive air showers (EASs), namely, electrons and hadrons. Full-scale air shower simulations that would include Monte Carlo simulations of \taon decays is beyond the scope of this paper. Instead, for the purposes of modeling the \taon air showers, we scale the energy of the shower as a fraction of the tau energy. To that end, we performed Monte Carlo simulations of tau decays in PYTHIA 8 and determined that the mean fraction of the tau energy that is given to showering daughters is approximately $\sim 50$\% (see Fig. \ref{fig:eshr} in Appendix C), so for all
of the results shown below, $E_{\rm shr}=0.5\, E_\tau $.

\subsection{Air shower modeling}
\label{subsec:air_shower_model}

Once the emergent \taon energy is determined, the resultant EAS needs to be generated based on the location of the \taon decay in the atmosphere.
We neglect the decay channel $\tau \rightarrow \mu \nu_\mu \nu_\tau$, with a branching fraction of 17\%. The muonic channel, while interesting, has different shower characteristics than the hadronic and electromagnetic induced extensive air showers modeled in this paper \citep{Stanev:1989bc}.

The EAS modeling philosophy we employ uses the established technique of using a parametric model to describe the average EAS development based on shower universality arguments \cite{Hillas1,Hillas2,Nerling-2006-APP-24-421}.
We develop the shower and generate the Cherenkov light in a modeled atmosphere and take into account the dominant attenuating atmospheric processes when propagating the signal to the detector. The unique nature of modeling the optical Cherenkov signal induced by upward-moving EASs and measuring the signal using a space-based instrument motivates this philosophy as a well-defined initial simulation.  This traditional modeling approach also offers a method to assess the use of simulation packages such as CORSIKA \cite{CORSIKA} that were developed for ground-based measurements but require significant modifications to adapt their use for upward-moving rather than down-going EAS modeling \cite{Ahnen:2018ocv,Otte:2018uxj}.

A detailed atmospheric model is required to define the EAS development, the beamed Cherenkov light emission and the Cherenkov light attenuation based on the optical depth between the EAS and observation point.
We employ a static, baseline model for the definition of the atmospheric index of refraction (N$_{\rm Air}$) as a function of altitude based on that given by Hillas \cite{Hillas1}, which provides N$_{\rm Air}$ as a function of temperature and atmospheric overburden, g/cm$^2$. We use the model of Shibata \citep{Shibata} to define the overburden.

The Cherenkov light attenuation includes the effects of Rayleigh scattering \cite{Sokolsky1989}, Mie (aerosol) scattering and ozone absorption. A model for calculating the wavelength dependence of Mie scattering uses the data presented by Elterman in
Ref. \citep{Elterman}, which also defines the atmospheric aerosol profile.  The Earth's ozone layer efficiently attenuates optical signals at shorter wavelengths ($\lambda \lsim 330$ nm). An ozone attenuation model \citep{Krizmanic1999} is used with an altitude dependent profile derived from Total Ozone Mapping Spectrometer (TOMS) measurements \citep{McPeters}. This composite parametric atmospheric model easily accommodates the underlying spherical geometry.

For the results presented in this paper, the EAS are modeled using the Greisen parameterization \cite{Hillas1} with the optical Cherenkov signal calculated over the 200 - 900 nm wavelength band \cite{Fernow1986}.  The EAS and Cherenkov signal development is generated in 100 meter linear increments as a function of Earth-emergence angle ($\beta_{\rm tr}$) and altitude $a$ of the \taon decay using the inherent spherical geometry. The 3-dimensional nature of the EAS development is modeled using Hillas's parameterization of the angular and energy spectra of the EAS-electron distributions as a function of shower age \citep{Hillas2}, with the Cherenkov angles and energy thresholds based on the local index of refraction.  As noted above, the EAS development simulation results rely on shower universality arguments. CONEX simulations \cite{Bergmann:2006yz} have been used to test our approach for upward EAS modeling, and the comparison shows very good agreement up to an altitude of 15 km, above which CONEX modestly enhances the particle yields.

\begin{figure}[t]
	\begin{center}
		\includegraphics[width=0.8\columnwidth]{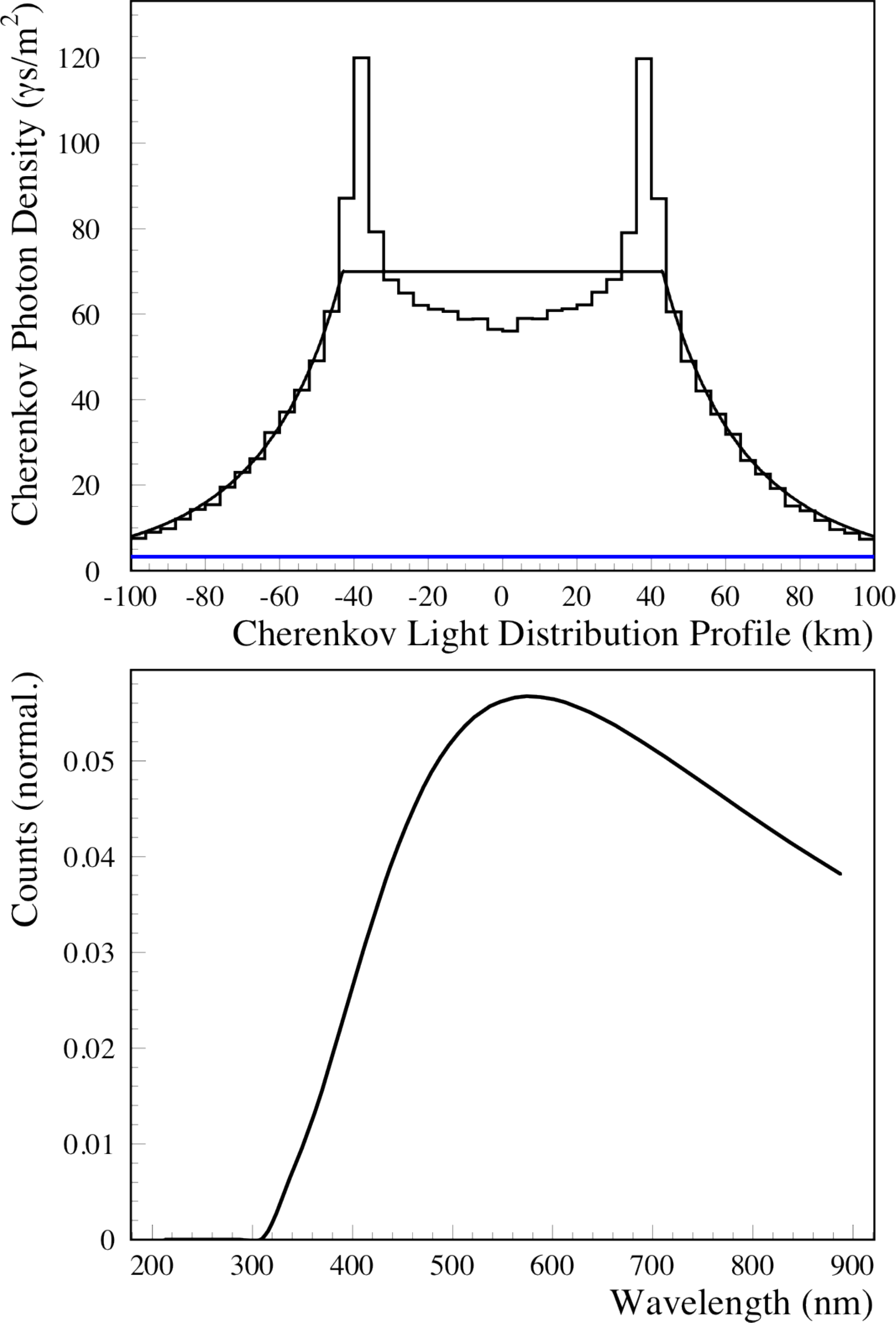}
	\end{center}
	\vspace{-0.75 cm}
	\caption{\small \textit{Upper:} The spatial profile of the Cherenkov signal
      (photons/m$^2$) at 525~km altitude for a 100~PeV upward EAS with
      a $10^\circ$ Earth emergence angle initiated at sea level. \textit{Lower:} The simulated Cherenkov spectrum observed at 525 km altitude for the EAS. }\label{CherProfFig}	
\end{figure}

Fig.~\ref{CherProfFig} shows the Cherenkov profile (upper figure), in photons/m$^2$, for a 100 PeV EAS initiated at sea level with an Earth-emergence angle of $\beta_{\rm tr}=10^\circ$ and the attenuated Cherenkov light spectrum (lower figure), both calculated at 525 km altitude. The photon density is approximately constant within the Cherenkov ring (horn-like structure in the profile), but for the Cherenkov light that falls outside the ring, the flux drops off as a power law with radius.  Note that at 525 km altitude, the Cherenkov ring spans $\sim$ 100 km and the photon density is quite substantial for a 100 PeV EAS. The Cherenkov ring is very pronounced in the profile due to fact that for modest $\beta_{\rm tr}$ the EAS completes its development at low altitudes, $< 5$ km in this case where the Cherenkov emission angle varies only by $\sim 0.2^\circ$.  The top panel in Fig.~\ref{CherProfFig} also shows a profile function \cite{ANeronovCom} that  describes the Cherenkov spatial profile as a flat top with an exponential falloff. This function is used to describe the Cherenkov signal intensity in the \taon EAS simulation for the results presented in this paper.
Details of this implementation are in Appendix C.

\begin{figure}[t]
	\begin{center}
		\includegraphics[width=0.8\columnwidth]{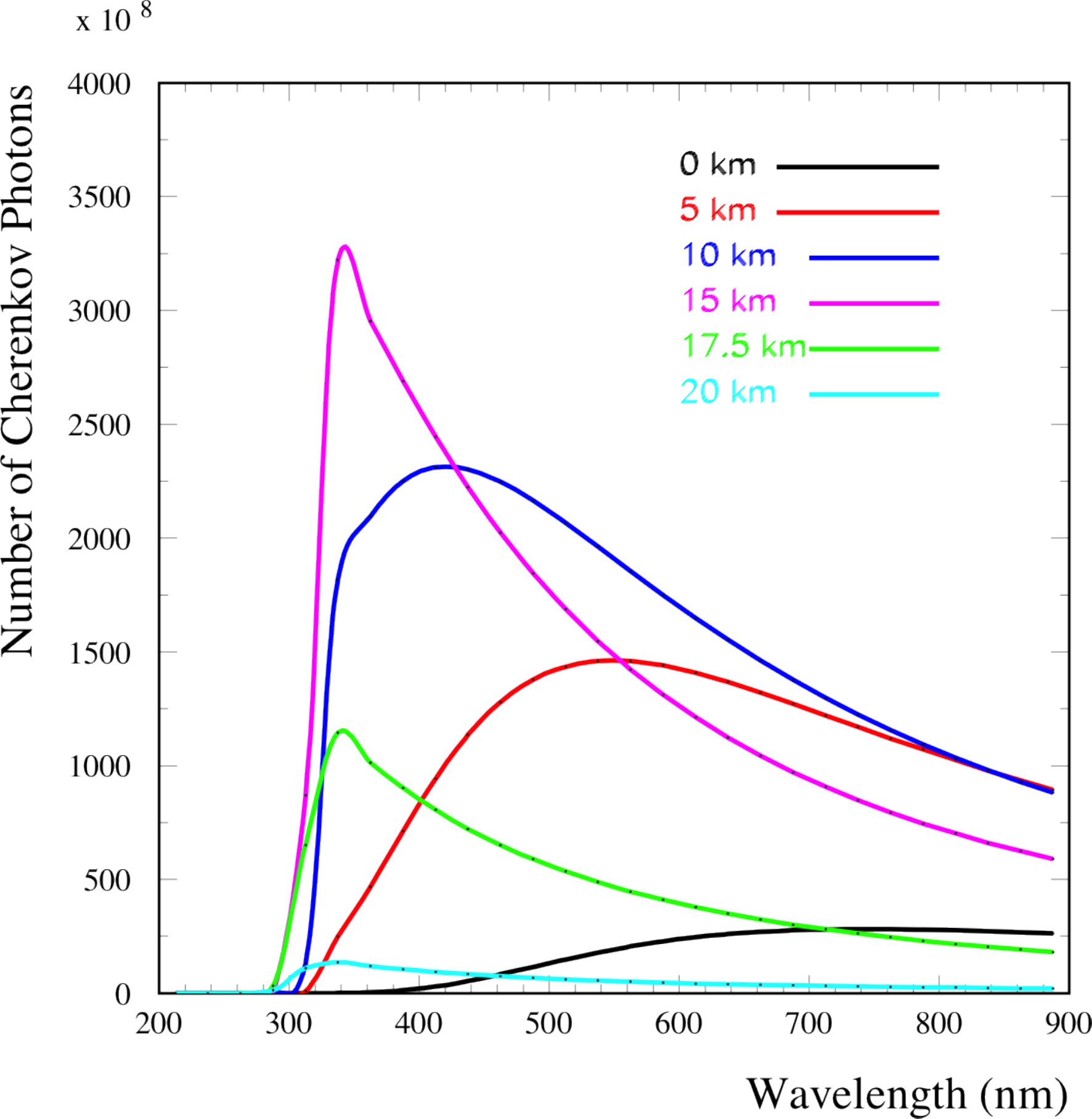}
	\end{center}
	\vspace{-0.75 cm}
	\caption{\small The intensity and wavelength dependence of the Cherenkov signal for 100~PeV upward-moving EASs for 5$^\circ$ Earth emergence angle 
as a function of EAS starting altitude.}\label{CherAltFig}	
\end{figure}

The intensity and spectrum of the Cherenkov light delivered to a specific altitude is a function of the \taon energy, Earth-emergence angle $\beta_{\rm tr}$, and attenuation in the atmosphere, which can be severe for small values $\beta_{\rm tr}$ mainly due to the effects of the low-altitude aerosols. The interplay between the EAS development and Cherenkov light attenuation is shown in Fig.~\ref{CherAltFig} where the simulated intensities and Cherenkov spectra are shown for 100 PeV EAS with $\beta_{\rm tr}=5^\circ$ but with the EAS initiated at different altitudes.  The EAS energy 100 PeV is chosen for comparison purposes. 
At the lowest altitudes, aerosol absorption decimates the Cherenkov intensity and pushes 
the spectrum towards the longest wavelengths. 
However, the exponential nature of both the aerosol layer ($\sim$ 1 km scale height) and atmosphere itself ($\sim$ 8 km scale height)
leads to signals with higher Cherenkov intensities and spectra peaked at lower wavelengths fairly 
quickly as a function of EAS starting altitude, due to the nature of the upward-moving \taon EASs. 

Eventually the atmosphere becomes too 
rarefied for complete EAS development at altitudes $\gsim$ 17 km. The Cherenkov angle becomes significantly reduced at higher altitudes and the Cherenkov threshold energy is increased due to the index of refraction of air (N$_{\rm air}$) approaching unity. These combine to lead to a significant reduction in the Cherenkov intensity for 
EAS that have a large fraction of their development above $\sim$ 17 km altitudes. The energy scale for a \taon to have a non-negligible decay probability at these high altitudes depends on $\beta_{\rm tr}$, but for $\beta_{\rm tr}=5^\circ$ and $E_\tau = 2.5$ EeV, 25\% of the \taons will decay above and altitude of 17 km.

\begin{figure}[t]
	\begin{center}
		\includegraphics[width=0.8\columnwidth]{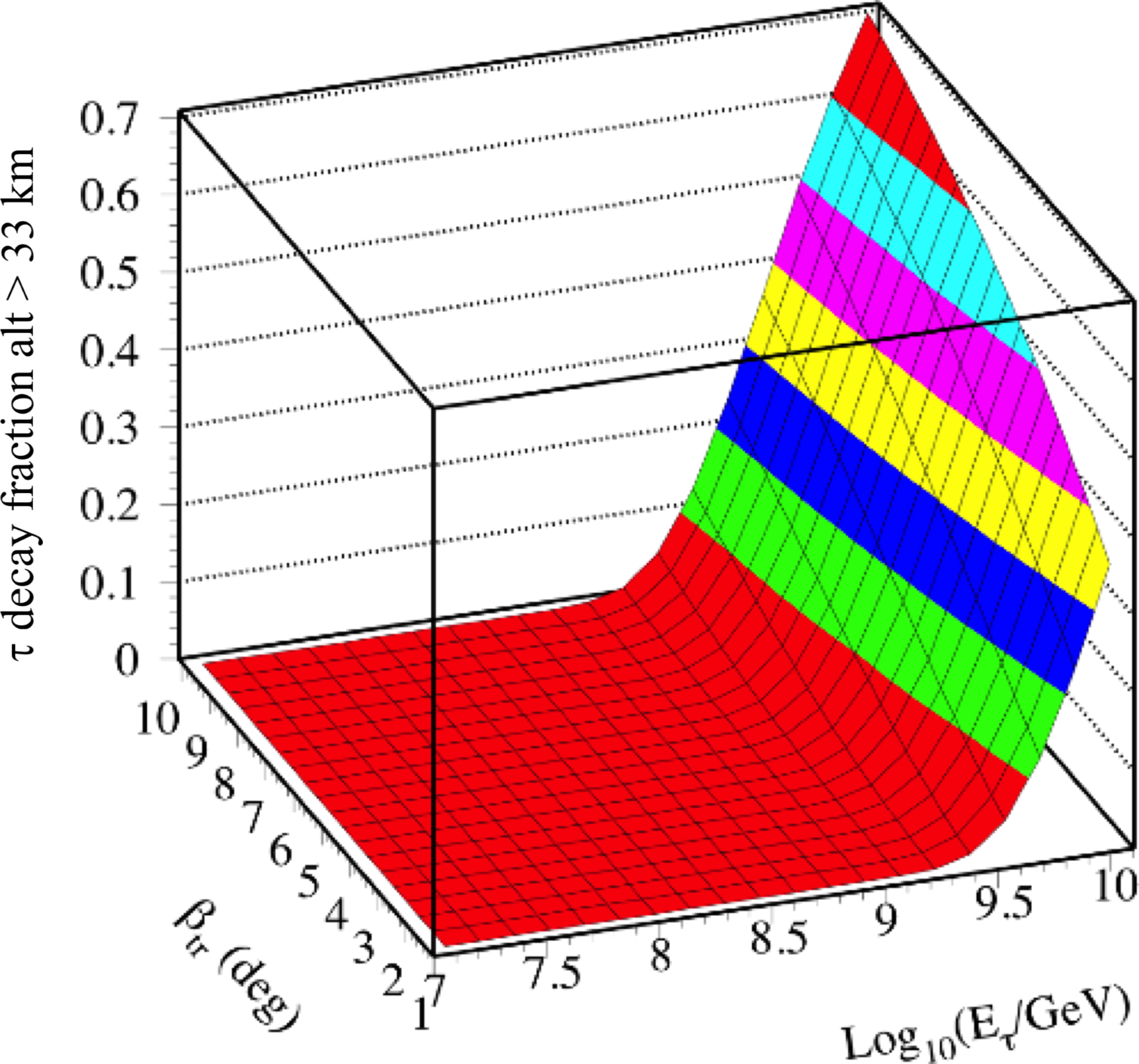}
	\end{center}
	\vspace{-0.75 cm}
	\caption{\small The fraction of taus that decay at an altitude larger than 33 km, as a function of 
	$\beta_{\rm tr}$ and $\log_{10}(E_\tau/{\rm GeV})$. The colored bands show 0.1-increments of the tau
	decay fraction. }
	\label{fig:airshralt}	
\end{figure}

The fact that UHE \taons can decay at altitudes comparable to that used by balloon-borne experiments is an interesting phenomena. Initial studies indicate that instruments on scientific balloons at altitudes $a\simeq 33$ km could be in the electromagnetic part of the EAS itself for \taon energies above $\sim$ EeV for Earth emergence angles below 
$\beta_{\rm tr}\sim 10^\circ$.  
For $\beta_{\rm tr}=1^\circ$ and $E_\tau=10^{10}$
GeV, at an altitude of $33$ km, the
probability for a \taon decay above that altitude is 0.33, while for $\beta_{\rm tr}=10^\circ$ for the same \taon energy, the decay probability above $a=33$ km is 0.70.
This is illustrated in 
Fig. \ref{fig:airshralt},  which shows the fraction of \taon decays at an
altitude larger than 33 km as a function of \taon energy and $\beta_{\rm tr}=1^\circ-10^\circ$, with colored bands marking increments of 0.1 in the tau decay fraction. For small $\beta_{\rm tr}$, the line of sight distance $v$ is large, so except for the highest energies, almost all of the \taons decay below $a=33$ km. On the other hand, for larger $\beta_{\rm tr}$, $v$ is shorter, so there is the possibility for more \taons to decay at higher altitudes.
The accurate calculation of the Cherenkov signal for this case requires a 3-dimensional particle-cascade simulation, e.g., CORSIKA \cite{CORSIKA} or Cosmos \cite{Cosmos} but with modifications for the modeling of the Cherenkov signal for upward-moving EASs. However, our simulation approach is valid for balloon altitudes ($\sim$ 33 km) for below an EeV, where the \taon decays below $\sim$ 5 km.

For space-based observations using an $\sim 0.1^\circ$ focal plane pixel field-of-view (FoV), a 1-dimensional treatment of the EAS signal is sufficient. 
This can be understood by considering the relevant distance scales. Assuming the EAS width is defined by a Moli\`ere radius value 8.83 g/cm$^2$ for air at STP
\cite{PDG}, near sea level the EAS radius is $\sim$ 100 m. From the view of the EAS from 525 km altitude, the 100 m radius is well contained in a single 0.1$^\circ$ pixel, even for nadir viewing. For viewing a highly inclined EAS originating near the Earth's limb, the distance to shower maximum is $> 1000$ km (assuming a 525 km orbit) for the Earth-emergence angles ($\beta_{\rm tr}$) with reasonable \taon exit probabilities.  This distance scale includes those $> 1$ EeV \taons that can decay at altitudes $\sim$ 20 km. While the the EAS radius will widen to $\sim$ 1 km at an altitude of 20 km ($\sim$ 10\% atmospheric pressure), the width of the visible portion of the EAS is still well contained in a $0.1^\circ$ pixel. In contrast, for observations on balloon-borne experiments ($\sim$ 33 km altitude) or on a mountain-top, such as Trinity ($\sim$ 3 km altitude) the width of an \taon EAS can be large compared the pixel FoV and a 3-dimensional EAS cascade development model is more appropriate.

Thus for the calculation of the Cherenkov signal intensity, spatial extent, and spectrum for low-Earth orbits, we use a parametric model based on our EAS 3-dim Cherenkov approach, which is much more computationally efficient when sampled in a Monte Carlo.  The Cherenkov intensities and angles as functions of $\beta_{\rm tr}$ and EAS decay altitude, are tabulated for a fixed, 100 PeV EAS energy in a library format. A profile function fit is used, shown in Fig.~\ref{CherProfFig}, to describe the beamed Cherenkov ``flattop'' signal within the Cherenkov cone, ignoring the horns. As discussed in Appendix C, we scale the intensity as a function of \taon energy and use a mathematical function to account for the increase in the effective Cherenkov acceptance angle for bright signals that place portions of the power-law part of the Cherenkov profile (outside the Cherenkov ring) above the detection threshold of the instrument. This models the increase in acceptance solid angle for brighter EASs.

\begin{figure}[t]
	\begin{center}
		\includegraphics[width=0.8\columnwidth]{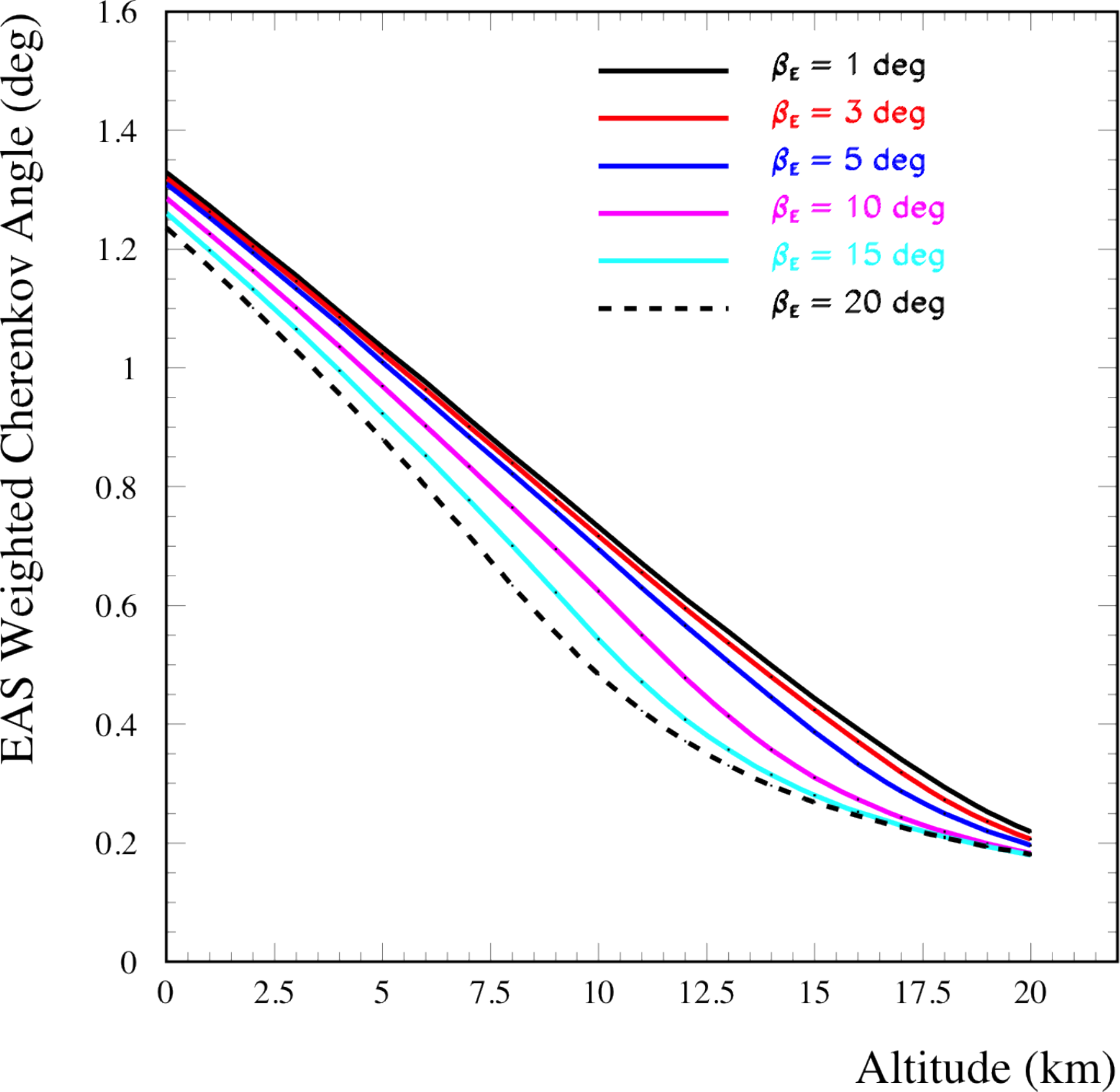}
		\includegraphics[width=0.8\columnwidth]{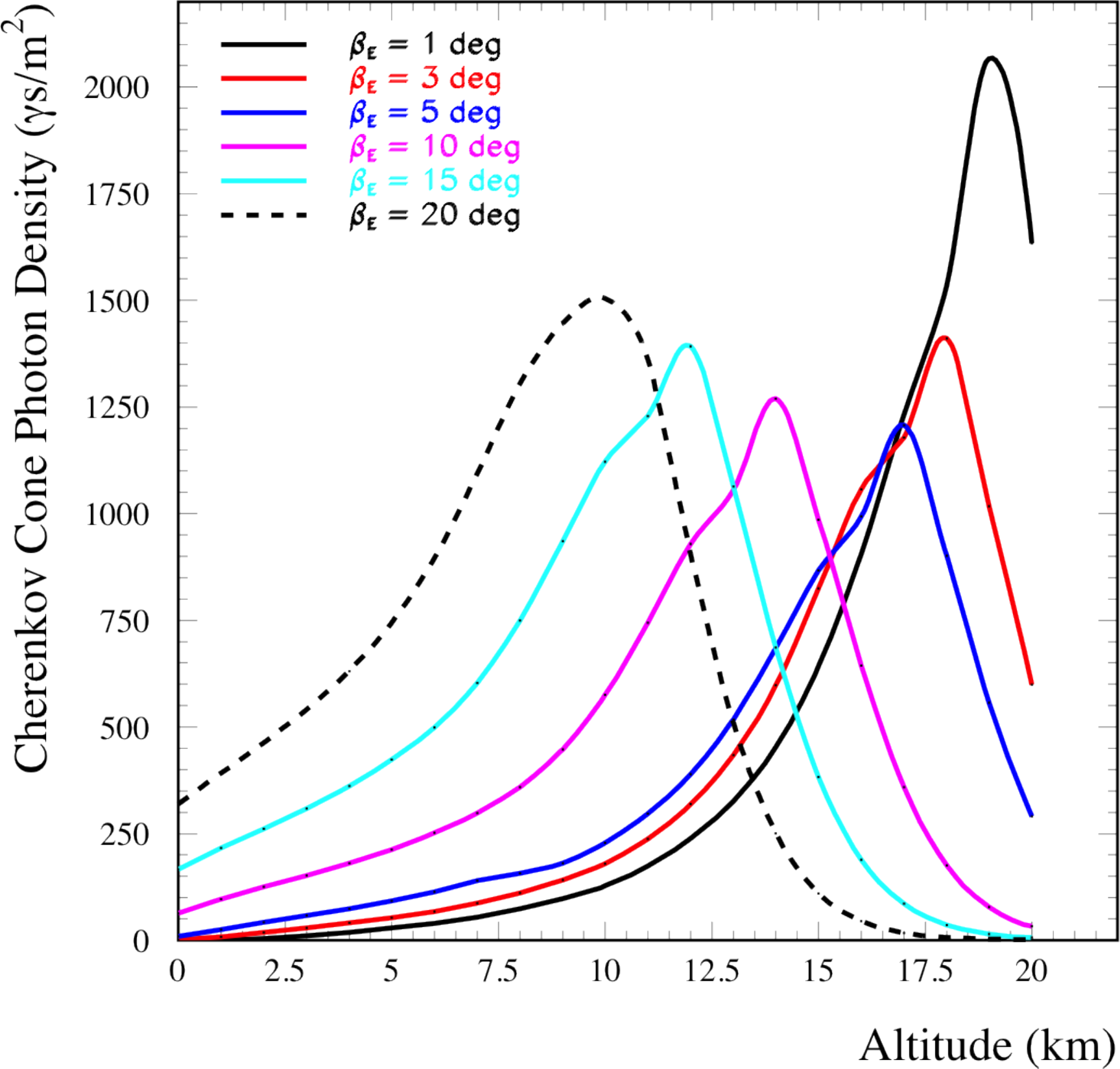}		
	\end{center}
	\vspace{-0.75 cm}
	\caption{\small
	\textit{Upper:} The Cherenkov angle $\theta_{\rm Ch}^0$ as a function of starting altitude for a 100 PeV air shower from a tau decay from the 1-dimensional Cherenkov EAS model.
	\textit{Lower:} Cherenkov cone photon distribution as a function of starting altitude and Earth-emergence angle for a $100$ PeV air shower from the 1-dimensional Cherenkov EAS model.}\label{cherenfig}	
\end{figure}

The Cherenkov angle $\theta_{\rm Ch}^0$ as a function of starting altitude, for a 100 PeV showers, is shown by the upper figure of Fig. \ref{cherenfig},  based on an evaluation of 3-dimensional EAS Cherenkov simulations. Showers that start at lower
altitudes have a Cherenkov angle between $\sim 1.2^\circ
-1.3^\circ$. The
Cherenkov angle reduces with altitude due to the reduction of the atmospheric index of refraction.
The detection of the air shower depends on the photon density at the detector, which in turn depends on the altitude of the detector, the altitude of the start of the air shower, and the Earth-emergence angle. For our evaluation of the sensitivity of instruments with POEMMA performance, we consider a detector at an altitude of $h=525$ km.
The photon density within the Cherenkov cone that arrives at such a detector is shown in the lower panel of Fig. \ref{cherenfig}. 

The photon density for $E_{\rm shr}=100$ PeV = $10^8$ GeV is the starting point for the photon density at other energies. We
approximate the photon density at POEMMA as a function of energy to be
\begin{eqnarray}
\nonumber
    \rho_\gamma(a,\beta_{\rm tr},E_{\rm shr}) &=& 
    \rho_\gamma(a,\beta_{\rm tr},E_{\rm shr}=10^8\ {\rm GeV})\\
    &\times &
    \frac{E_{\rm shr}}{10^8 \ {\rm GeV}}\ .
\end{eqnarray}
We discuss below how large photon densities effectively increase
the Cherenkov signal acceptance angle, or solid angle, in our evaluation of the POEMMA tau neutrino sensitivity.

We have not included geomagnetic effects in our EAS modeling. The angular spread of electrons in an EAS is due to Coulomb scattering. At the higher elevations, the rarified atmosphere leads to an longer Coulomb scattering length and thus geomagnetic bending can lead to an appreciable enhancement of the
angular distributions of the shower \cite{PhysRev.93.646.2,PhysRev.95.1705.4}.
We find that the effective Cherenkov acceptance angle enhancement may be as much as a factor of $\sim 2.2$ larger than
what we use in our model. The fact that this is a high altitude effect means that the enhancement is for showers
with energies higher than $\sim$EeV energies, most of which are already well above detection thresholds. As indicated in Fig. \ref{fig:taunormcompare}, the geomagnetic enhancement has the most impact for $\beta_{\rm tr} \lsim 10^\circ$. Thus, our
sensitivity calculations above the EeV scale would be only modestly corrected by including magnetic field effects. Without geomagnetic effects, our sensitivity results are somewhat conservative.

\section{Aperture, sensitivity and event rates for POEMMA}\label{sec:four}

\subsection{Aperture}\label{subsec:aperture}

\begin{figure}[ht]
\centering
\includegraphics[width=\columnwidth, trim = 0.1cm 8cm 0.1cm 9cm, clip]
{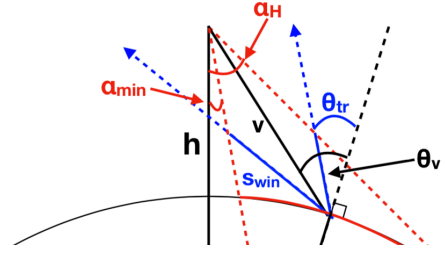}
\caption{\small As in Fig. \ref{fig:geometry-poemma}, with an exaggerated
difference between $\theta_v$ and $\theta_{\rm tr}$. Tau
decays in the atmosphere outside the observation window of
the detector (outside the dashed red lines) cannot be detected, as discussed in Sec. \ref{subsec:aperture}.}
\label{fig:geometry-poemma-swin}
\end{figure}


\begin{figure}[t]
	\begin{center}
		\includegraphics[width=1.1\columnwidth, trim=0 1.5cm 0 8cm, clip]{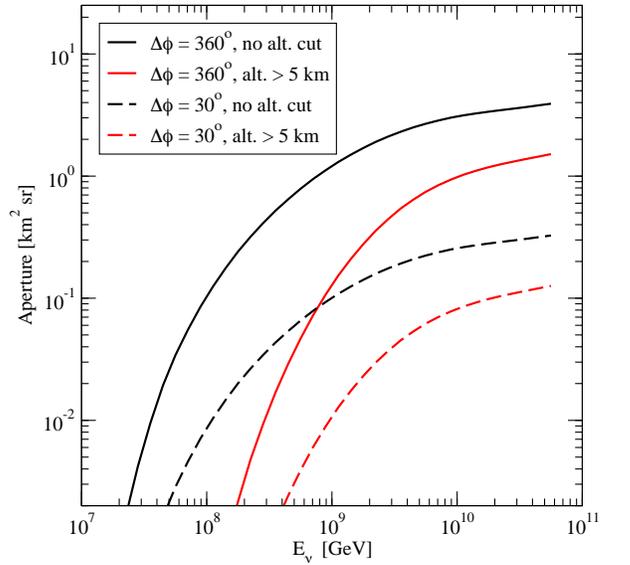}
	\end{center}
	\vspace{-0.75 cm}
	\caption{\small Aperture as a function of tau neutrino energy for the POEMMA $360^{\circ}$ (solid) and $30^\circ$ (dashed) configurations from Monte Carlo integration of eq. (\ref{apereqn}) with $N_{\rm PE}^{min}=10$. The black curves show
	the aperture for altitudes of decay between $0-20$ km, while the red curves restrict the altitudes of decay to $5-20$ km.}\label{apertmvfig}	
\end{figure}

A given instrument's capability to detect Earth-skimming tau neutrinos is determined by its aperture (or acceptance). Following \citet{Motloch:2013kva}, the detector aperture is given by 
\begin{equation}\label{apereqn}
\left<A\Omega\right> \left(E_{\nu_{\tau}}\right)= \int_{S}\int_{\Delta\Omega_{\rm tr}} P_{\rm obs}\ \hat{r}\cdot\hat{n}\, dS \, d\Omega_{\rm tr}\ ,
\end{equation}
where $S$ is the area of the observation region on the Earth, $\hat{r}\cdot\hat{n} = \cos\theta_{\rm tr}$ is the cosine of the angle between the \taon trajectory and the local zenith (see Fig. \ref{fig:geometry-poemma}), and $\Omega_{\rm tr}$ is the solid angle of the particle trajectories,
of interest in the observable solid angle $\Delta\Omega_{\rm tr}$ around the line of sight. Space-based and balloon borne detectors point near the Earth's limb for viewing air showers from 
Earth-skimming tau neutrinos. The observation region is determined by the altitude of the detector and the minimum nadir angle at which the detector can view air showers. 
For the calculation for POEMMA, we assume a configuration where POEMMA views the Earth $\Delta \alpha \simeq 7^\circ$ below the limb.
At an orbit altitude
of $h=525$ km, this translates to
$\theta_{\rm tr}^{\rm min}=90^\circ-\beta_{\rm tr}^{\rm max}\simeq 70^\circ$.
Accounting for just the area of the zone observed 
on the surface of the Earth over the full $2\pi$ azimuth, the effective area $G$ and geometric
aperture $\langle A\Omega\rangle_{\rm geo}$ are
\begin{eqnarray}
\label{eq:gfac}
    G &=& \int_S \ \hat{r}\cdot\hat{n}\,  dS =\int_S \cos\theta_{\rm tr}\, dS \\
    \label{eq:AOgeo}
    \langle A\Omega\rangle_{\rm geo} &=& \int_S\int_{\Delta \Omega_d} \cos\theta_{\rm tr} \,
    dS\, d\Omega_{\rm tr}\\
    \nonumber&\simeq & \pi\sin^2\theta_d G\ . 
\end{eqnarray}
where the last approximation is for a fixed  maximum detection angle $\theta_d$ away from the viewing angle, here, the Cherenkov angle.
The geometric aperture of this zone 
is $\langle A\Omega\rangle_{\rm geo}
=4,072$ km$^2$sr for the detection angle 
$\theta_d = 1.5^\circ$, and $\Delta \alpha=7^\circ$ at altitude $h=525$ km. The full surface cap under the detector, with $\alpha=0\to \alpha_H$, has a geometric aperture approximately twice the zone
geometric aperture.
For
reference, we include in Appendix A the details of the viewing geometry at $h=525$ km
as for POEMMA,
and for detectors at $h=33$ km and 1000 km.

The geometric aperture is modified by $P_{\rm obs}$, the probability that a tau neutrino with energy $E_{\nu_{\tau}}$ produces a shower that would be detectable. 
The observation probability
$P_{\rm obs}$ is given by 
\begin{align}\label{eqnpobs}
P_{\rm obs} &= \int p_{\rm exit}\left(E_{\tau}|E_{\nu_{\tau}},\beta_{\rm tr}\right) \nonumber \\
& \times \left[\int ds'\, {p}_{\rm decay}(s') p_{\rm det}\left(E_{\tau},\theta_{E},\beta_{\rm tr},s'
\right)\right]\, dE_{\tau}\, ,
\end{align}
as also discussed in the context of ANITA in ref. \cite{Romero-Wolf:2017ope}.
In Eq. (\ref{eqnpobs}),
$p_{\rm exit}$ is the differential probability that a \taon of energy $E_{\tau}$ emerges from the Earth given a parent tau neutrino energy of $E_{\nu_{\tau}}$ and an Earth-emergence angle of $\beta_{\rm tr}$,
as described in Sec. \ref{subsec:tauprop}. Our default
energy loss for determining $p_{\rm exit}$ uses the ALLM electromagnetic structure function for photonuclear energy loss.
The
differential decay probability $p_{\rm decay}\left(s \right)$ is for
a tau to decay a
distance $s$ from the Earth along its trajectory, discussed in Sec. \ref{subsec:taudecay}. The quantity $p_{\rm det}$ is the probability that the emerging tau produces an air shower that would be detected by
space-based detector. It depends on the shower energy ($E_\tau/2$ here), the position angles of the point
of emergence on the Earth $\theta_{E}$ (related to $\theta_v$) and $\phi_{E}$, and $\beta_{\rm tr}=\pi/2-\theta_{\rm tr}$ and $\phi_{\rm tr}$. Fig. \ref{fig:geometry-poemma-swin} shows the geometry, with an exaggerated difference between $\theta_{\rm tr}$ and  $\theta_v$
to show the distinction. 

For the detection probability, we approximate $p_{\rm det}$, 
\begin{equation}\label{eqnprobdet}
p_{\rm det} = H\left[\theta_{\rm Ch} - \theta\right]
H\left[s_{\rm win} - s \right]
H\left[N_{\rm PE} - N_{\rm PE}^{\rm min}\right]
\ ,
\end{equation}
in terms of the Heaviside function $H(x)$:
\[ H\left(x\right) = \left\{  \begin{array}{ll}
                         0 & \mbox{if $x < 0$};\\
                         1 & \mbox{if $x \geq 0$}. \end{array} \right. \]
The angle $\theta$ is the angle between the \taon trajectory and the line of sight to the detector labeled by $v$ in Fig. \ref{fig:geometry-poemma-swin}.

The \taon that decays a distance $s$ from its exit point must be within an ``observing cone" of the detector. A two-dimensional projection of the observing cone is shown by the dashed red lines in Fig. \ref{fig:geometry-poemma-swin}. To be observed, the
tau must decay before it passes outside of the observing cone. The maximum path length
for detection
of the \taon emerging from the Earth is  $s_{\rm win}$, labeled for one of the \taon trajectories in
Fig. \ref{fig:geometry-poemma-swin}. The value of  $s_{\rm win}$ depends on $\theta_{\rm tr}$.

The signal in an instrument is given by the number of photoelectrons which is evaluated from the number density of
photons in the Cherenkov cone, multiplied by the area of the detector times the quantum efficiency of the photo-detector for Cherenkov photons. For POEMMA, we assume
$A=2.5$ m$^2$ for the
effective optical collecting area and 0.2 for the quantum efficiency, based on the average Cherenkov-spectra-weighted photon detection efficiency (PDE) of an typical silicon photomultier (SiPM) \cite{HamamatsuPDE}:
\begin{equation}
    N_{\rm PE}= \rho_\gamma(a,\beta_{\rm tr},E_{\rm shr})\times 2.5\ 
    {\rm m}^2\times 0.2\ .
\end{equation}

For the results shown here, we take $N_{\rm PE}^{\rm min}=10$ with $E_{\rm shr}=0.5\, E_\tau$. This choice for
$N_{\rm PE}^{\rm min}$ follows from considerations of the night-time air glow which could give false signals of neutrino events,
an estimate of the temporal width of the Cherenkov signal based on a geometrical calculation \cite{CherenkovTime}, and the largest viewing angles away from the EAS trajectory that leads to measurable signals based on our POEMMA performance model. Note that we have not included the effects of the point-spread-function of POEMMA optics. We assume the Cherenkov signal is effectively delivering into a single 0.084$^\circ$ pixel.
A model of the air glow background 314 - 900 nm band \cite{MackovjakAirGlo} is used based on VLT/UVES measurements \cite{2003A&A...407.1157H,2006JGRA..11112307C} and the van Rhijn enhancement \cite{1921PGro...31....1V, 1955ApJ...122..530R, 1965P&SS...13..855B}.  This model yields an background-spectrum-weighted average PDE of 0.1 using the same SiPM performance for the Cherenkov signal $\langle{\rm PDE}\rangle$.
With a collecting area of
$A=2.5$ m$^2$ for POEMMA and a 60 ns coincidence window for neutrino events with stereo viewing, the false positive rate due to air glow background in the 314-900 nm band is effective eliminated for $N_{\rm PE}^{\rm min} \gsim 10$.

We use an effective Cherenkov angle $\theta_{\rm Ch}$  that depends on
$\beta_{\rm tr}$, altitude and number of photons.
The results we show below in Figs. \ref{apertmvfig}, \ref{sentmvfig} and \ref{sensourcestmvfig} have the full integration using Eq. 
(\ref{eqnprobdet}) to determine the observation probability.
To integrate Eq. (\ref{apereqn}), we have developed a code that performs the integration via Monte Carlo using importance sampling. 
More details of the numerical evaluation are discussed in Appendix C.
A reasonable approximation overall for the sensitivity is to take
$\beta_{\rm tr}=\beta_v$ in $P_{\rm obs}$ and integrate $d\Omega_{\rm tr}$ independently to a maximum angle equal to the effective
Cherenkov angle. In our consideration of variations to energy loss and neutrino cross section inputs, we use this approximation. 

The tau neutrino aperture for POEMMA is shown in  Fig. \ref{apertmvfig}. The solid black curve shows the aperture for a $360^\circ$ configuration (POEMMA360) with $N_{\rm PE}^{\rm min}=10$. For a configuration with $\Delta\phi=30^\circ$ (POEMMA30), the aperture is reduced by a factor of $12$. This is shown with the dashed black curve. 

The above calculations do not account for the loss of aperture due to cloud coverage. We do not model the effects of clouds here. As an approximate worst-case scenario of dense, optically opaque clouds over the entire field of view and below an altitude
of 5 km, we can reevaluate the effective aperture.
Mathematically, this involves multiplying eq. (\ref{eqnprobdet}) by another Heaviside function: $H\left[a_{\rm decay} - 5\mbox{ km}\right]$. The resulting aperture curves are plotted as red lines in the  Fig. \ref{apertmvfig}, with the solid red curve for $\Delta\phi=360^\circ$ and dashed curve for $\Delta\phi=30^\circ$.

\subsection{Sensitivity}

The tau neutrino aperture as a function of neutrino energy permits us to evaluate the sensitivity for POEMMA at $h=525$ km altitude to an isotropic tau neutrino flux. The sensitivity over a decade in energy for $N_\nu=3$ flavors is given by
\begin{equation}
\label{eq:sens}
F_{\rm sens}\left(E_{\nu_{\tau}}\right) = \frac{2.44\times N_\nu}{\ln(10)\times E_{\nu_{\tau}} \times \left<A\Omega\right> \left(E_{\nu_{\tau}}\right) \times t_{\rm obs}}\,,
\end{equation}
where the factor of $2.44$ events arises from the unified confidence upper limit, (i.e., the upper edge of the two-sided interval for which the lower limit is $0$) at the $90\%$ confidence level \cite{Feldman:1997qc}. 
The unified confidence upper limit includes all hypothetical Poisson means for which $n = 0$ observed events would be a reasonable realization (i.e., $n = 0$ is within the $90$\% acceptance interval of observed numbers of events) when drawing from a given Poisson distribution within the unified confidence interval. As such, for signals that are expected to fluctuate about their true values, our use of the unified confidence interval ensures that possible realizations in that observed number of events will be ``covered'' to the desired confidence level, in this case $90$\% (i.e., ``coverage probability'' of $90$\%).\footnote{Note that the value of $2.3$ that is often used in the literature excludes values in the interval $\left[2.3,2.44\right]$ for which $n = 0$ is a reasonable realization to within $90$\% and hence, does not fully cover the $90$\% confidence region. In this case, the coverage probability would in fact be less than $90$\%. For more in depth discussions, we refer the reader to Refs.~\cite{Feldman:1997qc,Tanabashi:2018oca}.}
For the results shown here, we take
$t_{\rm obs} = 0.2 \times \mbox{5 years}$  assuming a twenty percent duty cycle over five years. The assumption for the twenty percent duty cycle is motivated by the relatively large $N_{\rm PE} \gsim 10$ threshold needed to eliminate the effects of the large air glow background in the 314-900 nm range, e.g., some modest amount of moonlight can be tolerated.

\begin{figure}[t]
	\begin{center}
          \includegraphics[width=1.1\columnwidth, trim=7cm 2.5cm 4cm 3cm, clip]{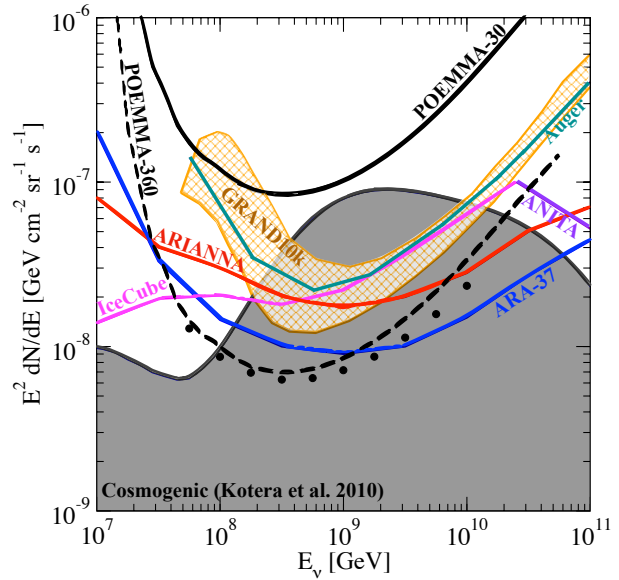}          
	\end{center}
	\vspace{-0.75 cm}
	\caption{\small All-flavor sensitivity scaled by neutrino energy squared, as a function of neutrino energy, assuming an operating time of five years and a duty cycle of 20\% percent, for showers produced at all altitudes (black curves and markers).  	The solid (dashed) curves follow from eq. (\ref{eq:sens}) for $\Delta\phi=30^\circ\ (360^\circ)$. The closed markers follow from eq. (\ref{eq:decade}) with $\Delta\phi=360^\circ$. The 90\% CL upper limits from Auger \cite{Aab:2019auo} (scaled for sliding decade-wide neutrino energy bins), 
	IceCube \cite{Aartsen:2018vtx},  and ANITA
\cite{Allison:2018cxu} are shown along with projected sensitivities of ARIANNA \cite{Barwick:2014pca}, ARA-37
\cite{Allison:2011wk} and 
GRAND10k \cite{Fang:2017mhl}, for the all flavor limits.}
	\label{sentmvfig}	
\end{figure}

The resulting three-flavor sensitivity curves $E^2 F_{\rm sens}$ are plotted as black lines in Fig. \ref{sentmvfig}, the dashed curve for $\Delta\phi=360^\circ$ and solid curve for $\Delta\phi=30^\circ$. The closed circular markers come from evaluating an
integral flux scaling like $E_\nu^{-\gamma}$ for $\gamma=2$ that yields 2.44 events per neutrino flavor for a given decade of energy centered (on the $\log_{10}$ scale) at the energy of the marker
for $\Delta\phi=360^\circ$. Thus, we find the normalization $F_0$ of
\begin{equation}
F_\nu(E_\nu) = F_0\times \Biggl(\frac{\rm GeV}{E_\nu}\Biggr)^\gamma
\end{equation}
from 
\begin{equation}
\label{eq:decade}
    N_{\rm evts}^{\nu_\tau}=\int_{10^{-0.5}E_\nu}^{10^{0.5}E_\nu}dE \frac{F_\nu (E)}{N_\nu}\langle A\Omega\rangle (E) \, t_{\rm obs} = 
    2.44\, .
\end{equation}
Plotted in Fig. \ref{sentmvfig} with
the closed circular markers are the values of $E^2 F_0$. There is some variation in the sensitivity to the spectral index $\gamma$, on the order of $\sim \pm 20\%$ for $\gamma=1.5-2.5$.

A range of cosmogenic fluxes from Ref. \cite{Kotera:2010yn} are shown in Fig. \ref{sentmvfig}, with the top of the shaded region bounded by the prediction labeled flux 4 in sec. II.F. We also show 90\% CL upper limits for Auger \cite{Aab:2019auo}, 
IceCube \cite{Aartsen:2018vtx},  and ANITA
\cite{Allison:2018cxu}, and the predicted sensitivities for ARIANNA \cite{Barwick:2014pca}, ARA-37 
\cite{Allison:2011wk} and 
GRAND10k \cite{Fang:2017mhl,Alvarez-Muniz:2018bhp}.

\begin{figure}[t]
	\begin{center}
          \includegraphics[width=1.1\columnwidth, trim=7cm 2.5cm 4cm 3cm, clip]{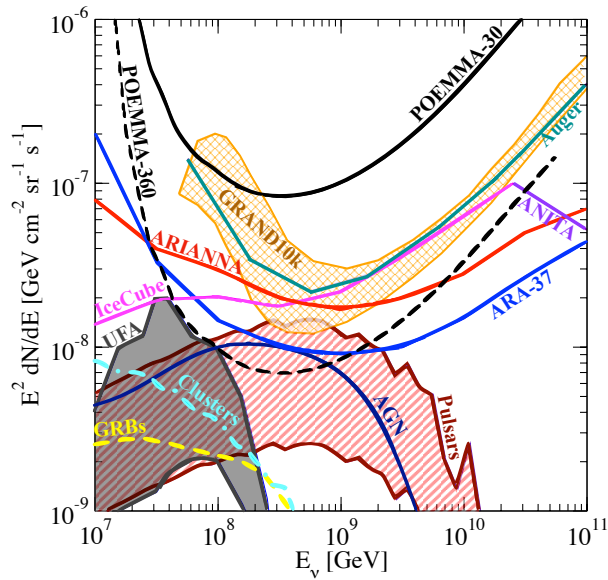} 	
	\end{center}
	\vspace{-0.75 cm}
	\caption{\small All-flavor sensitivity scaled by neutrino energy squared, as a function of neutrino energy, assuming an operating time of five years and a duty cycle of 20\% percent, for showers produced at all altitudes (black curves), as in fig. \ref{sentmvfig}.
The solid (dashed) black curves follow from eq. (\ref{eq:sens}) for $\Delta\phi=30^\circ\ (360^\circ)$. 
	Curves and bands for diffuse all-flavor neutrino fluxes are shown for newborn pulsar sources \cite{Fang:2013vla}, AGNs \cite{Murase:2015ndr}, galactic clusters with central sources \cite{Murase:2008yt,Fang:2017zjf}, late flares and prompt emission from GRBs \cite{Murase:2007yt} and from UHECR photodisintegration within a source (labeled UFA) \cite{Unger:2015laa}.
	Observational sensitivities are shown as in Fig. \ref{sentmvfig}.}
	\label{sensourcestmvfig}	
\end{figure}

Fig. \ref{sensourcestmvfig} shows our POEMMA sensitivity calculation as compared to some source models, summed over sources to get a diffuse neutrino flux. Shown with the red band are the all flavor neutrino predictions from newborn pulsar sources in ref. \cite{Fang:2013vla}. The dark blue curve
labeled AGN is a prediction from active galactic nuclei sources \cite{Murase:2015ndr}. Neutrinos from galactic clusters with central sources \cite{Murase:2008yt,Fang:2017zjf} are shown with the light
blue curve. The yellow curve shows a prediction from late flares and late prompt emission from gamma ray bursts \cite{Murase:2007yt}. Finally, the curve labeled UFA shows a neutrino flux prediction that comes from UHECR photodisintegration within a source from \citet{Unger:2015laa}.

The comparison of the sensitivities shows that an azimuthal
coverage of $360^\circ$ would be required for a 5 year sensitivity to be competitive with other detectors. The POEMMA360 sensitivity in 5 years would probe cosmogenic fluxes in the upper range of predictions in Ref.
\cite{Kotera:2010yn} and diffuse astrophysical fluxes from a range of models, e.g., pulsar models of Ref. \cite{Fang:2013vla}. The POEMMA360 sensitivity illustrates the benefits of full azimuthal coverage and demonstrates the potential of using the optical Cherenkov signal from upward-moving \taon EAS induced from tau neutrino interactions in the Earth. 

\subsection{Flux dependent event rates}

In addition to computing the energy-dependent aperture and sensitivity, we also calculate the event rate for a given flux of tau neutrinos above a specified energy. We follow the same procedure as for
the flux independent results, however, with a factor of the tau flux
and an integration over the
energy of the tau (twice the shower energy),
\begin{eqnarray}
\nonumber
N_{\rm evts} \left(> E_{\tau}^{\rm min}\right)&=&  \Delta t_{\rm obs} \int_{E_{\tau}^{\rm min}} dE\, dS\, d\Omega_{\rm tr}\\
&\times& P_{\rm obs}\, \hat{r}\cdot \hat{n}\, F_\tau(E, \beta_{\rm tr})
\, .
\end{eqnarray}
The tau flux is determined from the transmission functions discussed
in sec. II.F.

For the POEMMA $360^{\circ}$ configuration and an observation time of five years with a duty cycle of $\sim 20$\%, we find the expected number of events above $E_{\nu_{\tau},{\rm min}} = 10^{7}$ GeV is $\sim 0.2$ events in the flux 1 scenario, representing the prediction with a uniform distribution of sources with no evolution, a mixed composition of UHECRs and maximum cosmic ray energy of $E_{\rm max} = 100$ EeV.
For the flux 4 scenario that has a source evolution following that of Fanaroff-Riley Type II active galactic nuclei and a pure proton UHECR composition with $E_{\rm max} = 3160$ EeV, the number of events in the five-year, 20\% duty cycle time frame is $\sim 13$ events. 
A restriction to decays above an altitude of $a=5$ km reduces the number of events, for example, to $\sim 3$ event for Flux 4 in the same time
period. 

\section{Discussion}\label{sec:five}

How the tau neutrino and \taon interactions are modeled affects the sensitivity of POEMMA, as does the density model.
We have examined some of these features using a simplified evaluation of the aperture and sensitivity for POEMMA, setting $\beta_{\rm tr}=
\beta_{v}$ so $H[s_{\rm win}-s_{d}]=1$ in eq. (\ref{eqnprobdet}). Numerically, the simplified evaluation gives results very close to the full Monte Carlo evaluation, so we used this simplification to study variations in the predictions due to these effects.

The relative benefits of observing upward-going air showers over land and water have been discussed by \citet{PalomaresRuiz:2005xw}. They argue that the very high energy shower rate is significantly enhanced over water compared to over rock. Tau energy loss in water is less than in rock because of the different densities, but the density of rock favors neutrino interactions.
In the results shown thus far, the sensitivity is evaluated 
assuming the final density shell of the Earth is water, according to the PREM model. We can do the same evaluation of the sensitivity assuming the final density shell of the Earth is standard rock. 
A similar evaluation has been performed in Ref. \cite{Alvarez-Muniz:2017mpk}.
We find that most of the energy range to which POEMMA is sensitive is not high enough for the onset of an enhancement of the over-water event rate, in qualitative agreement with
the results presented in Ref. \cite{Alvarez-Muniz:2017mpk}. The water versus land effect
can be understood by considering the distance scales as a function of energy.

The angles $\beta_{\rm tr}=1^\circ-20^\circ$ 
correspond to a range of chord lengths in the final density 
shell, for example, the whole trajectory of  222 km for $\beta_{\rm tr}=1^\circ$ to a final 10.3 km in
the outer shell for $\beta_{\rm tr}=20^\circ$. For $E_{\tau}=10^8$
GeV, the time dilated decay length is 5 km. At energies $E_\nu \lsim 10^8$ GeV, the produced tau's lifetime, not energy loss, determines the tau range in the final density shell. To first approximation, there is a benefit to a rock target in the last density shell rather than a water target, since the column depth for neutrino interactions is
$X\sim \rho \gamma c \tau$ and the exit probability is $P^{\rm exit}_{\tau}\sim X/\lambda_\nu$ for 
the interaction length (in units of column depth) 
$\lambda_\nu=(N_A \sigma_{\nu N})^{-1}$, so $P^{\rm exit}_{\tau}$ is larger for higher density $\rho$.

\begin{figure}
    \centering
    \includegraphics[width=0.9\columnwidth]{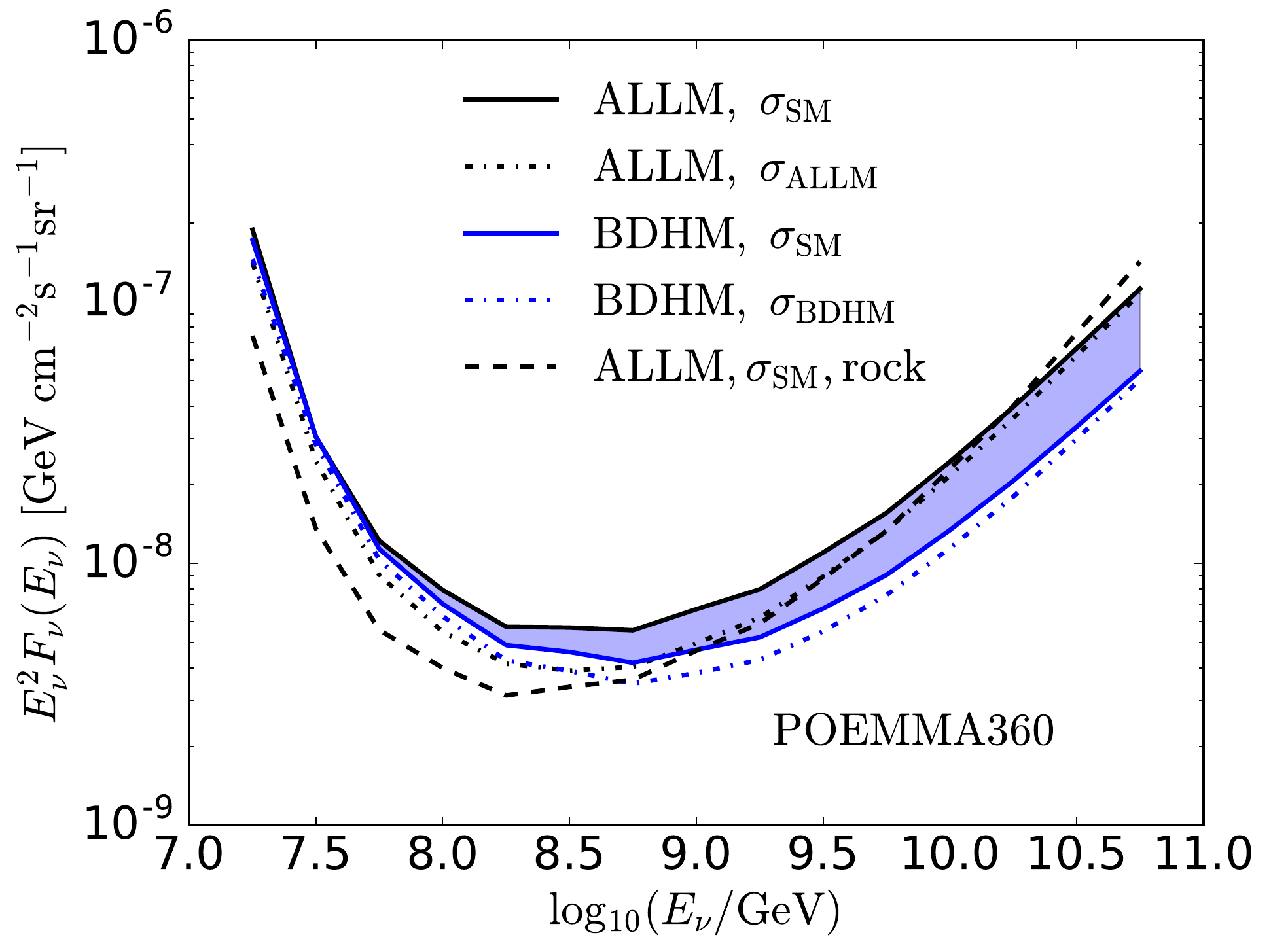}
    \caption{\small Three flavor sensitivity for the standard Earth density model of Table I for the ALLM (solid black curve) and BDHM (solid blue curve) tau energy loss with the standard model neutrino nucleon cross section.
    The ALLM energy loss with the outermost shell density is
    set to $\rho_{\rm rock}=2.65$ g/cm$^3$ is shown with the black dashed curve. The black (blue) dot-dashed curve shows $E_\nu ^2F_{\rm sens}$ using ALLM (BDHM) for both the \taon energy loss and $\sigma_{\nu N}$.}
    \label{fig:sensitivity-compare}
\end{figure}

As the neutrino energy increases, the time-dilated lifetime increases and tau electromagnetic energy loss becomes important. Fig. \ref{fig:beta} shows that the tau energy loss parameter is smaller for water
than for rock for the $e^+e^-$ pair production process.
In addition, the column depth in water is smaller than in rock for the final shell, so $\Delta E_\tau\sim b_\tau \Delta X$ is smaller for water (by a factor of $\rho_{\rm water}/\rho_{\rm rock}$) than for rock, allowing more taus to emerge from water than from rock. An evaluation of the sensitivity for an Earth model with the outer shell density set to $\rho_{\rm rock}$, in the approximation that $\beta_{\rm tr}=\beta_v$, is shown in Fig. \ref{fig:sensitivity-compare} by the dashed black curve, compared to an evaluation with the same approximation with the
final density shell with $\rho_{\rm water}$, as in Table I (solid black curve). For showers from taus that emerge from rock,
the sensitivity is lower (better) by a factor of $1/2.5$ for $E_{\nu_\tau}=10^7$
GeV. For $E_{\nu_\tau}=10^8$ GeV, the reduction is a factor of 1/2. When
$E_{\nu_\tau}=10^{10}$ GeV, the sensitivities are equal, whether the showers occur over rock or water. For POEMMA detector thresholds, energies above $E_{\nu_\tau}=10^{10}$ GeV show only a modest (up to $\sim 30\%$) advantage for
observations over water compared to rock.

The different extrapolations of the electromagnetic structure function needed for photonuclear energy loss also shift
the sensitivity curve. Our default choice is the ALLM extrapolation of $F_2(x,Q^2)$ in the evaluation of $b_\tau^{\rm nuc}$ with a neutrino nucleon cross section that relies on the nCTEQ-1 PDFs ($\sigma_{\rm SM}$), which also extrapolate structure functions. The blue solid curve in Fig. \ref{fig:sensitivity-compare} shows the sensitivity with the BDHM
extrapolation of $F_2(x,Q^2)$, keeping the neutrino nucleon cross section evaluated with nCTEQ-1 PDFs. The blue shaded region
is indicative of the uncertainty associated with the $F_2$ extrapolation, keeping the neutrino-nucleon cross section fixed. 

Again, at low energy where tau energy loss is not very important, the 
ALLM and BDHM energy loss evaluations yield nearly identical results.
As the energy increases, the smaller value of $b_\tau^{\rm nuc}$
with the BDHM extrapolation means less tau energy loss, a larger aperture and a lower sensitivity. The sensitivity curve from the BDHM evaluation is a factor of two lower than the ALLM evaluation at the highest energy shown in Fig. \ref{fig:sensitivity-compare}.

Changing the neutrino cross section to evaluations using ALLM (BDHM) structure function extrapolation instead of using nCTEQ15-1 in the next-to-leading order QCD calculation, along with the respective $b_\tau^{\rm nuc}$, gives results shown with the black (blue) dot-dashed lines. At the highest energy,
the change is modest. For $E_\nu=10^9$ GeV, the lowest curve is a factor of 1.75 below the default curve (the black
curve) with ALLM used for energy loss and the standard model 
neutrino cross section.

To what degree do the POEMMA360 detection characteristics limit the sensitivity? The sensitivity for POEMMA viewing for angles within $\Delta\alpha\sim 7^\circ$ below the horizon is shown in Figs.
\ref{sentmvfig} and \ref{sensourcestmvfig}. Increasing the viewing to $\Delta \alpha=9^\circ$ only marginally improves the sensitivity. For 
$\Delta\alpha=15^\circ$, corresponding to $\beta_{\rm tr}\lsim 31^\circ$,
the sensitivity at $E_{\nu}=10^{7}$ GeV is $\sim 10^{-7}$ for $E_\nu^2 dN/dE_\nu$ in Fig. \ref{sentmvfig}. The sensitivity for POEMMA360 would reduce by a factor of 0.42 for $E_\nu=2\times 10^7$ GeV, and only by a factor of 0.87 for $E_\nu=10^8$ GeV. The larger values of $\beta_{\rm tr}$ favor the lower neutrino energies relative to higher energies because the lower energy neutrino flux is less attenuated, so a relatively higher fraction of taus emerge to shower at angles that are better detected at low energies than for small $\beta_{\rm tr}$ where the distance $v$ from the point of emergence and the detector is much larger.

The threshold for the number of photoelectrons detected is an important feature of the sensitivity. If $N_{\rm PE}^{\rm min}=5$, or alternatively, if the detection area times quantum efficiency were a factor of two higher, we find that the sensitivity is a factor of 0.10 of the black curve for POEMMA360 in Fig. \ref{sentmvfig} for $E_\nu=2\times 10^7$ GeV, a factor of 0.43 lower for $E_\nu=10^8$ GeV and lower by a factor
of $0.60$ for $E_\nu=10^9$ GeV, for example. The photoelectron threshold is set for POEMMA to achieve a negligible false positive neutrino rate from the night-time air glow background. Lowering the photoelectron threshold may be possible with a restricted wavelength range, however, further work is needed to assess whether or not the sensitivity would be lowered.

The results presented here for the detection of skimming tau neutrinos via upward-going air showers from tau decays, with POEMMA detection as a specific example, come from a Monte Carlo evaluation of neutrino interaction and tau energy loss, then a one-dimensional model of the extensive air shower from the tau decay. A broader program to simulate signals of skimming tau neutrinos is underway to address some of the approximations used here, and  to compare with approximations in other approaches. For example, in Ref. \cite{Alvarez-Muniz:2017mpk}, tau energy loss is treated as continuous, whereas we use stochastic energy loss here.  

An evaluation of timing of the extensive air showers shows that for detectors more than 16 km away from an extensive air shower with a radius of 100 m, our one-dimensional approximation for the air shower modeling is reliable. A full three-dimensional shower simulation will be useful for showers at larger elevation angles for balloon detection. 

The air shower modeling uses the 83\% of non-muonic tau decays, with half of the tau energy going into the shower. Incorporation of the energy distribution of $E_{\rm shr}$ is planned. An additional consideration is whether or not air showers associated with high energy muons are detectable \cite{Stanev:1989bc}. 
The particle yield from HE muon showers is suppressed compared to that of a similar energy electron (or photon) EAS. The length of a muonic air shower has a much longer slant depth due to this and the long lifetime of the muon. From a practical standpoint, the propagation of the muonic shower could yield an EAS signature at much high altitudes, even for $\tau$-lepton decays that initiate near the Earth's surface.

To summarize, we have presented a
new calculation of the flux and spectrum of Earth emerging \taon 
from an isotropic flux of cosmic neutrinos, then applied 
our results to a space-based experiment with a performance modeled on the POEMMA mission. We have illustrated many features of neutrino and tau propagation and tau shower detection. 
Optical Cherenkov signals from upward-going air showers show promise for detecting cosmic neutrinos, especially below 100 PeV.
We find that a POEMMA-like instrument requires a $360^\circ$ azimuthal optical Cherenkov coverage to be competitive with other detectors current or planned. While our focus has been on an isotropic diffuse flux, the calculational tools developed here can be applied
to searches for individual neutrino sources. For a detector like POEMMA, the ability to quickly reorient the detectors will permit tracking of  target-of-opportunity neutrino sources. An assessment of POEMMA's sensitivity to these sources appears in Ref. \cite{Venters:2019xwi}. 

\acknowledgments

We thank Y. Akaike, L. Anchordoqui, D. Bergman, C. Gu\'epin, S. Mackovjak, A. Neronov, A. Olinto, P. Privitera, A. Romero-Wolf,  F. Sarazin and E. Zas for discussions, and the POEMMA collaboration for motivating this analysis. Our work was supported in part by a US Department of Energy grant DE-SC-0010113, NASA grant 17-APRA17-0066
and NASA award NNX17AJ82G.

\appendix
\section{Geometry of Earth at altitude}\label{sec:app_a}

The detection of tau neutrino induced tau air showers relies on the Earth as a neutrino converter. In this appendix, we show the angle relations needed to evaluate the geometric aperture in terms of angles labeled in Fig. \ref{fig:geometry-poemma}.
Air shower detectors at altitude $h$ above the Earth will point near the limb, which is at a viewing angle $\alpha_H$ away from the nadir. 
For POEMMA, we take $h=525$ km. Emergence
angles at the limb are related by 
\begin{eqnarray}
\sin \alpha_H &= & \frac{R_E}{R_E+h}\\
\theta_E^H &\equiv& \frac{\pi}{2} - \alpha_H
\end{eqnarray}
while more generally,
\begin{eqnarray}
\frac{\cos\beta_v}{R_E+h} &=& \frac{\sin\alpha}{R_E}\\
v\sin\alpha &=& R_E\sin \theta_E
\end{eqnarray}
where $v$ is the path length in the atmosphere,
\begin{equation}
\label{eq:v2}
v^2 = (R_E+h)^2 +R_E^2 - 2R_E(R_E+h)\cos\theta_E\ ,
\end{equation}
and $\theta_{E}$ is the polar angle of the given position on the Earth.
These relations are used to find the difference in the viewing angle $\alpha$ relative to the angle to the limb, $\Delta \alpha \equiv
\alpha_H-\alpha$.
For $h=33,\ 525,\ 1000$ km, $\alpha_H=84.2^\circ, \ 67.5^\circ,\ 59.8^\circ$,
respectively. Table \ref{table:deltaa} shows $\Delta \alpha = \alpha_H-\alpha$ as a function of $\beta_E$ for these three altitudes. The planned POEMMA
Cherenkov detector will have a detection viewing angle range 
of $\Delta\alpha\simeq 7^\circ$, when pointed near the limb, so the angular coverage for $h=525$ km
is for $\beta_v\sim 0\to 20^\circ$. In fact, very close to the limb, the signal will be overcome by background, but for $\Delta\alpha>1 ^\circ$, the backgrounds are significantly reduced.

The viewing angle relative to the local zenith, $\theta_{v}$, is given by
\begin{equation}
\cos\theta_{v} = \frac{\left(R_{E} + h\right)\cos\theta_{E} - R_{E}}{\sqrt{\left(R_{E} + h\right)^2 + R_{E}^2 - 2\left(R_{E} + h\right)R_{E}\cos\theta_E}}\ .
\end{equation}
The angles $\theta_v=90^\circ-\beta_v$, $\alpha$
and $\theta_E$ are related by
\begin{equation}
    \alpha+\beta_v+\theta_E = 90^\circ\ .
\end{equation}

\begin{table}
\centering
\begin{tabular}{| l | c| c| c|}
\hline
$\Delta \alpha $ & $\beta_v(33\ {\rm km})$ & $\beta_v (525\ {\rm km})$ & $\beta_v(1000\ {\rm km})$ \\ \hline
1 & 3.6 & 7.0 & 8.2 \\ \hline
2 & 5.2 & 10.0 & 11.7 \\ \hline
3 & 6.6 & 12.3 & 14.5 \\ \hline
4 & 7.9 & 14.4 & 16.9 \\ \hline
5 & 9.1 & 16.2 & 19.0\\ \hline
6 & 10.3 & 18.0 & 21.0 \\ \hline
7 & 11.4 & 19.6 & 22.8 \\ \hline
8 & 12.6 & 21.2 & 24.6 \\ \hline
9 & 13.6 & 22.6 & 26.3 \\ \hline
10 & 14.7 & 24.1 & 27.9 \\ \hline
15 & 20.0 & 30.8 & 35.4 \\ \hline
20 & 25.2 & 37.0 & 42.2 \\ \hline
\end{tabular}
\caption{For a given $\Delta\alpha = \alpha_H - \alpha$ as measured from altitude
$h=33,\ 525$ and 1000 km, the viewing angle relative to the horizon at Earth $\beta_v$, all in degrees.}
\label{table:deltaa}
\end{table}

Absent detector field of view considerations and in the approximation that $\theta_{\rm tr}\simeq \theta_v$, the accessible flux from the area of a cap of the spherical Earth below a detector at altitude $h$ comes from a  surface area $S$ from $\theta_E^{min}=0^\circ $ to $\theta_E^{max}= \theta_E^H$.  
The effective area for air showers from taus emerging from the Earth
with angle $\theta_v$ relative to the normal to the cap surface is
\cite{Motloch:2013kva}
\begin{eqnarray}
G& =& \int \cos\theta_v dS=\int R_E^2 \cos\theta_v d\Omega_E  \\
\nonumber
&=&  2\pi R_E^2 \int^{\theta^{max}_E}_{\theta^{min}_E} \cos{\theta_v} d\cos{\theta_E} \\
\nonumber
&=& 2\pi R_E^2 \int^{\theta^{max}_E}_{\theta^{min}_E} \frac{(R_E+h) \cos{\theta_E}-R_E}{v}d\cos{\theta_E} \ .
\end{eqnarray}
The effective area of the cap below the detector can be written as 
\begin{equation}\label{eq:gcap}
G = \frac{2 \pi}{3(h+R_E)}\Bigl( (h(h+2R_E))^\frac{3}{2} - h^2(h+3R_E)\Bigr)\, .
\end{equation}
For a field of view characterized by $\Delta \alpha=7^\circ$, the effective area is reduced. The accessible area is a band around the cap. For $\Delta \alpha=7^\circ$ and $h=525$ km, $\theta_E^{min}=9.9^\circ$. 
We refer to this as the effective area of the ``zone,'' smaller than the ``cap'' described by Eq. (\ref{eq:gcap}). 

\begin{figure}{r}
\vspace{-5mm}
  \begin{center}
    \includegraphics[width=0.9\columnwidth,trim=0 1.5cm 0 1cm]{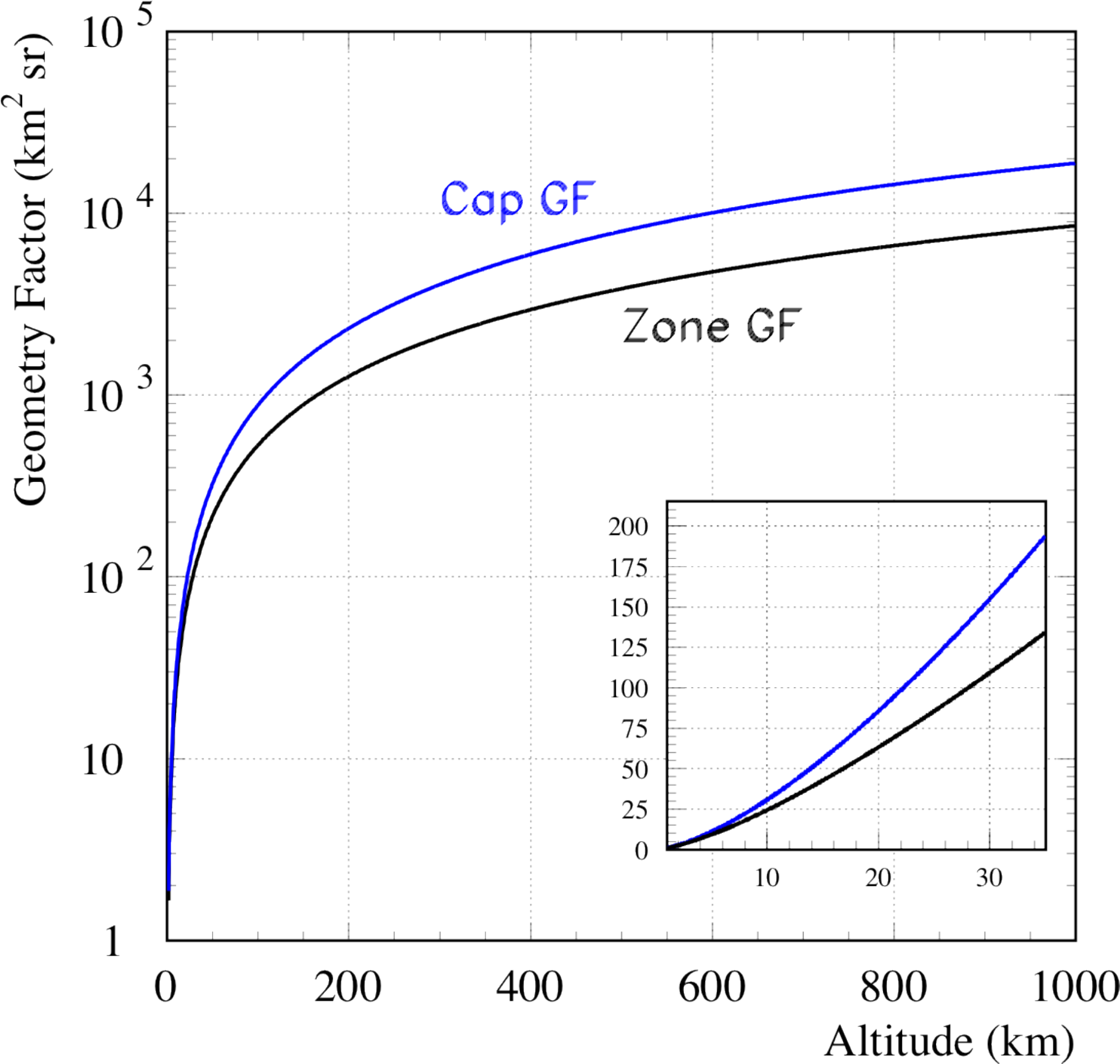}
   \end{center}
 \caption{\small Comparison of the geometric aperture from the Earth cap (upper blue curve) versus that for the Earth zone (lower black curve, defined by $\Delta\alpha=7^\circ$) for
 $\theta_d=1.5^\circ$, as a function of altitude. The inset shows the calculation on a linear scale from 0 to 35 km altitude.}
 \label{fig:geometryfactor}
\end{figure}

The geometric aperture (geometry factor) $\langle A\Omega\rangle_{\rm geo}$ defined in
eq. (\ref{eq:AOgeo}), however, depends on $\hat{r}\cdot\hat{n}=\cos\theta_{\rm tr}$.
In terms of the angles $\delta$ and $\phi_\delta$ that the \taon trajectory makes with respect to the line of sight, 
and the angle $\theta_v$ the line of sight makes with respect to the local zenith,
\begin{equation}
\hat{r}\cdot \hat{n}=\cos\theta_{\rm tr}=\cos\theta_v\cos \delta - 
\sin\theta_v\sin\delta \cos\phi_\delta\  .
\end{equation}
Then for $d\Omega_{\rm tr}=\sin\delta\, d\delta\, d\phi_\delta$ with the full $\phi_\delta$ integral
over $2\pi$ and $\delta= 0\to \theta_d$, the geometric aperture
is 
\begin{equation}
    \langle A\Omega\rangle_{\rm geo}= \pi\sin^2\theta_d G\ .
\end{equation}
Table \ref{table:aperture} compares the cap and zone geometric
apertures
for several altitudes when $\theta_d=1.5^\circ$ and $\Delta 
\alpha = 7^\circ$.
For $h=525$ km, the ratio of apertures for the zone and cap is 0.48 with these assumptions.
Fig. \ref{fig:geometryfactor} shows the geometric aperture of the cap and of the zone 
(for $\Delta \alpha=7^\circ$) as a function of detector altitude for  $\theta_d=1.5^\circ$.

\begin{table}
\begin{center}
{\small
\begin{tabular}{|c|c|c|c|} \hline
Altitude $h$ [km] & Cap [km$^2$ sr] & Zone [km$^2$ sr] &  Zone/Cap \\ \hline
3 & 5.2 & 4.5 & 0.87 \\ \hline
4 & 7.9 & 6.8 &  0.85 \\ \hline
33 & 178 & 124 & 0.70 \\ \hline
525 & 8,480 & 4,072 & 0.48 \\ \hline
1000 & 18,857 & 8,538 & 0.45 \\ \hline
\end{tabular}
}
\end{center}
\caption{\small Comparison of the cap $\langle A_{\rm cap}\Omega\rangle_{\rm geo}$ and zone $\langle A_{\rm zone}\Omega\rangle_{\rm geo}$ geometric apertures for several altitudes $h$ when $\theta_d=1.5^\circ$ and $\Delta \alpha=7^\circ$. The final
column is the ratio $\langle A_{\rm zone}\Omega\rangle_{\rm geo}/
\langle A_{\rm cap }\Omega\rangle_{\rm geo}$.}
\label{table:aperture}
\end{table}

\section{Tau exit probability and energy distributions}

\begin{figure}[t]
\centering
	\includegraphics[width=0.9\columnwidth]{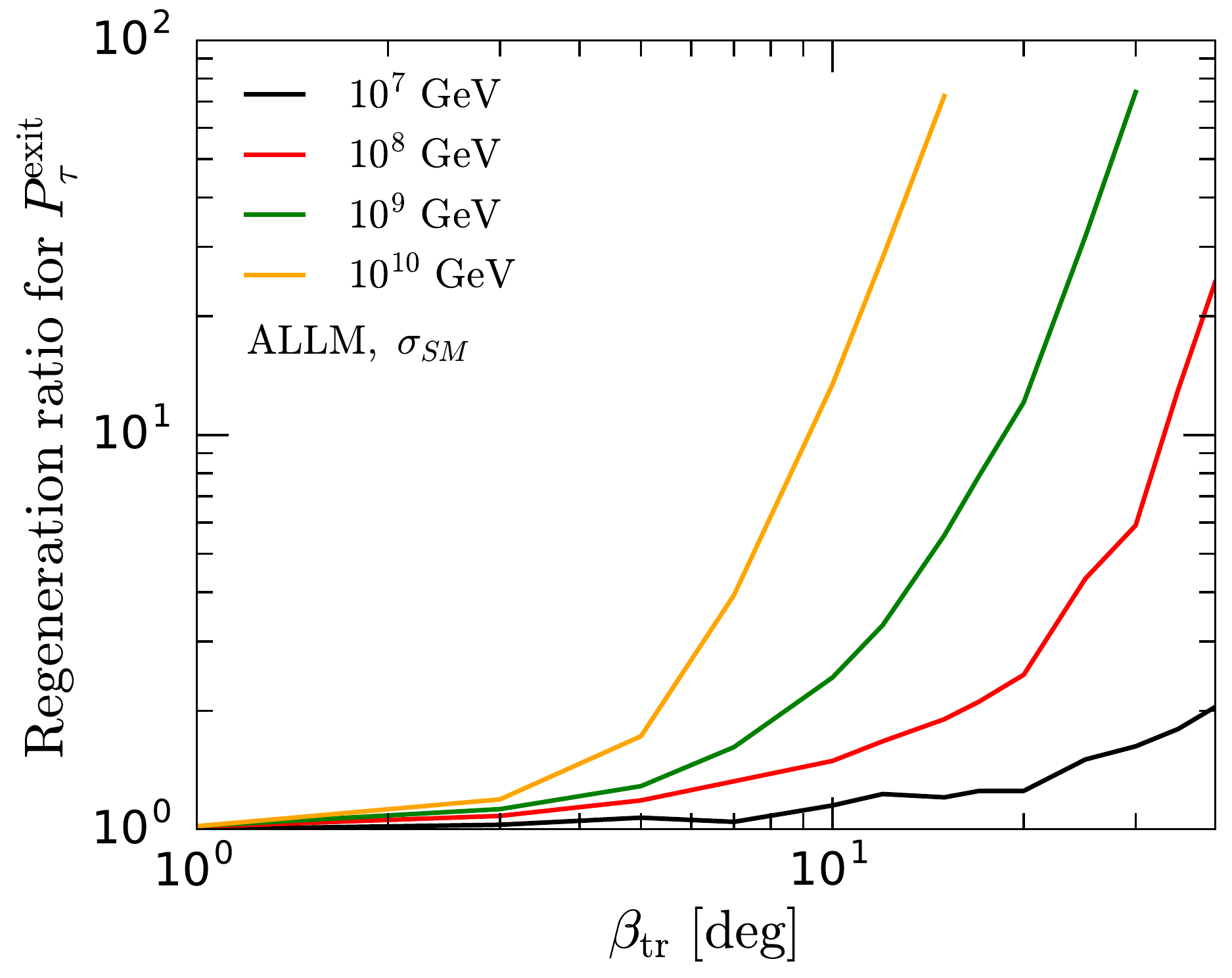}	
	\caption{\small Ratio of tau exit probability with regeneration (up to 5 charged-current interactions) to no regeneration (one charged-current interaction) as a function of $\beta_{\rm tr}$ for incident $E_{\nu_\tau}=10^7$, $10^8$, $10^9$ and $10^{10}$ GeV evaluated using the 
	ALLM photonuclear energy loss and the nCTEQ-1 neutrino cross section ($\sigma_{SM}$).}
	\label{fig:proballm-ratio}
\end{figure}

 In this section, we include tables for the \taon exit probabilities
 for a given tau neutrino energy and tables for the exiting tau energy given
 a fixed tau neutrino energy and $\beta_{\rm tr}$.
 Table \ref{tab:pallm} lists the exit probabilities for fixed
 energies $E_{\nu_\tau}=10^7$, $10^8$, $10^9$ and $10^{10}$ GeV
 when the ALLM extrapolation is assumed for $b_\tau^{\rm nuc}$.
 The standard model neutrino cross section, as described in sec. II.B is assumed.
 For reference, we show in Fig. \ref{fig:proballm-ratio} the ratio of the exit
 probabilities in Table \ref{tab:pallm} to the exit probabilities without regeneration,
 namely, assuming a single $\nu_\tau\to \tau$ conversion. The probabilities by themselves
 do not reflect the shift in energy from the multiple interactions when regeneration is important.
 
\begin{table}[htbp!]
    \centering
    \begin{tabular}{|c|c|c|c|c|}
    \hline
$\beta_{\rm tr} $ [$^\circ$] & $10^7$ GeV& $10^8$ GeV & $10^9$ GeV& 
$10^{10}$ GeV\\ \hline     
     1 &   2.89e-05 &   7.41e-04 &   6.85e-03 &   2.54e-02\\ \hline
    3 &   2.48e-05 &   5.55e-04 &   3.27e-03 &   5.15e-03\\ \hline
    5 &   2.07e-05 &   3.92e-04 &   1.49e-03 &   1.23e-03\\ \hline
    7 &   2.01e-05 &   2.50e-04 &   5.79e-04 &   3.36e-04\\ \hline
   10 &   1.29e-05 &   1.48e-04 &   2.20e-04 &   1.38e-04\\ \hline
   12 &   1.49e-05 &   1.06e-04 &   1.35e-04 &   8.20e-05\\ \hline
   15 &   9.30e-06 &   6.42e-05 &   6.81e-05 &   5.48e-05\\ \hline
   17 &   8.90e-06 &   4.99e-05 &   5.40e-05 &   3.77e-05\\ \hline
   20 &   7.90e-06 &   3.63e-05 &   3.31e-05 &   2.59e-05\\ \hline
   25 &   4.70e-06 &   1.57e-05 &   1.43e-05 &   1.31e-05\\ \hline 
   30 &   2.44e-06 &   6.73e-06 &   5.92e-06 &   5.85e-06\\ \hline
   35 &   1.44e-06 &   3.13e-06 &   2.84e-06 &   2.67e-06\\ \hline   
   40 &   9.00e-07 &   1.95e-06 &   1.83e-06 &   1.59e-06     

\\ \hline
    \end{tabular}
    \caption{The tau exit probability for $E_{\nu_\tau}=10^7$, $10^8$, 
    1$0^9$ and $10^{10}$ GeV assuming the ALLM structure function
    extrapolation for the photonuclear energy loss parameter, as 
    a function of $\beta_{\rm tr}$. The standard model cross section for neutrino-nucleon interactions is assumed.}
    \label{tab:pallm}
\end{table}

\begin{figure}[t]
	\begin{center}
		\includegraphics[width=0.9\columnwidth]{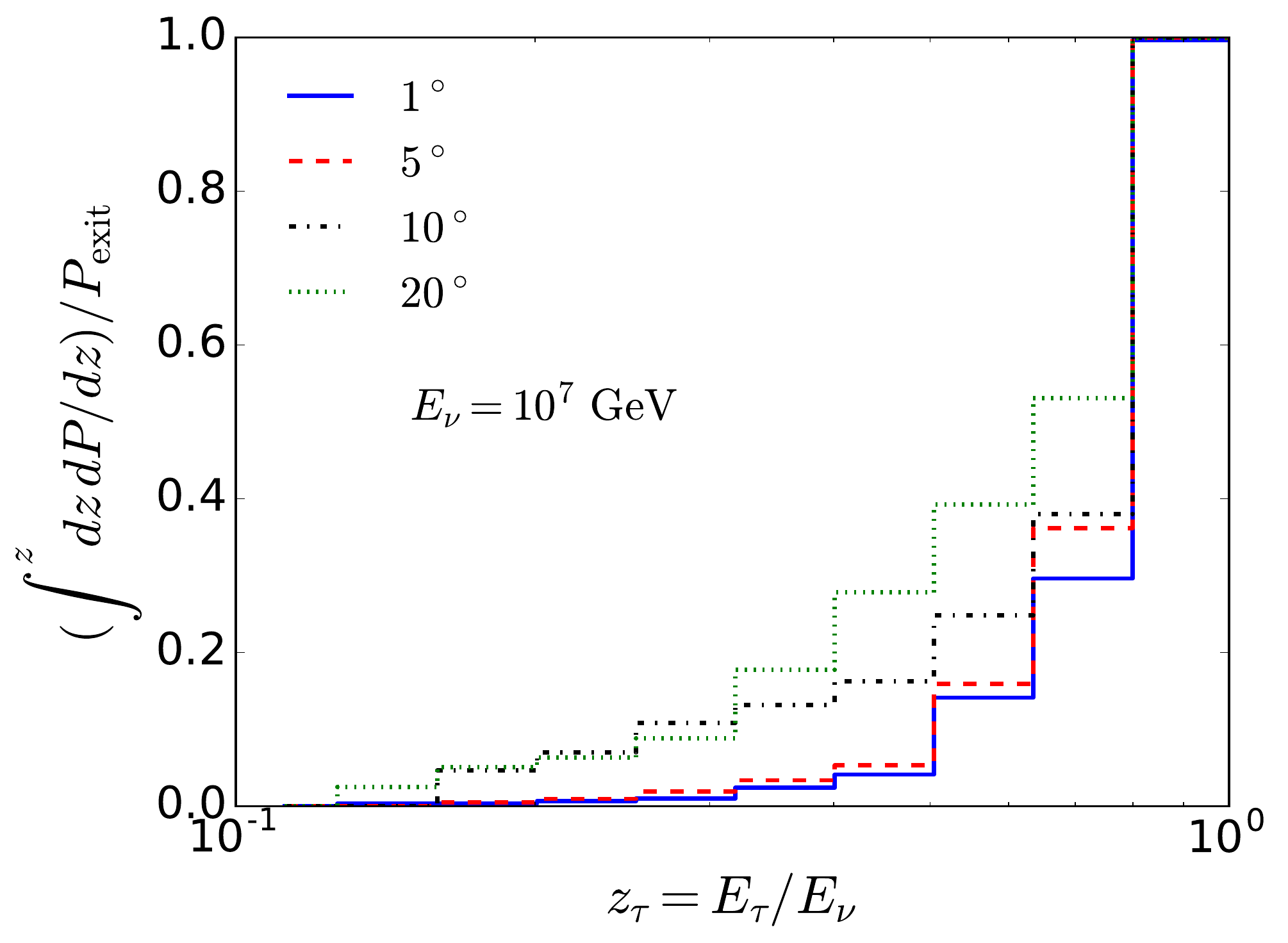}	\end{center}
	\vspace{-0.75 cm}
	\caption{\small For angles $\beta_{\rm th}=1^\circ,\ 5^\circ,\ 10^\circ$ and $20^\circ$, the cumulative distribution function for the relative tau exit probability for $E_{\nu_\tau}=10^7$ GeV, as a function of
	$z=E_\tau/E_{\nu_\tau}$.  The ALLM small-$x$ extrapolation of the electromagnetic structure function in $b_\tau^{\rm nuc}$ has been used.}
	\label{fig:E7-cume}	
\end{figure}

\begin{figure}[t]
	\begin{center}
		\includegraphics[width=0.9\columnwidth]{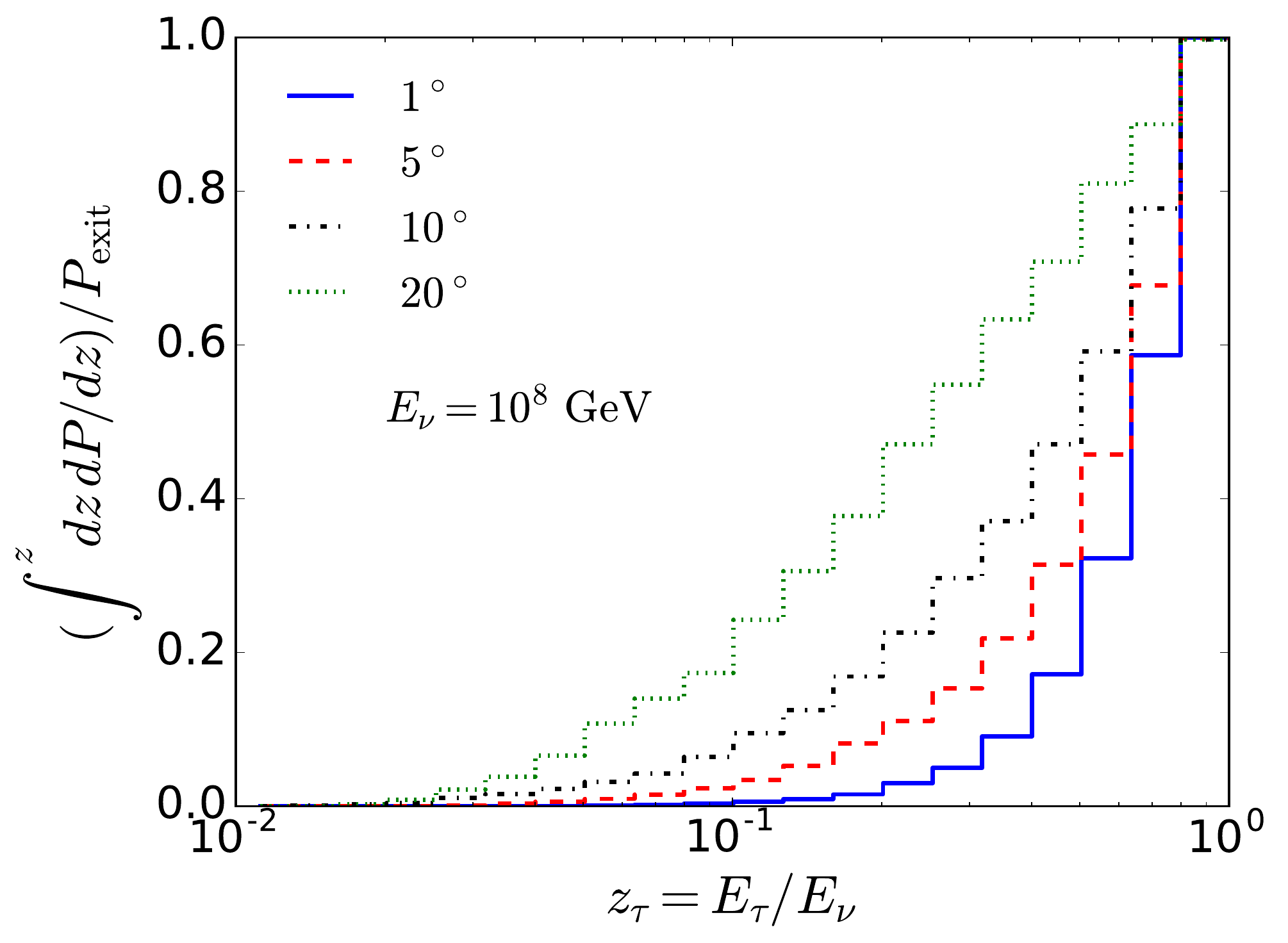}	\end{center}
	\vspace{-0.75 cm}
	\caption{\small For angles $\beta_{\rm th}=1^\circ,\ 5^\circ,\ 10^\circ$ and $20^\circ$, the cumulative distribution function for the relative tau exit probability for $E_{\nu_\tau}=10^8$ GeV, as a function of
	$z=E_\tau/E_{\nu_\tau}$.  The ALLM small-$x$ extrapolation of the electromagnetic structure function in $b_\tau^{\rm nuc}$ has been used.}
	\label{fig:E8-cume}	
\end{figure}

\begin{figure}[t]
	\begin{center}
		\includegraphics[width=0.9\columnwidth]{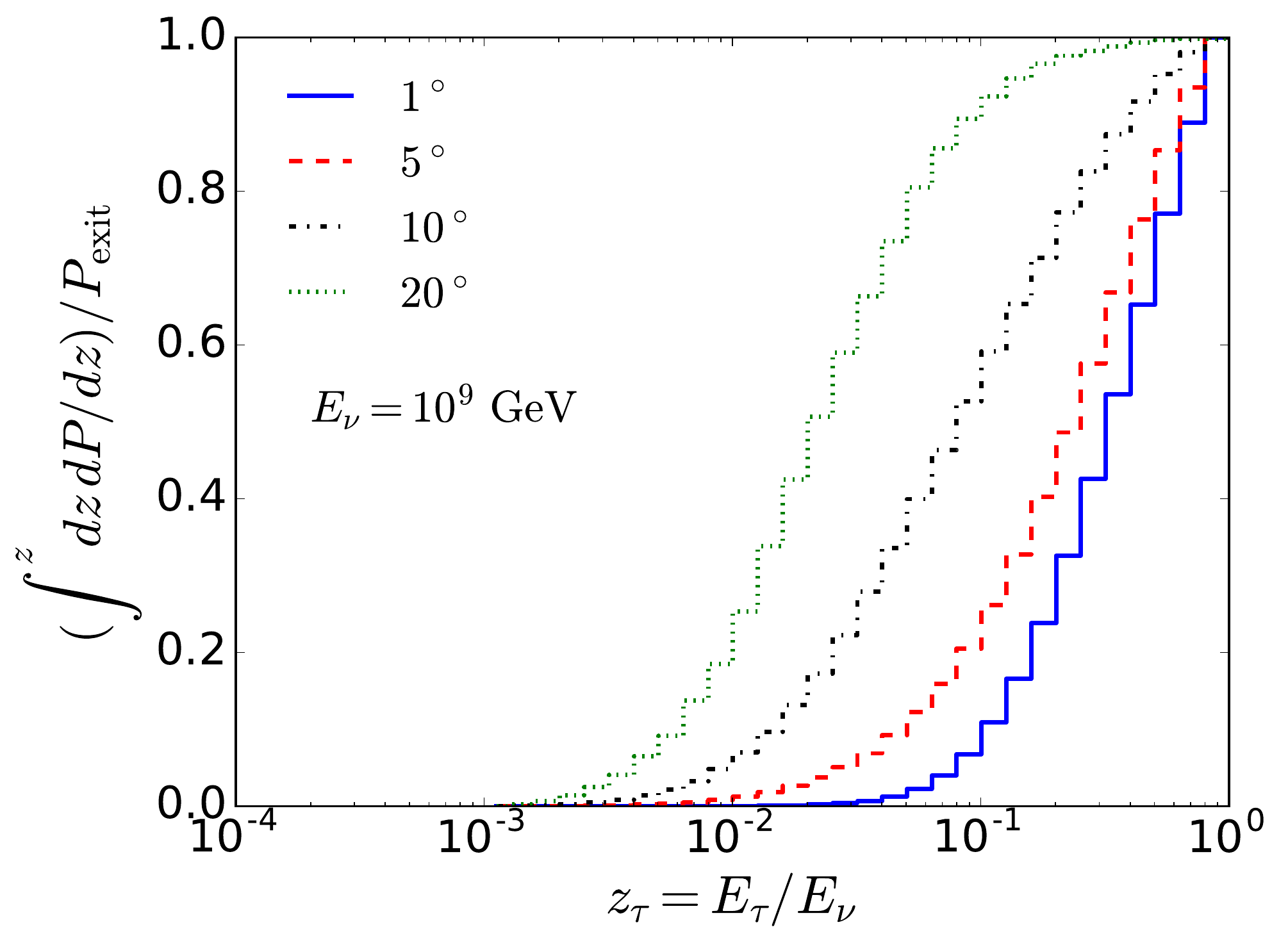}	\end{center}
	\vspace{-0.75 cm}
	\caption{\small For angles $\beta_{\rm th}=1^\circ,\ 5^\circ,\ 10^\circ$ and $20^\circ$, the cumulative distribution function for the relative tau exit probability for $E_{\nu_\tau}=10^9$ GeV, as a function of
	$z=E_\tau/E_{\nu_\tau}$.  The ALLM small-$x$ extrapolation of the electromagnetic structure function in $b_\tau^{\rm nuc}$ has been used.}
	\label{fig:E9-cume}	
\end{figure}

\begin{figure}[t]
	\begin{center}
		\includegraphics[width=0.9\columnwidth]{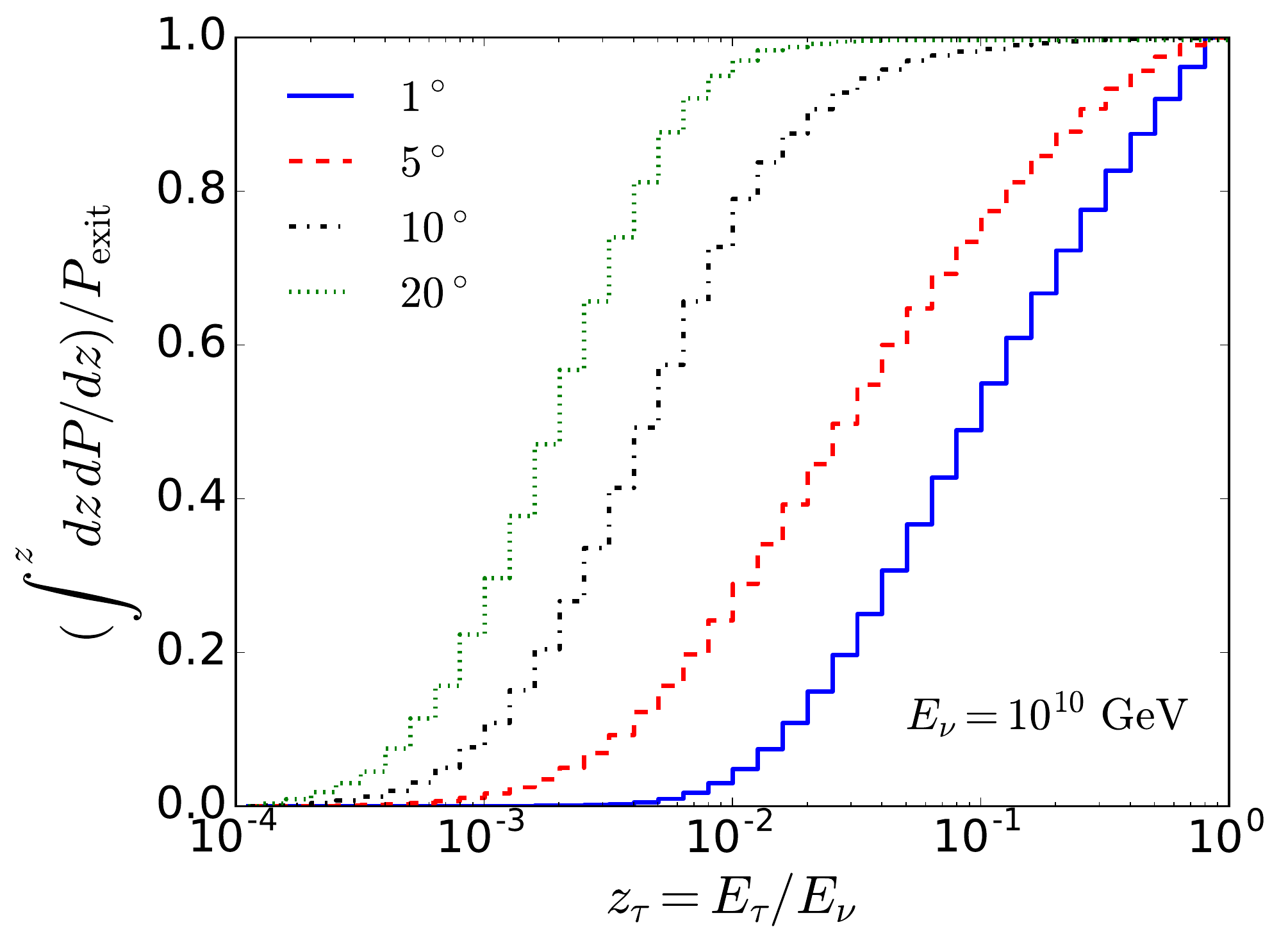}	\end{center}
	\vspace{-0.75 cm}
	\caption{\small For angles $\beta_{\rm th}=1^\circ,\ 5^\circ,\ 10^\circ$ and $20^\circ$, the cumulative distribution function for the relative tau exit probability for $E_{\nu_\tau}=10^{10}$ GeV, as a function of
	$z=E_\tau/E_{\nu_\tau}$.  The ALLM small-$x$ extrapolation of the electromagnetic structure function in $b_\tau^{\rm nuc}$ has been used.}
	\label{fig:E10-cume}	
\end{figure}

Given an exit probability for a given neutrino energy,
the outgoing tau energy
distribution depends on elevation angle $\beta_{\rm tr}$, as shown, for a few energies, in Fig. \ref{fig:gridsN}. In our evaluation of the aperture and sensitivity, instead of the distributions like those in Fig. 
\ref{fig:gridsN}, we use the cumulative distribution functions,
\begin{eqnarray}
\nonumber
    f(E_{\nu_\tau},E_\tau, \beta_{\rm tr})& = &\frac{1}{P_\tau^{\rm exit}(E_{\nu_\tau},\beta_{\rm tr})}\\
    &\times &
    \int_{E_\tau^{\rm min}}^{E_\tau} dE \frac{d P_\tau^{\rm exit}(E_{\nu_\tau},E,\beta_{\rm tr})}{dE}\ .
\end{eqnarray}
Figs. \ref{fig:E7-cume}, \ref{fig:E8-cume}, \ref{fig:E9-cume}, and \ref{fig:E10-cume} show the cumulative
distribution function as a function of the scaled energy
$z_\tau = E_\tau/E_{\nu_\tau}$. Tables \ref{tab:E07-cume}-\ref{tab:E10-cume} list numerical values
for four tau neutrino energies and $\beta_{\rm tr}=1^\circ$, $5^\circ$,
$10^\circ$ and $20^\circ$.

The average exiting tau energy decreases with an increase in $\beta_{\rm tr}$. Multiple interactions and energy loss are responsible for the
shift to a lower exiting tau energy as the incident neutrino energy increases. This means that even though regeneration significantly
enhances the exit probability, the energy of the exiting tau
is lower, so regeneration does not necessarily translate to a better sensitivity. For example, with the POEMMA360 detection characteristics
modeled here, the correction to the sensitivity due to regeneration is at most a $\sim 20\%$ effect. 

\begin{table}[htbp!]
    \centering
    \begin{tabular}{|c|c|c|c|c|}
    \hline
$z_\tau $ & $1^\circ $& $5^\circ$ & $10^\circ $& $20^\circ$  \\ \hline
  1.41e-01 &   3.45e-03 &   0.00e+00 &   0.00e+00 &   2.53e-02\\ \hline
  1.78e-01 &   3.45e-03 &   4.83e-03 &   4.65e-02 &   5.06e-02\\ \hline
  2.24e-01 &   6.90e-03 &   9.66e-03 &   6.98e-02 &   6.33e-02\\ \hline
  2.82e-01 &   1.03e-02 &   1.93e-02 &   1.09e-01 &   8.86e-02\\ \hline
  3.55e-01 &   2.41e-02 &   3.38e-02 &   1.32e-01 &   1.77e-01\\ \hline
  4.47e-01 &   4.13e-02 &   5.31e-02 &   1.63e-01 &   2.78e-01\\ \hline
  5.62e-01 &   1.41e-01 &   1.59e-01 &   2.48e-01 &   3.92e-01\\ \hline
  7.08e-01 &   2.96e-01 &   3.62e-01 &   3.80e-01 &   5.31e-01\\ \hline
  8.91e-01 &   9.96e-01 &   1.00e+00 &   1.00e+00 &   9.99e-01\\ \hline
    \end{tabular}
    \caption{The cumulative distribution function displayed in Fig.
    \ref{fig:E7-cume} for $E_{\nu_\tau}=10^7$ GeV and $\beta_{\rm tr}=1^\circ$, $5^\circ$, $10^\circ$ and $20^\circ$, as a function  of $z_\tau=E_\tau/E_{\nu_\tau}$.}
    \label{tab:E07-cume}
\end{table}

\begin{table}[htbp!]
    \centering
    \begin{tabular}{|c|c|c|c|c|}
    \hline
$z_\tau $ & $1^\circ $& $5^\circ$ & $10^\circ $& $20^\circ$  \\ \hline    1.41e-02 &   0.00e+00 &   0.00e+00 &   1.35e-03 &   0.00e+00\\ \hline 
  1.78e-02 &   1.35e-04 &   0.00e+00 &   2.70e-03 &   2.75e-03\\ \hline
  2.24e-02 &   1.35e-04 &   5.10e-04 &   4.05e-03 &   8.26e-03\\ \hline
  2.82e-02 &   1.35e-04 &   7.65e-04 &   1.08e-02 &   2.21e-02\\ \hline
  3.55e-02 &   1.35e-04 &   3.32e-03 &   1.55e-02 &   3.86e-02\\ \hline
  4.47e-02 &   2.70e-04 &   5.62e-03 &   2.23e-02 &   6.61e-02\\ \hline
  5.62e-02 &   8.10e-04 &   9.70e-03 &   3.18e-02 &   1.07e-01\\ \hline
  7.08e-02 &   2.16e-03 &   1.53e-02 &   4.26e-02 &   1.40e-01\\ \hline
  8.91e-02 &   3.64e-03 &   2.35e-02 &   6.42e-02 &   1.74e-01\\ \hline
  1.12e-01 &   6.07e-03 &   3.42e-02 &   9.53e-02 &   2.42e-01\\ \hline
  1.41e-01 &   9.58e-03 &   5.28e-02 &   1.25e-01 &   3.06e-01\\ \hline
  1.78e-01 &   1.55e-02 &   8.21e-02 &   1.69e-01 &   3.77e-01\\ \hline
  2.24e-01 &   3.01e-02 &   1.11e-01 &   2.26e-01 &   4.71e-01\\ \hline
  2.82e-01 &   4.98e-02 &   1.53e-01 &   2.97e-01 &   5.48e-01\\ \hline 
  3.55e-01 &   9.11e-02 &   2.18e-01 &   3.71e-01 &   6.34e-01\\ \hline 
  4.47e-01 &   1.71e-01 &   3.14e-01 &   4.71e-01 &   7.08e-01\\ \hline 
  5.62e-01 &   3.22e-01 &   4.57e-01 &   5.92e-01 &   8.10e-01\\ \hline 
  7.08e-01 &   5.86e-01 &   6.77e-01 &   7.78e-01 &   8.87e-01\\ \hline 
  8.91e-01 &   1.00e+00 &   9.99e-01 &   9.98e-01 &   9.97e-01\\ \hline 
    \end{tabular}
    \caption{The cumulative distribution function displayed in Fig.
    \ref{fig:E8-cume} for $E_{\nu_\tau}=10^8$ GeV and $\beta_{\rm tr}=1^\circ$, $5^\circ$, $10^\circ$ and $20^\circ$, as a function  of $z_\tau=E_\tau/E_{\nu_\tau}$.}
    \label{tab:E08-cume}
\end{table}

\begin{table}[htbp!]
    \centering
    \begin{tabular}{|c|c|c|c|c|}
    \hline
$z_\tau $ & $1^\circ $& $5^\circ$ & $10^\circ $& $20^\circ$  \\ \hline  
  1.41e-03 &   0.00e+00 &   3.36e-05 &   5.37e-04 &   2.60e-03 \\ \hline
  2.24e-03 &   8.74e-06 &   3.83e-04 &   2.60e-03 &   1.46e-02\\ \hline
  3.55e-03 &   4.23e-05 &   1.26e-03 &   8.33e-03 &   4.06e-02\\ \hline
  5.62e-03 &   1.24e-04 &   3.43e-03 &   2.17e-02 &   9.14e-02\\ \hline
  8.91e-03 &   3.05e-04 &   8.29e-03 &   4.87e-02 &   1.85e-01\\ \hline
  1.41e-02 &   7.71e-04 &   1.82e-02 &   9.69e-02 &   3.39e-01\\ \hline
  2.24e-02 &   2.24e-03 &   3.73e-02 &   1.73e-01 &   5.06e-01\\ \hline
  3.55e-02 &   6.95e-03 &   6.89e-02 &   2.79e-01 &   6.64e-01\\ \hline
  5.62e-02 &   2.28e-02 &   1.22e-01 &   4.00e-01 &   8.05e-01\\ \hline
  8.91e-02 &   6.79e-02 &   2.05e-01 &   5.27e-01 &   8.94e-01\\ \hline
  1.41e-01 &   1.66e-01 &   3.28e-01 &   6.54e-01 &   9.47e-01\\ \hline
  2.24e-01 &   3.26e-01 &   4.86e-01 &   7.73e-01 &   9.76e-01\\ \hline
  3.55e-01 &   5.36e-01 &   6.68e-01 &   8.74e-01 &   9.88e-01\\ \hline
  5.62e-01 &   7.71e-01 &   8.53e-01 &   9.52e-01 &   9.97e-01\\ \hline
  8.91e-01 &   1.00e+00 &   1.00e+00 &   1.00e+00 &   9.98e-01\\ \hline
    \end{tabular}
    \caption{The cumulative distribution function displayed in Fig.
    \ref{fig:E9-cume} for $E_{\nu_\tau}=10^9$ GeV and $\beta_{\rm tr}=1^\circ$, $5^\circ$, $10^\circ$ and $20^\circ$, as a function  of $z_\tau=E_\tau/E_{\nu_\tau}$.}
    \label{tab:E9-cume}
\end{table}

\begin{table}[htbp!]
    \centering
    \begin{tabular}{|c|c|c|c|c|}
    \hline
$z_\tau $ & $1^\circ $& $5^\circ$ & $10^\circ $& $20^\circ$  \\ \hline
  1.41e-04 &   7.86e-07 &   2.45e-05 &   5.88e-04 &   3.09e-03\\ \hline
  2.24e-04 &   6.69e-06 &   3.51e-04 &   4.04e-03 &   1.82e-02\\ \hline
  3.55e-04 &   1.77e-05 &   1.60e-03 &   1.24e-02 &   4.53e-02\\ \hline
  5.62e-04 &   5.46e-05 &   4.41e-03 &   3.13e-02 &   1.15e-01\\ \hline
  8.91e-04 &   1.36e-04 &   1.13e-02 &   7.64e-02 &   2.24e-01\\ \hline
  1.41e-03 &   3.43e-04 &   2.47e-02 &   1.51e-01 &   3.77e-01\\ \hline
  2.24e-03 &   9.75e-04 &   5.02e-02 &   2.67e-01 &   5.68e-01\\ \hline
  3.55e-03 &   2.90e-03 &   9.29e-02 &   4.14e-01 &   7.40e-01\\ \hline
  5.62e-03 &   9.70e-03 &   1.57e-01 &   5.74e-01 &   8.77e-01\\ \hline
  8.91e-03 &   2.99e-02 &   2.42e-01 &   7.28e-01 &   9.50e-01\\ \hline
  1.41e-02 &   7.46e-02 &   3.41e-01 &   8.38e-01 &   9.83e-01\\ \hline
  2.24e-02 &   1.49e-01 &   4.45e-01 &   9.07e-01 &   9.92e-01\\ \hline
  3.55e-02 &   2.50e-01 &   5.49e-01 &   9.47e-01 &   9.96e-01\\ \hline
  5.62e-02 &   3.67e-01 &   6.47e-01 &   9.70e-01 &   9.97e-01\\ \hline
  8.91e-02 &   4.89e-01 &   7.34e-01 &   9.81e-01 &   9.97e-01\\ \hline
  1.41e-01 &   6.10e-01 &   8.12e-01 &   9.90e-01 &   9.97e-01\\ \hline
  2.24e-01 &   7.23e-01 &   8.77e-01 &   9.95e-01 &   9.97e-01\\ \hline
  3.55e-01 &   8.27e-01 &   9.33e-01 &   9.98e-01 &   9.97e-01\\ \hline
  5.62e-01 &   9.20e-01 &   9.75e-01 &   9.99e-01 &   9.97e-01\\ \hline
  8.91e-01 &   1.00e+00 &   1.00e+00 &   1.00e+00 &   9.97e-01\\ \hline
    \end{tabular}
    \caption{The cumulative distribution function displayed in Fig.
    \ref{fig:E10-cume} for $E_{\nu_\tau}=10^{10}$ GeV and $\beta_{\rm tr}=1^\circ$, $5^\circ$, $10^\circ$ and $20^\circ$, as a function  of $z_\tau=E_\tau/E_{\nu_\tau}$.}
    \label{tab:E10-cume}
\end{table}

\section{Detection probability}

In this appendix, we give more details for our evaluation of the detection 
probability in our Monte Carlo computer program.
The probability of detecting the \taon shower depends on the shower energy, the altitude of the tau decay and the detection angle.  In 
the full Monte Carlo simulation calculating the probability that the emerging tau produces a detectable air shower, $p_{\rm det}$, we include the following requirements: 
\begin{enumerate}
\item 
The trajectory of the parent neutrino, assumed to be collinear with the emerging tau lepton, must be appropriately aligned with the line of sight between the detector and the point of emergence.
We require the angle between the trajectory of the neutrino and the line of sight of the detector, $\theta=\lvert \theta_v-\theta_{\rm tr}\rvert$ in Fig. \ref{fig:geo-decay}, to be less than the effective Cherenkov angle, $\theta_{\rm Ch}$.
\item
The tau must decay before it leaves the observation window, the three-dimensional zone that is visible to the detector denoted by the red lines in Figure \ref{fig:geometry-poemma-swin}.
\item  The shower from the tau neutrino must be able to trigger the detector, namely, the number of photoelectrons in the detector generated by light from the shower, $N_{\rm PE}$, must be greater than a threshold value, taken to be $10$ for POEMMA. 
\end{enumerate}
For the purposes of calculation, we model each requirement using a Heaviside function. 

The detection window is determined by the Cherenkov angle, discussed
in Sec. \ref{subsec:air_shower_model}. 
The Cherenkov angle for $E_{\rm shr} = 10^8$ GeV is approximated by the results shown in the lower plot of Fig. \ref{cherenfig}. For showers $N_{\rm PE}\gg N_{\rm PE}^{\rm min}$,
the effective Cherenkov angle is larger than what is shown in the upper panel of Fig. \ref{cherenfig}. The tails of the Cherenkov photon density in the upper panel of Fig. \ref{cherenfig} show that
a Cherenkov angle based on the plateau of the photon density underestimates the width of the Cherenkov signal.
We use an effective Cherenkov angle
\begin{equation}
    \theta_{\rm Ch}={\rm max}
    \Bigl(\theta_{\rm Ch}^0,\theta_{\rm Ch}^0\times \sqrt{2\ln(N_{\rm PE}/N_{\rm PE}^{\rm min})}\,\Bigr)\ .
\end{equation}
This comes from assuming the one-dimensional profile of the shower is approximately Gaussian and scaling by the half-width at $f=N_{\rm PE}/N_{\rm PE}^{\rm min}$ times the maximum. 

The observation window requires the pathlength of the \taon from exit point to decay $s_d$ to be less than $s_{\rm win}$.
To compute $s_{\rm win}$, we consider three cases, where the  
zenith angle $\theta_{\rm tr}$ of the trajectory of the tau lepton is less than, greater than, or equal to the zenith angle $\theta_v$ of the detector line of sight (black line segment labelled ``v'' in Fig. \ref{fig:geometry-poemma-swin}). 
Considering all 
three cases, $s_{\rm win}$ is given by
\[ s_{\rm win} = \left\{ \begin{array}{ll}
                       \sin\left(\alpha_H - \alpha\right){v}/{\sin \xi} & \mbox{if $\theta_{\rm tr} < \theta_{v}$};\\
                       \sin\left(\alpha - \alpha_{\rm min}\right){v}/{\sin \xi} & \mbox{if $\theta_{\rm tr} > \theta_{v}$};\\
                       v & \mbox{if $\theta_{\rm tr} = \theta_{v}$}, \end{array} \right. \]
where $v$ is defined in eq. (\ref{eq:v2}).
The quantity  $\alpha_{\rm min}$ is the minimum nadir angle of the detector viewing zone, and $\xi$ is the angle opposite the detector's line of sight, given by
\[ \xi = \left\{ \begin{array}{ll}
		\pi - \left(\left(\alpha_H - \alpha\right) + \left(\theta_v - \theta_{\rm tr}\right)\right) & \mbox{if $\theta_{\rm tr} < \theta_{v}$};\\
		\pi - \left(\left(\alpha - \alpha_{\rm min}\right) + \left(\theta_{\rm tr} - \theta_v\right)\right) & \mbox{if $\theta_{\rm tr} > \theta_{v}$}. \end{array} \right. \]
The value of $\alpha$ can be found using the law of cosines:
\begin{equation}
\cos\alpha = \frac{2R_{E}h + h^2 +v^2}{2v\left(R_E+h\right)}\ .
\end{equation}

\begin{figure}[htb]
\centering
\vskip 0.1in
\includegraphics[width=0.99\columnwidth]
{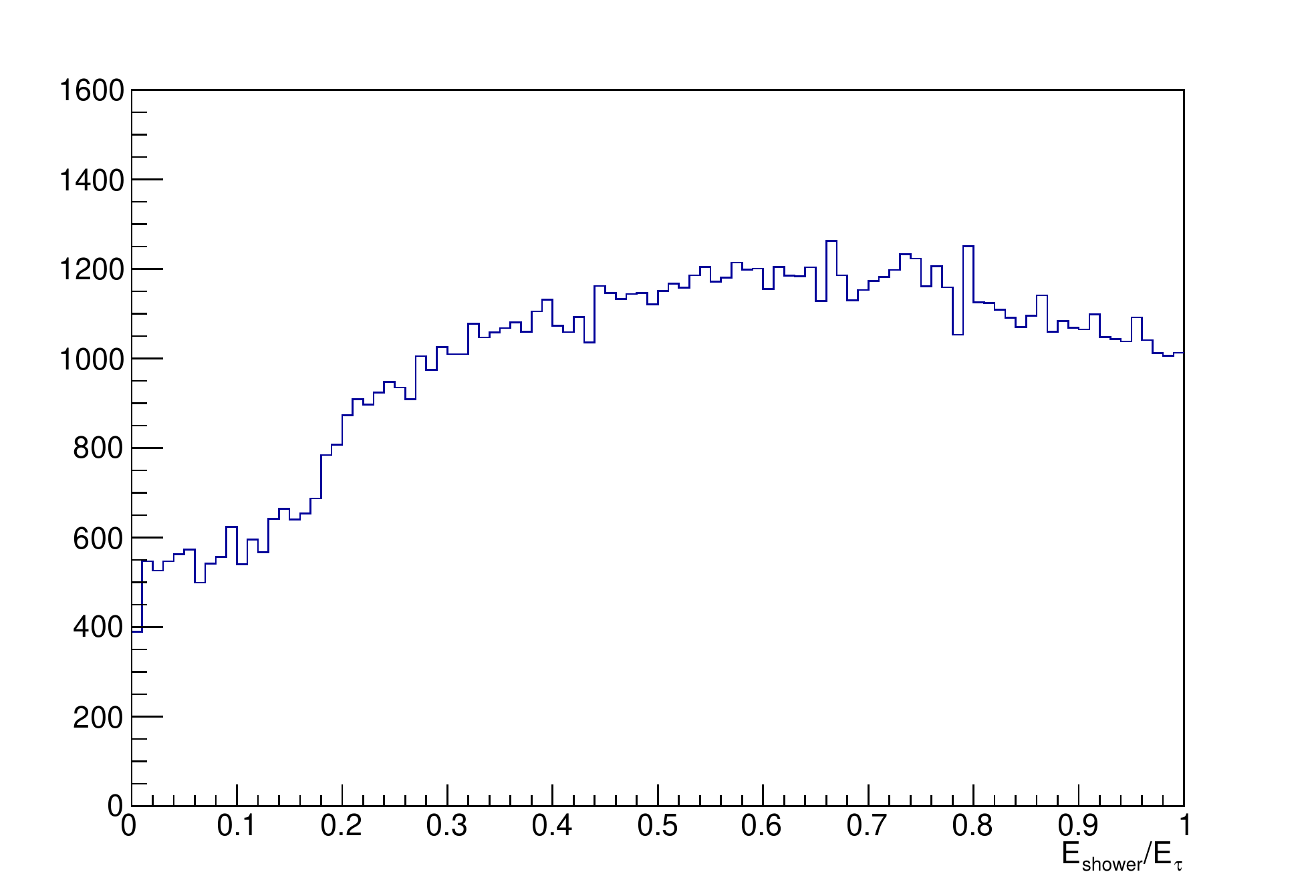}
\caption{\small The frequency as a function of the ratio of $E_{\rm shr}/E_\tau$ for tau decays from a PYTHIA8 simulation.}
\label{fig:eshr}
\end{figure}

As discussed in Sec. \ref{subsec:fluxindep}, the number of photoelectrons detected depends on the photon number density, the
elevation angle $\beta_{\rm tr}$ and the altitude of the decay, related
to $s_d$. In Fig. \ref{fig:eshr}, we show the frequency of a given $E_{\rm shr}/E_\tau$ from a PYTHIA8 
simulation of tau decays. Without the electron decay channel, the 
average energy of the shower is $\sim 0.6 E_\tau$, but including
the electron channel lowers the average to $\sim 0.5 E_\tau$.
The results in this paper 
use the approximation $E_{\rm shr}=E_\tau/2$.
We take $N_{\rm PE}^{\rm min}=10$. 

We integrate eq.~(\ref{apereqn}) via Monte Carlo Integration using importance sampling (also known as the inverse transform method) \cite{PDGMC}. In this sampling method, random variables are drawn from cumulative distribution functions (CDFs) constructed from selected probability density functions (PDFs). Ideally, the selected PDFs would be as similar to the functions being integrated as possible, though normalized, in order to minimize sample variance. In general, for a function, $f\left(x\right)$ being integrated and samples drawn from the selected PDF, $p\left(x\right)$, the Monte Carlo estimator is given by
\begin{equation}
F_{N} = \frac{1}{N}\sum^{N}_{i = 1} \frac{f\left(X_{i}\right)}{p\left(X_{i}\right)}\,,
\end{equation}
where $N$ is the number of samples and $X_{i}$ is the $i^{\rm th}$ drawn random variable. The formula for the Monte Carlo estimator can be verified by taking the expectation of $F_{N}$ over $p\left(x\right)$ over the interval of integration. 

For the full Monte Carlo integration, the integrand of eq.~(\ref{apereqn}) includes $\hat{r} \cdot \hat{n} = \cos \theta_{\rm tr}$. Putting $\hat{r}$ and $\hat{n}$ in a frame in which $v$ points in the $\hat{z}$ direction and taking the dot-product, we find that $\cos \theta_{\rm tr} = \cos \theta_v \cos \delta - \sin \theta_v \sin \delta \sin \phi_p$, where $\theta_v$ is the local zenith angle of the line-of-sight between the spot on the ground and the detector, $\delta$ is the angle between the particle's trajectory and the line-of-sight to the detector, $\phi_p$ is the azimuthal angle of the particle in the frame in which $v$ points in the $\hat{z}$ direction. Then, eq.~(\ref{apereqn}) becomes
\begin{align}
\left<A\Omega\right> &= 2\pi R^2_E \int\!\!\!\! \int\!\!\!\! \int\! \left(\cos \theta_v \cos \delta - \sin \theta_v \sin \delta \sin \phi_p\right) P_{\rm obs} \nonumber \\
& \times d\left(\cos \delta\right)\, d\phi_p\, d\left(\cos \theta_E\right) \,,
\end{align}
where $\theta_E$ is the zenith angle of the position on the surface of the Earth, $R_E$ is the radius of the Earth, and $P_{\rm obs}$ is the observation probability given by eq.~(\ref{eqnpobs}). For the full Monte Carlo integration, the chosen PDF is 
\begin{eqnarray}
\nonumber
p\left(\theta_E,\delta,\phi_p,s\right) &=&\cos \theta_v \cos \delta\,  p_{\rm decay}\left(s\right)\, \\
&\times & d\left(\cos \delta\right)\, d\phi_p\, d\left(\cos \theta_E\right) ds\ ,
\end{eqnarray}
where $p_{\rm decay}\left(s\right)$ is the probability that the $\tau$ decays after traveling a path length $s$. Then, the Monte Carlo estimator is given by
\begin{equation}
F_{N} = \mathcal{N}\frac{1}{N}\sum^{N}_{i = 1} \frac{P_{\rm obs} \cos \theta_{\rm tr}}{p\left(\theta_E,\delta,\phi_p,s\right)}\,,
\end{equation}
where $\mathcal{N}$ is a factor that includes the normalization of $p\left(\theta_E,\delta,\phi_p,s\right)$ and the factor $2\pi R^2_E$.

The POEMMA sensitivity is determined with the full Monte Carlo. 
A simpler formalism in which $\theta_{\rm tr} \rightarrow \theta_v$ and the integration is performed over $d\Omega_{\rm tr}$ yields sensitivities that are reasonably close to the full calculation.

\end{document}